\documentclass[%
 reprint,
amsmath,amssymb,
aps,
pra,
floatfix,
]{revtex4-2}

\usepackage{graphicx}
\usepackage{dcolumn}
\usepackage{bm}
\usepackage{hyperref}
\hypersetup{colorlinks=true, linkcolor=blue, citecolor=blue, filecolor=blue, urlcolor=blue}
\begin{document}

\preprint{APS/123-QED}

\title{Bound dark energy: Particle origin of dark energy with DESI BAO and DES supernova data }

\author{Jose Agustin Lozano Torres}
\email{jalozanotorres@gmail.com}
\author{Axel de la Macorra}%
 \email{a.macorra@fisica.unam.mx}
\affiliation{Instituto de Física, Universidad Nacional Autónoma de México, Ciudad de México, 04510, México
}%

\date{\today}

\begin{abstract}
\noindent The recent findings from the Dark Energy Spectroscopic Instrument (DESI) indicate a preference for dynamical dark energy at a significance level  above $2.5\sigma$, with baryon acoustic oscillation (BAO) combined with cosmic microwave background (CMB) data and Type Ia supernovae (SNe) data, favoring a time-dependent equation of state $w(z)$ rather than the cosmological constant ($w = -1$). We introduce the Bound Dark Energy (BDE) model, in which dark energy arises from the lightest meson field $\phi$ in a dark SU(3) gauge group, developing dynamically through non-perturbative interactions. Governed by an inverse power law potential $V(\phi) = \Lambda_c^{4+2/3}\phi^{-2/3}$, BDE features no dark energy free parameters: one less than $\Lambda$CDM--($\Lambda$) and three less than the $w_0w_a$CDM--($w_0,w_a,\Lambda$) models. By integrating DESI BAO measurements, CMB data and Dark Energy Survey SN Ia distance data collected during the fifth year, BDE demonstrates a reduction of $42\%$ and $37\%$ in the reduced $\chi^2_{\rm BAO}$ as well as lower AIC and BIC values compared to the $w_0w_a$CDM and $\Lambda$CDM models, respectively, while maintaining a comparable fit for both type Ia supernovae and the cosmic microwave background data. Although the ($w_0,w_a$) contour in BDE is 10,000 times smaller than that found in the $w_0w_a$CDM model, the BDE model suggests a dynamical dark energy scenario with precise values of $w_0=$-0.9301 $\pm$ 0.0004 and $w_a=$ -0.8085 $\pm$ 0.0053   while providing a consistency on the six Planck and derived parameters at the 1$\sigma$ level between BDE, $\Lambda$CDM and $w_0w_a$CDM models. The critical parameters—condensation energy scale $\Lambda_c = 43.806 \pm 0.190$ eV and epoch $a_c = (2.497 \pm 0.011) \times 10^{-6}$—are consistent with predictions from high-energy physics. Through a detailed analysis of our results, which are consistent with current observational data, the BDE model effectively elucidates the origins and dynamics of dark energy without free parameters in the dark energy sector, providing an interpretation from contemporary astrophysical experiments within a particle physics framework.
\end{abstract}

\maketitle


\section{\label{Introduction}Introduction}

 The cosmological constant $\Lambda$, responsible for the cosmic acceleration detected by independent studies of distant Type Ia supernovae (SNe) \cite{Riess_1998,Schmidt_1998,Perlmutter_1999}, has been proposed as the most direct candidate for dark energy up to know. For over two decades, the standard cosmological framework, composed of a cosmological constant $\Lambda$ and cold dark matter ($\Lambda$CDM), provides a remarkable data fit to many cosmological probes \cite{Bennett_2013, refId0}. However, the Dark Energy Spectroscopic Instrument (DESI) precise measurements of baryon acoustic oscillations (BAO) combined with \textit{Planck} CMB data and three different samples of Type Ia SNe distance moduli measurements (Pantheon-Plus\cite{Scolnic_2022}, Union3 \cite{rubin2025unionunitycosmology2000} and DESY5 \cite{descollaboration2024darkenergysurveycosmology}), reveal a preference --- though not yet firm evidence --- for dynamical dark energy (DDE), i.e., time-varying dark energy equation of state (DE EoS) parameter $w(z)$, at a statistical level from $2.5\sigma$ to $3.9\sigma$ away from the cosmological constant value ($w_0 \equiv -1$), depending on the combination of SNe data used \cite{desicollaboration2024desi2024vicosmological}. This finding comes from assuming a linear Chevallier-Polarski-Linder (CPL) parametrization $w(a)=w_0+w_a(1-a)$, where $w_0$ is the present-time DE EoS  and $w_a=-(\mathrm{d}w(a)/\mathrm{d}a)|_{a=a_{0}}$ which is the parameter quantifying the dynamical evolution. Given the potential implications that a high statistical preference for models beyond the standard model of cosmology, we present our dark energy model called "Bound Dark Energy" (BDE) presented in \cite{delamacorra2025bounddarkenergyparticles,PhysRevLett.121.161303,PhysRevD.99.103504}, where DE corresponds to the lightest meson field denoted by $\phi$, residing within a supersymmetric SU(3) dark gauge group (DGG) with $N_f=6$ flavors \cite{PhysRevLett.121.161303,PhysRevD.72.043508} and its progressive evolution is described by a scalar field $\phi$ following an inverse-power-law (IPL) potential $V(\phi)$ derived from the Affleck-Dine-Seiberg (ADS) superpotential \cite{AFFLECK1984493}. We also assume that the gauge coupling constant of the DGG is unified with the coupling of the Standard Model (SM) at the unification scale $\Lambda_{\mathrm{GUT}}$. This assumption permits us to reduce the number of free parameters describing the DE sector of our model. For energies below $\Lambda_{\mathrm{GUT}}$ the particles of the DGG and SM interact only through gravity. As the Universe continues expanding and temperature decreases at a rapid rate, the gauge coupling of the DGG becomes strong and the DGG particles form composite states whose masses are proportional to the condensation energy scale $\Lambda_{c}$, akin to the baryons and mesons in the SM, in which BDE model is described in the same way than the Standard Model (SM) gauge groups,  which encompasses the gauge interactions described by SU(3)$\times$SU(2)$\times$U(1) corresponding to the strong, weak and electromagnetic forces. Since the potential and initial conditions are derived quantities in BDE, our DE model predicts the amount of DE today and thus, has one less free parameter than $\Lambda$CDM model ($\rho_{\Lambda}$) and three less than $w_0w_a$CDM model ($\rho_{\mathrm{DE}},w_0,w_a$). Therefore, the BDE model offers a natural explanation of dark energy connecting particle physics and cosmology by proposing that dark energy represents the lightest meson field ($\phi$) within a supersymmetric dark sector \cite{Axel_de_la_Macorra_2003,PhysRevLett.121.161303} with null DE free parameters. 
\noindent By making use of the first-year data release from DESI, in combination with cosmic microwave background (CMB) data \cite{refId0} and Type Ia supernova distance measurements from the fifth year of the Dark Energy Survey program (DESY5) \cite{descollaboration2024darkenergysurveycosmology}—we obtain cosmological constraints on the cosmic key parameters as a consequence of the composite dark energy particle $\phi$. These parameters are compared against the estimations of the $\Lambda$CDM and the $w_0w_a$CDM models, as deeply analyzed by the DESI collaboration \cite{desicollaboration2024desi2024vicosmological}. This comprehensive investigation seeks to illuminate the intricate interplay between these cosmological scenarios, forging a path toward a deeper understanding of dark energy's elusive nature.

This paper is organized as follows. In Section \ref{Theory}, we present the theoretical foundations of the BDE model. We expose the origins of the composite particle $\phi$, providing a thorough derivation of its properties and exploring its implications for cosmological evolution across various epochs. Section \ref{methodology} details the data sources and methodologies employed in this analysis, including a description of the observational datasets from the Dark Energy Spectroscopic Instrument (DESI), Baryon Acoustic Oscillations (BAO), the Dark Energy Supernova Survey (DESY5), and cosmic microwave background (CMB) measurements. In Section \ref{results}, we present the cosmological constraints imposed on the BDE model derived from these observations. We analyze how these constraints shape our understanding of dark energy, its influence in the cosmic evolution of the universe and differences from $w_0w_a$CDM and $\Lambda$CDM models. Finally, in Section \ref{conclusions}, we summarize our primary findings and conclusions of our results.

\vspace{-0.5cm}
\section{Theory}\label{Theory}

The BDE model assumes that a lightest meson field, $\phi$, emerged dynamically and drives the late-time acceleration of the universe. This emergence is driven due to non-perturbative dynamics of the supersymmetric dark gauge group $SU(N_c=3)$ with $N_f=6$ flavors, at a strong  gauge coupling regime occurring at low energies \cite{PhysRevLett.121.161303}. As a result, a phase transition occurs at the condensation scale $\Lambda_c$ and scale factor $a_{c}$, where above $\Lambda_{c}$ the fundamental particles of the DGG are massless and below $\Lambda_{c}$ the original fundamental particles bind together forming neutral composite states e.g. dark mesons, as occurs in QCD theory, where mesons and baryons are generated from underlying quark constituents. The energy density evolution of the lightest meson field, denoted by $\phi$, is governed by a non-perturbative inverse-power-law (IPL) scalar potential $V(\phi)$, generated within a strong gauge coupling regime in a supersymmetric gauge group \cite{BURGESS1997181,AFFLECK1984493}. The resulting scalar potential $V(\phi)$ is described by an inverse law potential \cite{PhysRevD.72.043508,PhysRevLett.121.161303}:
\begin{equation}
V(\phi)=\Lambda_{c}^{4+n}\phi^{-n}=\Lambda_{c}^{4+2/3}\phi^{-2/3},
\end{equation}
\noindent for a  $SU(N_c=3)$ supersymmetric gauge group with particle content $N_f=6$ giving an inverse power law exponential potential  $n=2(N_c-N_f)/(N_c+N_f)=2/3$ for our BDE model \cite{2003JHEP...01..033D, PhysRevLett.121.161303}. The condensation energy scale $\Lambda_c$ determines the scale  where the $SU(3)$ gauge coupling constant becomes strong and neutral composite particles are dynamically formed. It is given by 
\begin{equation}
    \Lambda_c=\Lambda_{\mathrm{gut}} e^{-8\pi^2/(b_0 g _{\mathrm{gut}}^2)} = 34^{+16}_{-11}\mathrm{eV}
\end{equation}
with $b_0=3N_c-N_f=3$,  $g^2_{gut}= 4\pi/(25.83\pm 0.16)$  and  $\Lambda_{\mathrm{gut}}=(1.05\pm 0.07)\times 10^{16}$ GeV the strength of the coupling constant and the unification scale, respectively as given by \cite{Bourilkov_2015,ParticleDataGroup:2024cfk,PhysRevD.110.030001}. The condensation energy scale $\Lambda_{c}$ is no longer a free parameter in BDE model. It is noticed that the uncertainties in $\Lambda_{c}$ are due to a lack of high precision estimates of $\Lambda_{\mathrm{GUT}}$ and $g^{2}_{\mathrm{GUT}}$, derived from uncertainties of the QCD gauge coupling constant \cite{Bourilkov_2015}. This framework parallels the established SM of particle physics, encapsulated by $SU_{\text{QCD}}(N_{c}=3) \times SU(N_{c}=2)_{L}\times U_{Y}(N_{c}=1) $, which incorporates three generations of particles and dictates the strong, weak and electromagnetic interactions. We emphasize that the choice of gauge groups and particle content in the SM specify the theoretical constructs but do not arise from any underlying fundamental theory, setting an equivalent principle between BDE and the SM. The BDE energy density before the condensation scale ($a \leq a_{c} $) consists of relativistic particles and can be expressed as $\rho_{\mathrm{DE}}(a) = 3(a_{c} \Lambda_{c})^{4} a^{-4}$, before the phase transition at $a_c$. This equation encompasses two fundamental aspects of the BDE model: the energy scale $\Lambda_{c}$ and the scale factor \( a_{c} \), where condensation occurs. For $a<a_c$ the energy density of the DGG is expressed as: $\rho_\mathrm{DG}(a)=\rho_\mathrm{DG}(a_c)(a_c/a)^{4} = 3\Lambda_c^4  (a_c/a)^{4}$, with $\rho_\textrm{DG}(a_c) =3\Lambda_c^4$. Therefore we have $\rho_\mathrm{DG}(a_c)/\rho_r(a_c)=3\Lambda_c^3/(\rho_{r0}a_c^{-4})=3(a_c\Lambda_c)^4/\rho_{r0}$, where $\rho_{r}$ accounts for relativistic particles with the present-day radiation energy density $\rho_{r0}=(\pi^{2}/15)g_{r}T^{4}_{0}$, $T_{0}=2.7255 \pm 0.0006 \textbf{ }\mathrm{K}$ \cite{Fixsen_2009}. Solving for $a_c\Lambda_c$, we get the constraint equation \cite{PhysRevD.99.103504,PhysRevLett.121.161303}:
\begin{align}\label{eq:bde_acLc_theory}
  \frac{a_c\Lambda_c}{\textrm{eV}} & = 1.0939 \times 10^{-4}.
\end{align}
\noindent The present-day radiation energy density is given by $\rho_{r0}=(\pi^{2}/15)g_{r}T^{4}_{0}$ and FIRAS measures $T_{0}=2.7255 \pm 0.0006 \textbf{ }\mathrm{K}$. The relative uncertainty in $\delta T_{0}/T_{0}=0.0006/2.7255=0.00022$. Since $\rho_{r0}\propto T^{4}_{0}$, the relative uncertainty in $\rho_{r0}$ is given by $\delta \rho_{r0}/\rho_{r0}=4(\delta T_{0}/T_{0})=0.00088$. From Eq. \ref{eq:bde_acLc_theory}, $a_{c}\Lambda_{c}\propto \rho_{r0}^{1/4}$, hence $\delta (a_{c}\Lambda_{c})/a_{c}\Lambda_{c}=1/4(\delta \rho_{r0}/\rho_{r0})=0.00022 \rightarrow \delta(a_{c}\Lambda_{c})=(1.0939\times10^{-4}) \cdot0.00022=2.40\times10^{-8}\text{ } \mathrm{eV} $. From the above, we see that FIRAS temperature uncertainty introduces a very tiny correction ($0.022\%$) to the $\Lambda_{c}-a_{c}$ relation, but it is irrelevant for our analysis. Therefore, we assume a fixed CMB temperature in the $a_{c}\Lambda_{c}$ relation described in Eq. \ref{eq:bde_acLc_theory}. At the condensation transition ($a_{c} $) the strong DG forms composite (bound) particles (similar as mesons in QCD) and its further evolution is described by a canonical scalar field whose density and pressure are given by the quintessence terms \cite{PhysRevLett.121.161303}
\begin{equation}
    \rho_{\mathrm{BDE}} = \dot{\phi}^{2}/2 + V(\phi),\quad  P_{\mathrm{BDE}} = \dot{\phi}^{2}/2-V(\phi).
\end{equation}
The DE EoS of BDE, $w_{\mathrm{DE}}$, is time-dependent and influenced by the weight of both kinetic $\dot{\phi}^{2}/2 $ and potential $V(\phi)$ terms. After passing through the condensation scale $a\geq a_{c}$, the dynamical evolution of the scalar field adheres to a Friedmann-Lemaître-Robertson-Walker (FLRW) metric and is determined by the Klein-Gordon equation
\begin{equation}
    \ddot{\phi} + 3H\dot{\phi} + \frac{\mathrm{d}V}{\mathrm{d}\phi} = 0,
\end{equation}
with the Hubble parameter defined as $H \equiv \dot{a}/{a} = \sqrt{8\pi G \rho_{\mathrm{tot}}/3}$, where the total amount of energy density of the universe is found to be $\rho_{\mathrm{tot}}(a)= \rho_{m0} a^{-3} + \rho_{r0} a^{-4} + \rho_{\mathrm{BDE}}$. From the above, the gauge coupling unification between the BDE and the SM gauge groups allow the BDE model to discard free parameters in the DE sectors, e.g., avoiding to introduce extra parameters compared to CPL or alternative parameterizations. Furthermore, all of the quantities that BDE introduced such as $\Lambda_{c}$, $a_{c}$, $n$ describe the aspects of the DGG and they are all fixed quantities without the opportunity to be varied. As we will discuss during the upcoming sections, our BDE  model provides a concise understanding of the origin and dynamics of dark energy based on gauge theories, with an excellent agreement with current DESI observations. Our BDE model does not provide predictions on other important cosmological parameters, including the specific quantities of dark matter and various exotic particles. These aspects are considered to be outside the intended scope of our research focused on dark energy, which aims to address the dynamics of the universe's expansion rather than the composition of its matter.

\section{Data and Methodology}\label{methodology}
The cosmological data employed to investigate the cosmological implications and parameter constraints of BDE, $\Lambda$CDM, and $w_0w_a$ CDM models come from a variety of state-of-the-art astrophysical experiments: BAO measurements from the first year of data release of the Dark Energy Spectroscopic Instrument (DESI) survey \cite{desicollaboration2024desi2024vicosmological}, the power spectra of temperature anisotropies (TT) and polarization (EE) auto-spectra, as well as their cross-spectra (TE), which were integrated through advanced likelihood analyses implemented in software packages such as \texttt{simall}, \texttt{Commander} (for multipoles with $\ell < 30$), and \texttt{plik} (for $\ell \ge 30$), all derived from the  \textit{Planck} cosmic microwave background (CMB) satellite data release \cite{refId0}. Moreover, our analysis was further enriched by distance measurements of Type Ia supernovae of the fifth year from the Dark Energy Survey (DESY5) \cite{descollaboration2024darkenergysurveycosmology}. This dataset, part of their Year 5 data release, contains 1829 SNe Ia, encompasses a remarkable spectrum of distances—of which 1635 are photometrically-classified objects characterized in the redshift range from $0.1$ to $1.3$, complemented with 194 low-redshift SNe Ia in the redshift range from $0.024$ to 0.1. To analyze and obtain the constraints on the cosmological parameters of the cosmological scenarios under study, we employed an advanced Markov Chain Monte Carlo (MCMC) analysis technique \cite{Lewis_2002}, using the cosmological codes CAMB \cite{Lewis_2000} and CosmoMC \cite{Lewis_2002} and modifying such codes to implement our BDE model. The Metropolis-Hastings MCMC sampler was leveraged to run four independent chains in parallel for each specific dataset and model combination, initiating the process with proposal covariance matrices derived from preliminary simulation runs. The chains were persistently executed until they satisfied the default convergence criteria of Gelman-Rubin statistic $R - 1 < 0.01$, thus ensuring the robustness and statistical reliability of the results obtained. The parameters that are left free in our MCMC runs are the usual 6 base-$\Lambda$CDM parameters $\{\Omega_{b}h^{2},\Omega_{c}h^{2},\tau,n_{s},100\theta_{\mathrm{MC}},\mathrm{ln}(10^{10}A_{s})  \}$  plus the condensation energy scale $\Lambda_{c}$ of the BDE model. The priors used are those the same that were set by DESI collaboration (see Table 2 in \cite{desicollaboration2024desi2024vicosmological}) and for $\Lambda_{c}$ the prior used is $\mathcal{U}[20,100]$. To present the confidence intervals and likelihoods distributions of the cosmological parameters from our MCMC chains, we use the Python package \textsc{getdist} \cite{2019arXiv191013970L} code. Furthermore, we obtained the best-fit values for each cosmological model utilizing the \texttt{Powell's 2009 BOBYQA} bounded minimization routine, as integrated into \texttt{CosmoMC}. This process involved executing four independent minimizations from various random starting points to ensure convergence and significantly enhance the reliability and integrity of our results.
\begin{figure*}
\hspace{-0.8cm}
  \begin{tabular}{c@{\hspace{0.01em}}c@{\hspace{0.01em}}c@{\hspace{0.01em}}c@{\hspace{0.01em}}c@{\hspace{0.01em}}c}
      \includegraphics[width=19em]{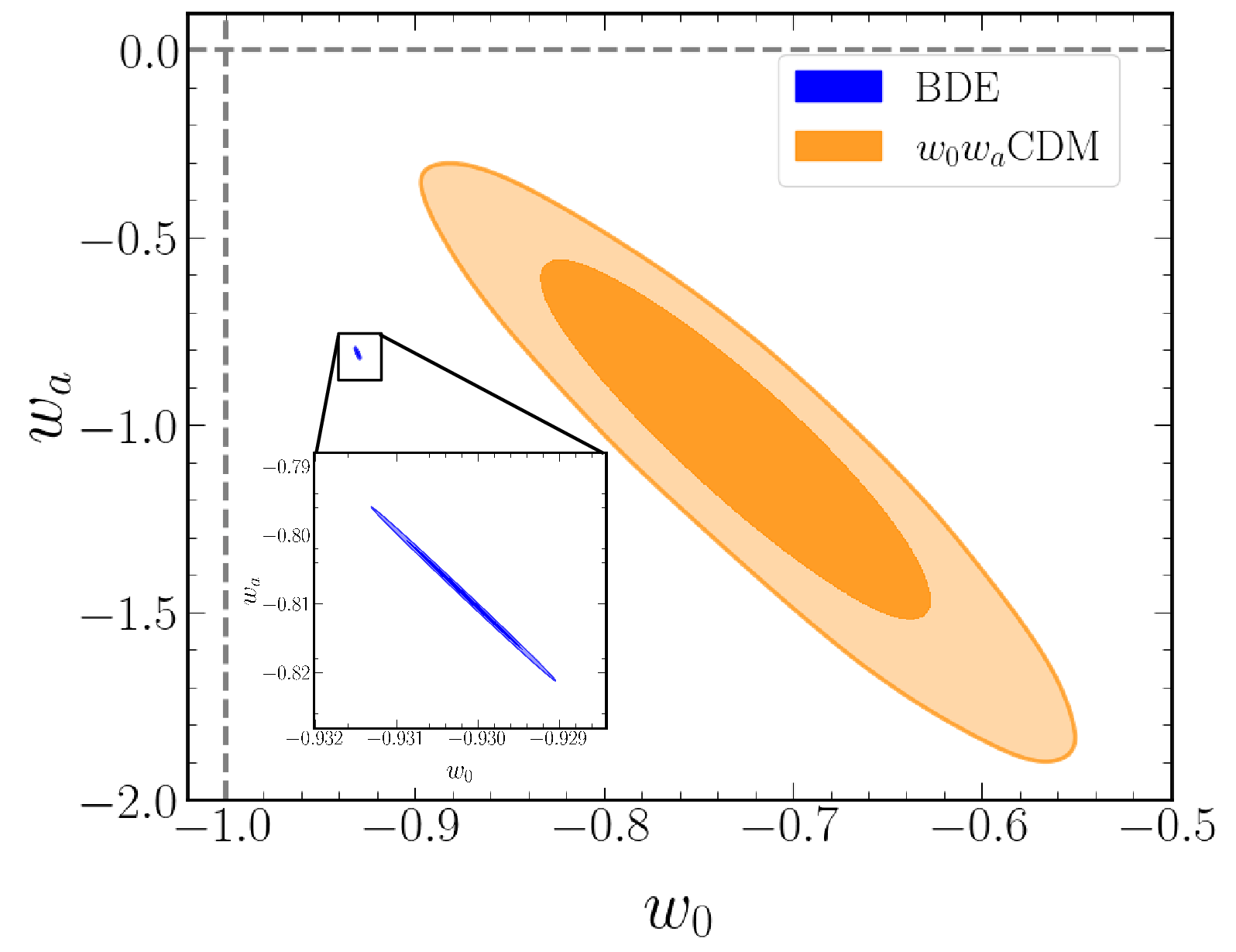} & 
      \includegraphics[width=19em]{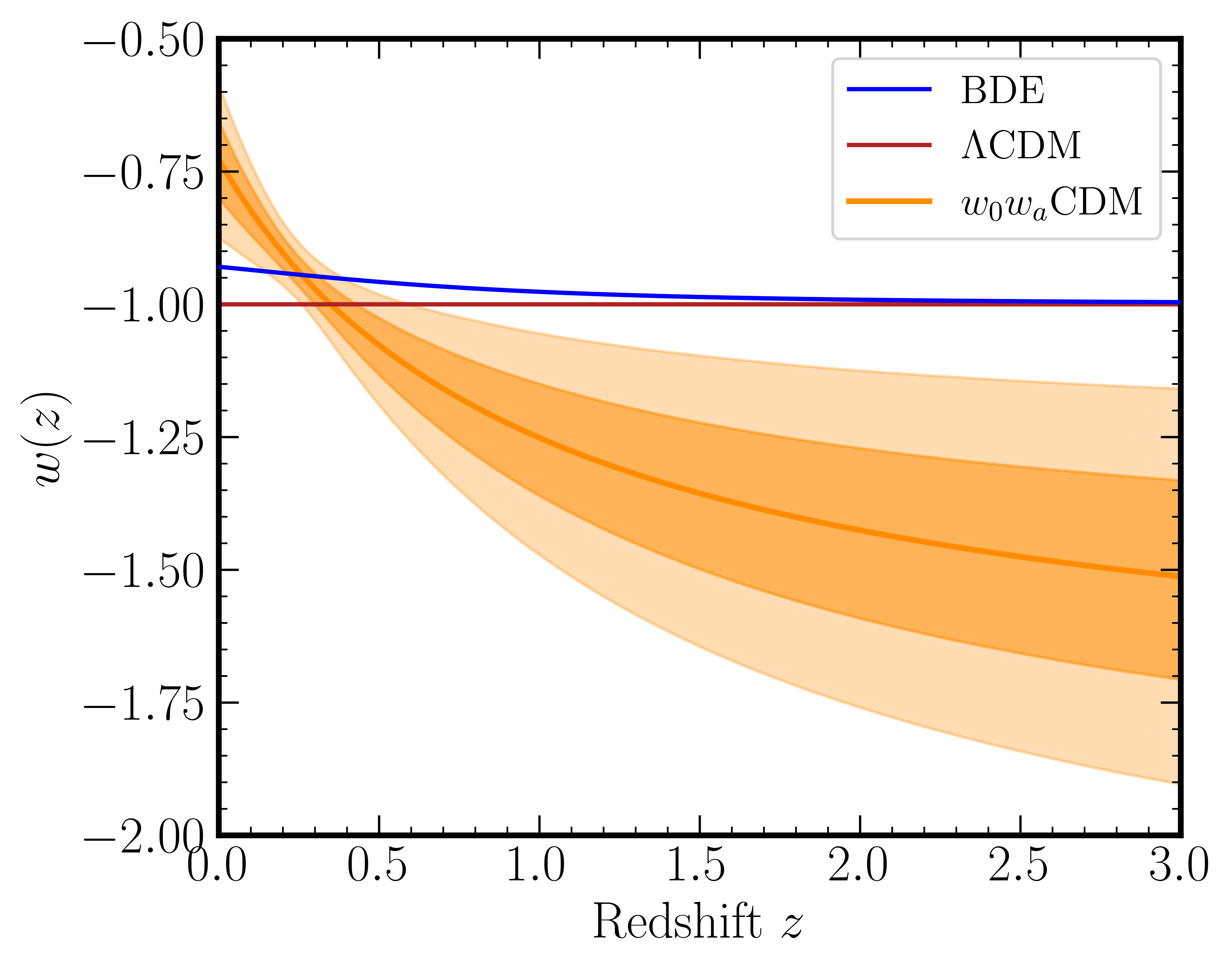}& 
      \includegraphics[width=19em]{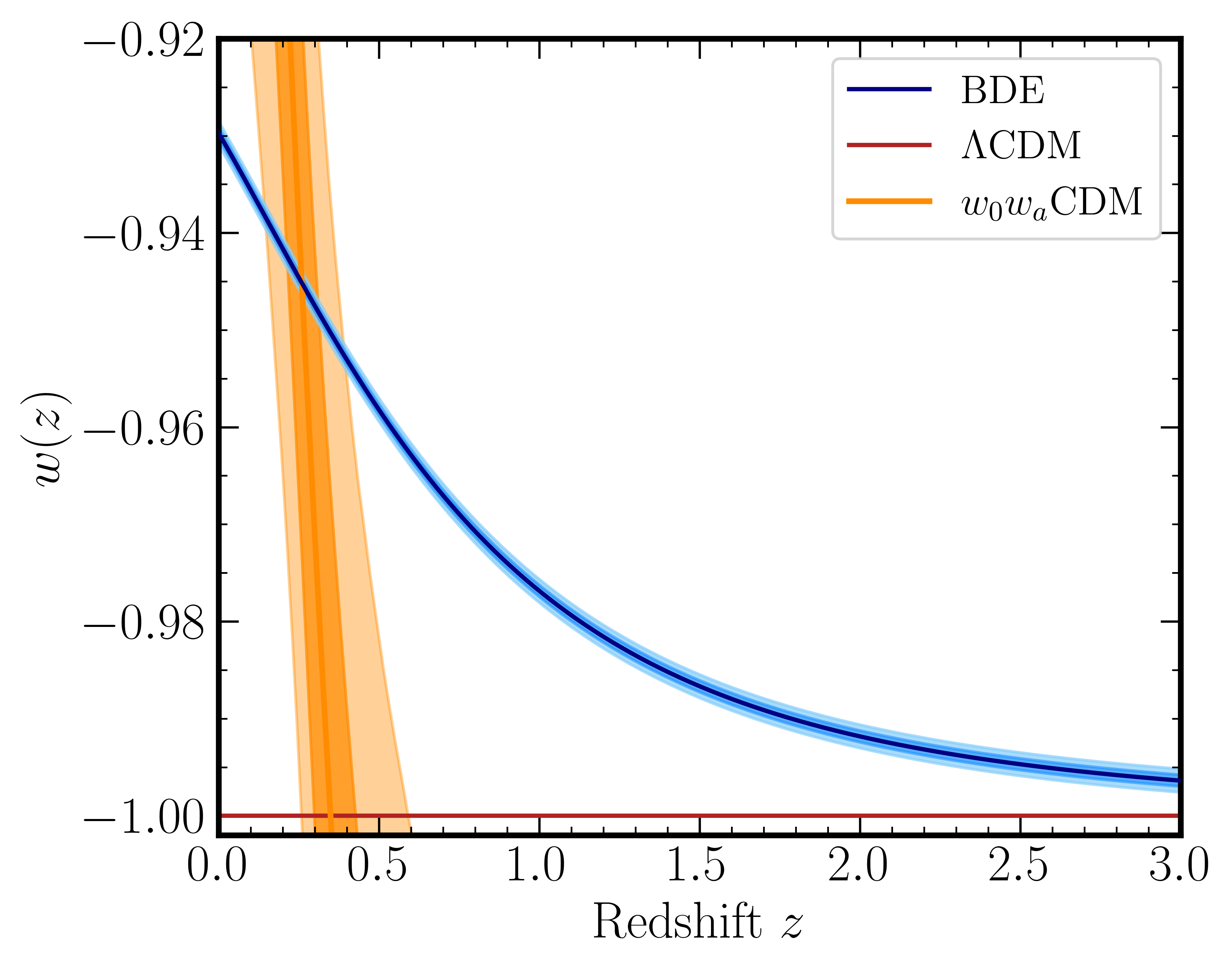} & \\
      \includegraphics[width=19em]{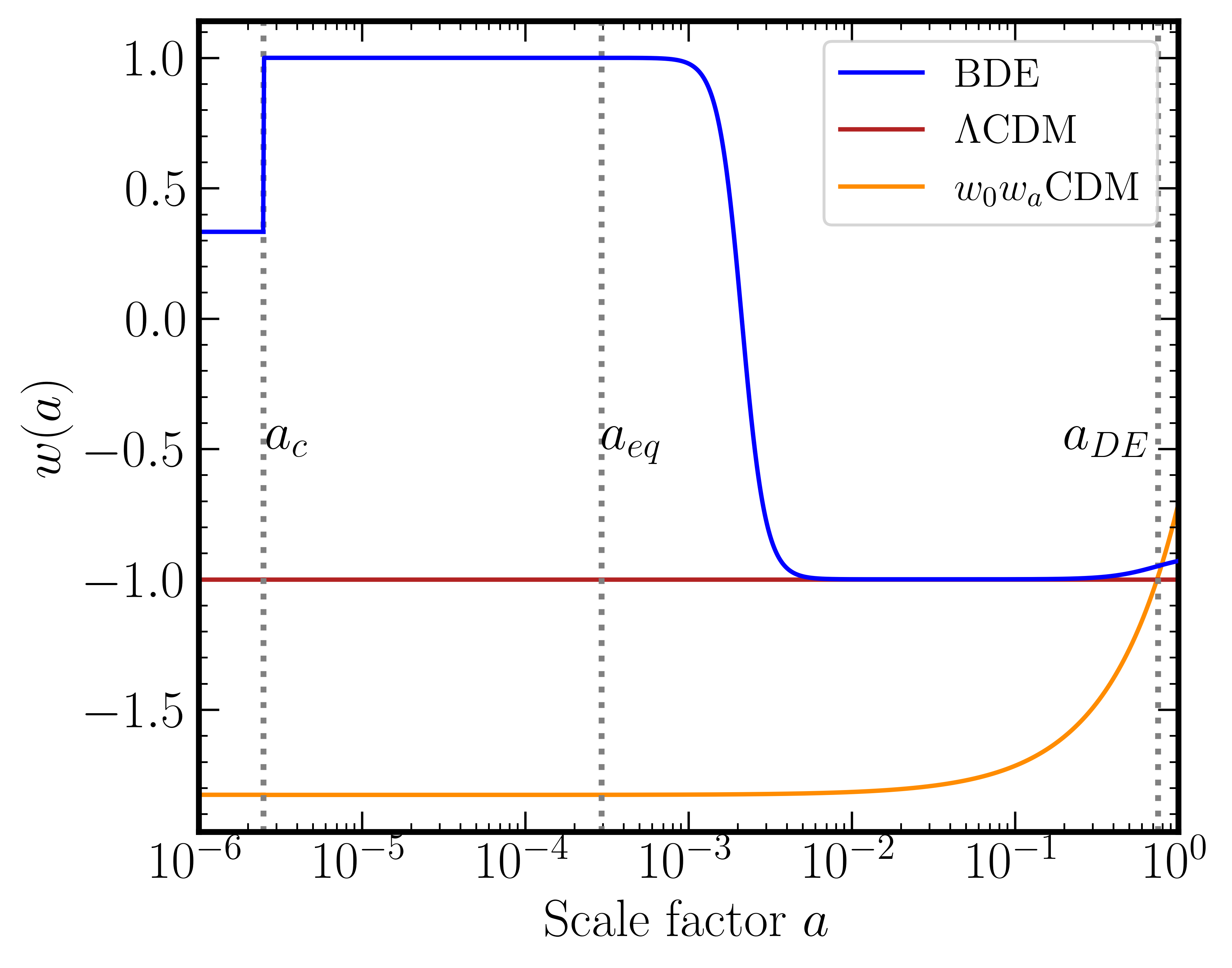} & 
      \includegraphics[width=19em]{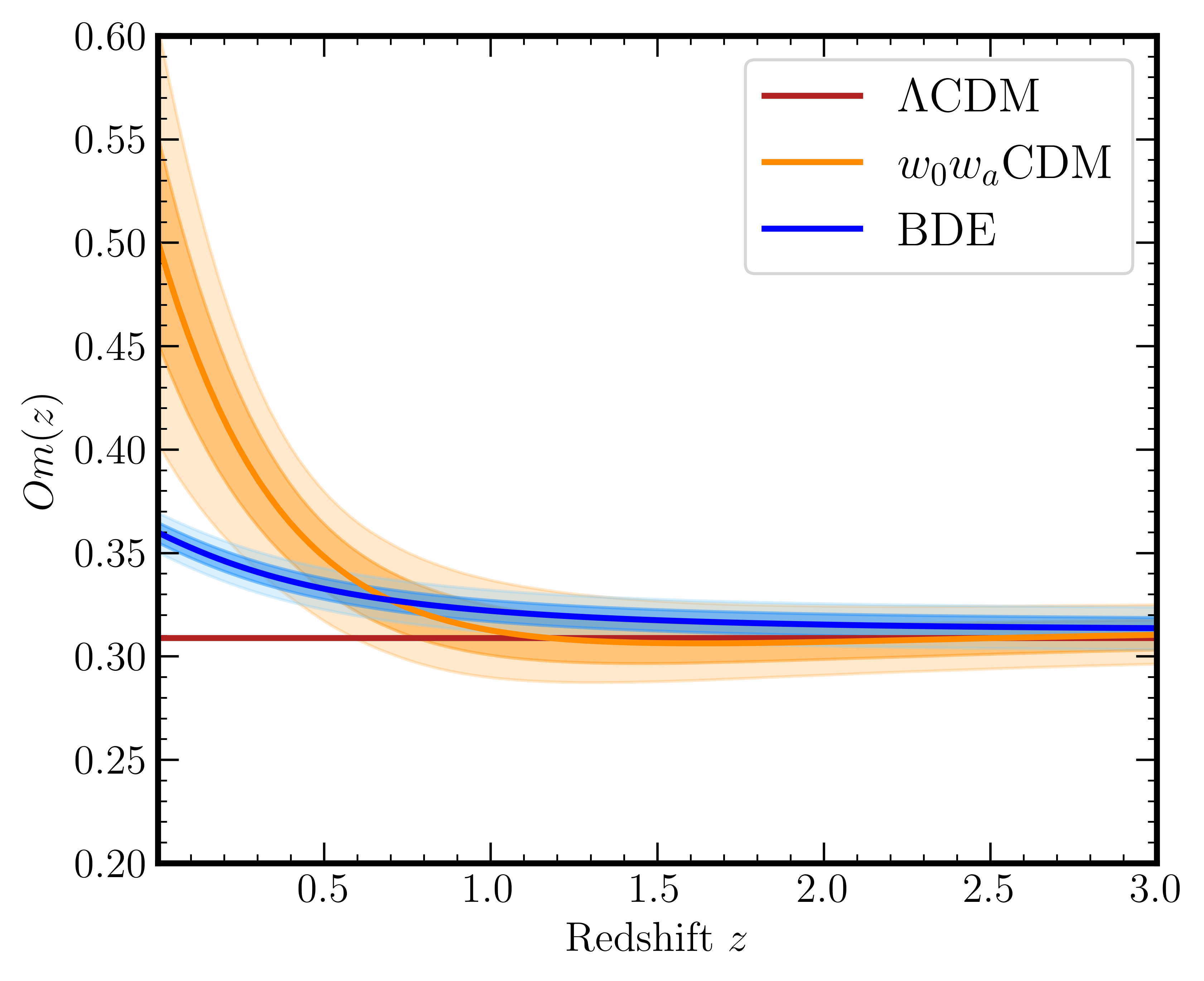} & 
      \includegraphics[width=19em]{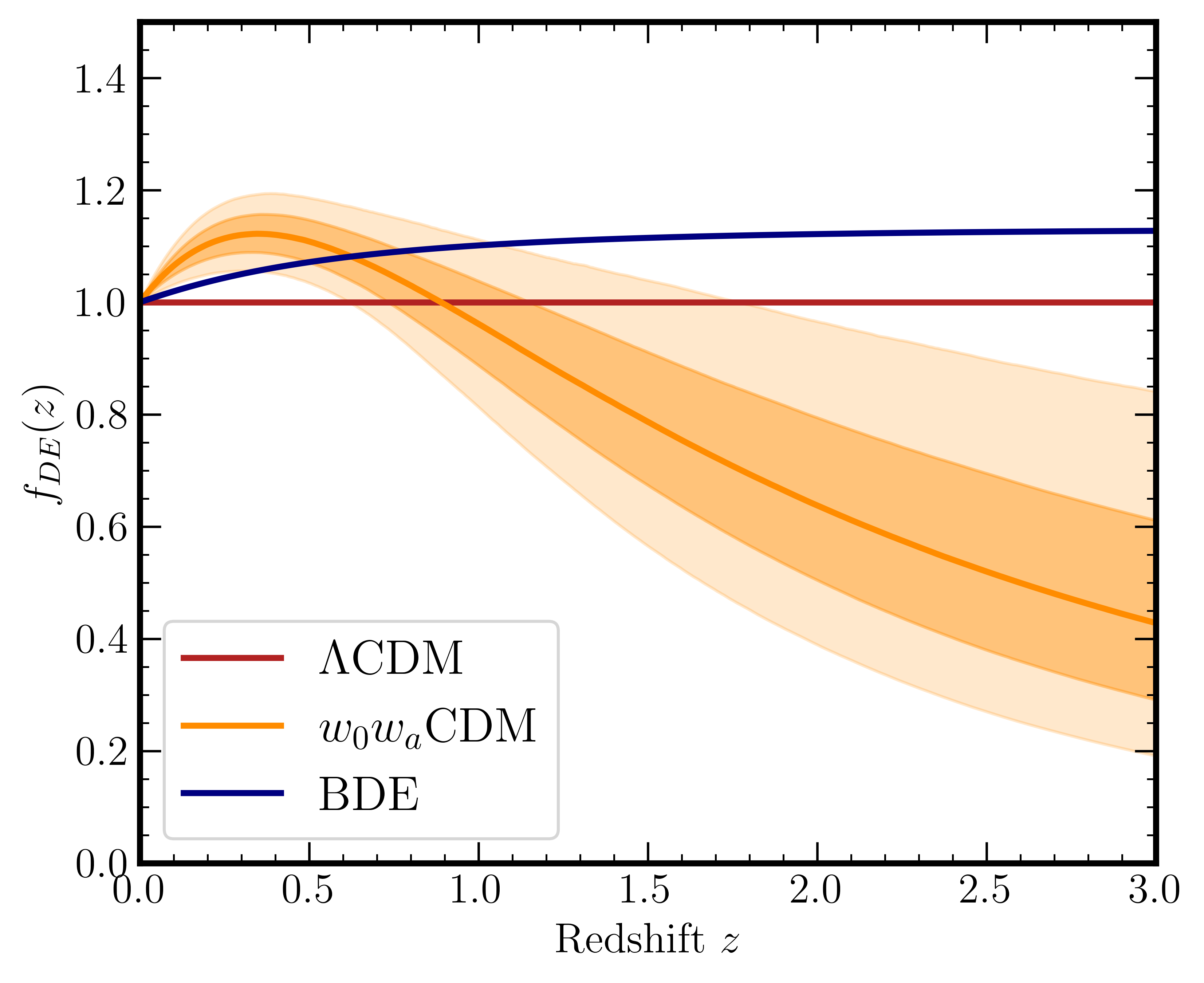} 
  \end{tabular}
  \vspace*{8pt}
  \caption{\label{Fig:1} \textit{Upper-left}: Marginalized posterior constraints in the $w_0-w_a$ plane are shown at $68\%$ and $95\%$ confidence levels for the BDE and $w_0w_a$CDM model models. These results stem from analyzing DESI BAO data alongside CMB and DES Supernova 5-Year (DESY5) datasets. The plot indicate $w_0 > -1$ and $w_a < 0$, highlighting a notable disagreement with the $\Lambda$CDM model above $2.5\sigma$. \textit{Upper-middle}: The equations of state (EoS) of $w_0w_a$CDM model (in orange) and BDE model (in blue). Contours around the best-fit lines illustrate the $68\%$ and $95\%$ confidence intervals, with the $\Lambda$CDM limit marked by a solid red line.  \textit{Upper-right}: Zoom in the EoS of BDE and $w_0w_a$CDM models. \textit{Lower-left}: Evolution of the EoS for BDE, $w_0w_a$CDM, $\Lambda$CDM models from early times ($a\sim 10^{-6}$) to present time $a=1$. \textit{Lower-middle}: Evolution of the $Om(z)$ diagnostic parameter as function of redshift in the BDE, $w_0w_a$CDM and $\Lambda$CDM models. \textit{Lower-right}: The normalized dark energy density $f_{\mathrm{DE}}(z)=\rho_{DE}(z)/\rho_{DE}(0)$ as a function of redshift in the BDE, $w_0w_a$CDM and $\Lambda$CDM models.  }
\end{figure*}
\section{Analysis of Cosmological Results}\label{results}

In this section, we present the observational constraints and the cosmological implications on the three DE models considered in this work. We discuss the results and make a comparison between the DE models. It must be remarked that we do not aim to restrict the $\Lambda$CDM or $w_0w_a$CDM model, as they are fundamentally different from the BDE model that we study here. Instead, our BDE model makes specific predictions without free parameters describing the DE sector and evaluates these predictions in comparison to both the $\Lambda$CDM and $w_0w_a$CDM models. Therefore, we highlight the distinctions of the given cosmological scenarios with either null free dark energy parameters (BDE) or additional dark energy free parameters ($\Lambda$CDM, $w_0w_a$CDM) along with their implications that can have.
\begin{figure*}
\includegraphics[width=\textwidth]{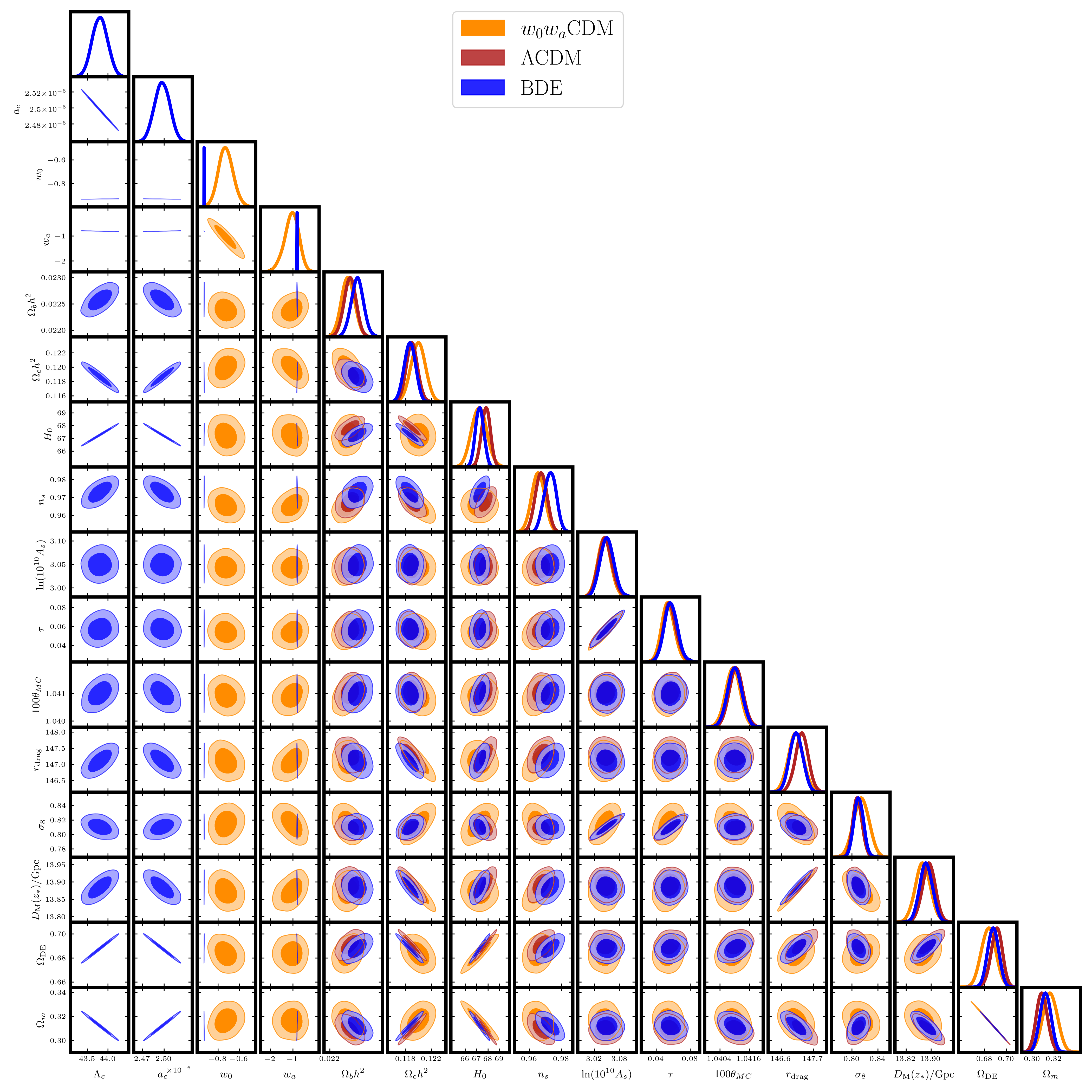}
\caption{\label{Fig:2} Marginalized distributions and $68\%$ and $95\%$ confidence contours of the six baseline \textit{Planck} and derived cosmological parameters according to the BDE (red), $\Lambda$CDM (green), and $w_0w_a$CDM models from the joint analysis of DESI DR1 BAO \cite{desicollaboration2024desi2024vicosmological}, CMB\cite{refId0}, and DESY5 \cite{descollaboration2024darkenergysurveycosmology}. }
\end{figure*}
\begin{table*}
\caption{\label{tab: Results of some parameters in the BDE, CPL, LCDM for the full dataset} The table shows some key cosmological parameter results from the joint analysis of the DESI DR1 BAO data \cite{desicollaboration2024desi2024vicosmological}, the full CMB dataset \cite{refId0} and the Dark Energy Survey SNeIa compilation from the fifth year \cite{descollaboration2024darkenergysurveycosmology} for BDE, $\Lambda$CDM, and $w_0w_a$CDM cosmological models. The results presented are the best fit, the marginalized means, and $68\%$ intervals in each case.}
\begin{ruledtabular}
\begin{tabular}{lp{0.4in}lllll}
& \multicolumn{2}{c}{BDE} & \multicolumn{2}{c}{$w_0w_a$CDM} & \multicolumn{2}{c}{$\Lambda$CDM}\\
 Parameters & Best fit & $68\%$ limits & Best fit & $68\%$ limits & Best fit & $68\%$ limits\\ \hline
 $10^{6} a_{c}$     & 2.486     & 2.497    $\pm$ 0.011     &    ---     &          ---          & ---   &  ---  \\
 $\Lambda_{c}$[eV ] & 43.998    & 43.806   $\pm$ 0.190     &    ---     &          ---          &  ---  & ---    \\
 $\Omega_{\mathrm{BDE}}(a_{c})$ &  0.1117   & 0.11172 $\pm$ 0.00001     &    ---     &          ---   &  ---  & ---    \\
 $\Omega_{b}h^2$    &   0.02268  &  0.02258  $\pm$ 0.00013     &   0.02240 &  0.02238  $\pm$ 0.00014     &   0.02242    & 0.02242 $\pm$ 0.00013    \\
 $\Omega_{c}h^2$    &   0.1178  &  0.1186  $\pm$ 0.0008    &   0.1200      &   0.1199    $\pm$ 0.0011       &   0.1188    & 0.1190 $\pm$ 0.0008    \\
 $\mathrm{ln}(10^{10}A_{s})$    &  3.115   & 3.050   $\pm$ 0.016   &  3.067 &    3.043 $\pm$ 0.015  &  3.042     & 3.045 $\pm$ 0.016    \\
 $\tau$             &  0.0887   &  0.0573  $\pm$ 0.0081    &  0.0634  &   0.0542 $\pm$ 0.0078          &   0.0561    & 0.0562 $\pm$ 0.0079    \\
 $n_{s}$             &   0.9759  & 0.9729   $\pm$ 0.0037     &  0.9660 &  0.9656  $\pm$ 0.0040        &  0.9684   & 0.9671 $\pm$ 0.0036    \\
 $100\theta_{MC}$   & 1.04117  & 1.04101 $\pm$ 0.00029  & 1.04094    & 1.04093 $\pm$ 0.00029 & 1.04110  & 104106 $\pm$ 0.00028\\
 $w_{0}$            & -0.9300  &-0.9301  $\pm$ 0.0004   & -0.7139    & -0.724 $\pm$ 0.071     &  -1      &  -1    \\
 $w_{a}$            & -0.8102  & -0.8085  $\pm$ 0.0053   & -1.1128    & -1.068$^{+0.35}_{-0.30}$ &  0     & 0    \\
 $H_{0}$            & 67.65    & 67.26   $\pm$ 0.36     & 67.27      & 67.20 $\pm$ 0.65         & 67.86    & 67.79 $\pm$ 0.40\\
 $\Omega_{m}$       & 0.307    & 0.312   $\pm$ 0.005    & 0.316      & 0.316 $\pm$ 0.006        & 0.308    & 0.309 $\pm$ 0.005\\
 $\Omega_{\mathrm{DE}}$      & 0.692    & 0.687   $\pm$ 0.005    & 0.683      & 0.683 $\pm$ 0.006        & 0.691    & 0.690 $\pm$ 0.005\\
 $\sigma_{8}$       & 0.805    & 0.810   $\pm$ 0.007    & 0.825      & 0.814 $\pm$ 0.012        & 0.807    & 0.808 $\pm$ 0.007\\
$D_{M}(z_{*})\mathrm{[Gpc]}$   & 13.89    & 13.88   $\pm$ 0.02   &  13.86      & 13.87 $\pm$ 0.02   & 13.89   & 13.89   $\pm$ 0.02 \\
$r_{*}$             & 144.64   & 144.50  $\pm$ 0.21   & 144.38      & 144.44 $\pm$ 0.25  & 144.68  & 144.64 $\pm$ 0.21\\
$z_{*}$             & 1089.81  & 1090.02 $\pm$ 0.21   & 1089.87     & 1089.92 $\pm$ 0.23 & 1089.74 & 1089.76 $\pm$ 0.20 \\
$r_{\mathrm{drag}}\mathrm{[Mpc]}$ & 147.21   & 147.11  $\pm$ 0.23   & 147.03      & 147.10 $\pm$ 0.25  & 147.33  & 147.29 $\pm$ 0.22 \\
$z_{eq}$            & 3371     & 3375    $\pm$ 21     &  3404       & 3401 $\pm$ 26      & 3376    & 3380 $\pm$ 21 \\
$k_{D}\mathrm{[Mpc^{-1}]}$     &  0.1405  &   0.1404 $\pm$ 0.0002     &   0.1409 & 0.1408 $\pm$ 0.0003      & 0.1406   & 0.1406 $\pm$ 0.0002 \\
$Y^{BBN}_{P}$       &   0.2588   &  0.25882   $\pm$ 0.00005      &  0.2467      & 0.24672  $\pm$ 0.00005    & 0.2467    & 0.24674 $\pm$ 0.00005 \\
$10^{5}D/H$         &   2.858   &  2.872   $\pm$ 0.027     &  2.578 &  2.583 $\pm$ 0.026      &  2.574   & 2.575 $\pm$ 0.023  \\
\end{tabular}
\end{ruledtabular}
\end{table*}

\subsection{The dynamics of the equation of state}

 Figure \ref{Fig:1} illustrates a detailed comparison between the analyzed cosmological models: BDE (in blue), $w_0w_a$CDM (in orange), and $\Lambda$CDM (in red) models, with marginalized posterior constraints at $68\%$ and $95\%$ confidence levels in terms of EoS--$w(z)$, $\mathcal{O}m(z)$, and normalized dark energy density--$f_{\mathrm{DE}}=\rho_{\mathrm{DE}}/\rho_{\mathrm{DE}}(0)$. The upper-left plot of figure \ref{Fig:1} shows the confidence contours for two models: the BDE model and the $w_0w_a$CDM model, at levels of $68\%$ and $95\%$ in the $w_0-w_a$ parameter space. One important finding is that the area covered by the confidence contours of the BDE model is about $10,000$ times smaller than those of the $w_0w_a$CDM model. This significant reduction in contour area highlights the robustness of the BDE model owing to its absence of dark energy free parameters, which is fundamentally derived, in contrast to the $w_0w_a$CDM model that has three additional free parameters: $(w_0,w_a,\rho_{\mathrm{DE}})$. According to observational data, at a $68\%$ confidence level, the values for $w_0=w(a_0)$ and $w_a=-(\mathrm{d}w(a)/\mathrm{d}a)|_{a=a_{0}}$ predicted by BDE are $w_0 = -0.9301 \pm 0.0004$ and $w_a = -0.8085 \pm 0.0053$. In contrast, for the $w_0w_a$CDM model, the values are $w_0 = -0.724 \pm 0.070$ and $w_a = -1.068^{+0.35}_{-0.30}$ as reported in \cite{desicollaboration2024desi2024vicosmological}. In particular, this combination of data favors a value of $w_0>-1$ and $w_a<0$, which reveals a significant discrepancy with the standard cosmological model equation of state ($w_0=-1$, $w_a=0$) at a confidence level above $3.9\sigma$. The upper-middle plot of figure \ref{Fig:1} illustrates a detailed comparison of the constraints on the DE EoS, with the $w_0w_a$CDM model (in orange) and the BDE model (in blue) in the redshift range $0<z<3$. The shaded bands around the best-fit from each model delineate the regions corresponding to the $68\%$ and $95\%$ confidence intervals, providing insight into the uncertainties associated with these parameters. The limit imposed by the $\Lambda$CDM model is indicated by the solid horizontal red line. The upper-right plot of figure \ref{Fig:1} offers a closer inspection of the differences in the equation of state of BDE and $w_0w_a$CDM models, revealing nuanced insights into their underlying dynamics. The lower-left plot of figure \ref{Fig:1} presents a more general view of the evolution of the dark energy equation of state $w(a)$ across the analyzed cosmological models as a function of scale factor $a$. In BDE model, before the onset of the condensation epoch ($ a_{c} $), the EoS is found in relativistic dominance, represented by $w=1/3$ in the early universe, particularly at $a \sim 10^{-5}$. At the condensation scale factor $a_c$, a notable phase transition takes place as the dark meson field $\phi$ condensates. This transition marks a shift in the dynamics of dark energy, leading to a non-asymptotic convergence towards a quintessence behavior with an EoS of $w_0=-0.9301 \pm 0.0004$ at present time. 
This analysis reveals that the EoS predicted by the BDE model exhibits a pronounced, sharp, and non-monotonic evolution, always remaining larger than $w=-1$ at all times. In contrast, the $w_0w_a$CDM model starts out "phantom" (with $w(z)<-1$) at high redshift, and cross into the $w(z)>-1$ regime at $z\lesssim 1$. This transition propels the model into a phantom-like regime, raising concerns over theoretical instabilities attributed to negative kinetic energy, which challenge the consistency of the underlying physical theories. This at face value implies two special features not expected in the standard $\Lambda$CDM model illustrated in the lower-right plot from Fig.\ref{Fig:1}, in which the DE density $\rho_{\mathrm{DE}}$ normalized to its present value is displayed. One feature seen in the lower-right plot, according to $w_0w_a$CDM, is a dark energy density that increases in time (in the phantom regime at high redshifts). The second feature is a dark energy density that decreases in time (in the $w(z)>-1$ regime). For BDE, the dark energy density tends to decrease in the $w(z)>-1$ regime as seen in the lower-right plot from Fig.\ref{Fig:1}. Furthermore, we test the DE models considered here throughout the important diagnostic quantity called $Om(z)$, which is sensitive to DE EoS and serves as a null test for the $\Lambda$CDM model. Specifically, $Om(z)$ is a geometric diagnostic that connects the Hubble parameter—the rate of the universe's expansion—with redshift. This allows $Om(z)$ to distinguish between DDE models and $\Lambda$CDM, independent of matter density. A constant value of $Om(z)$ across redshifts suggests dark energy behaves as a cosmological constant $\Lambda$. Conversely, a positive slope indicates phantom dark energy ($w < -1$), while a negative slope suggests quintessence ($w > -1$). In line with \cite{PhysRevD.78.103502, PhysRevLett.101.181301}, $Om(z)$ for a spatially flat Universe is defined as follows:
\begin{equation}
    Om(z) \equiv \frac{H^{2}(z)/H^{2}_{0}-1}{(1+z)^{3}-1},
\end{equation}
where $H_0$ is the present value of the Hubble parameter. The reconstructed $Om(z)$ in the lower-middle plot in figure \ref{Fig:1} shows a clear $(>2\sigma)$ deviation of $w_0w_a$CDM and BDE models  from the cosmological constant $\Lambda$ model in the range $0< z < 0.5$ and $0< z <1.5$, respectively, where the red line represents the best-fit $\Lambda$CDM value of $\Omega_m=0.308$, which shows that $Om(z)=\Omega_{m0}$ for $\Lambda$CDM, whereas $Om(z)>\Omega_{m0}$ for quintessence $(w>-1)$ and $Om(z)<\Omega_{m0}$ for phantom $(w<-1)$ models. This effect could correspond to dark energy decaying into an unknown substance. This trend in $Om(z)$ was previously observed with DESI DR1 and currently in DR2 data. In a cosmological context, a phantom equation of state ($w(z) < -1$) corresponds to a dark energy density that exhibits an increasing trend with cosmic expansion, indicated by $\mathrm{d}\rho_{DE}/\mathrm{d}a > 0$. This increase persists until it reaches a critical value at redshift $z_{c} \approx 0.35$, where the equation of state crosses the phantom threshold ($w(z_{c}) = -1$) observed in the lower-right plot of figure \ref{Fig:1}. Following this crossing, $\rho_{DE}$ begins to dilute as the universe continues to expand, as illustrated in the lower-right plot of figure \ref{Fig:1}. As for BDE, $\rho_{DE}$ remains nearly constant at high redshifts and starts to dilute at late times. It is important to note that the specific redshift of this crossing is contingent upon the particular combination of datasets analyzed.
\begin{figure*}
  \centering
  \begin{tabular}{c@{\hspace{1em}}c@{\hspace{1em}}c@{\hspace{1em}}c}
      \includegraphics[width=24em,height=30em]{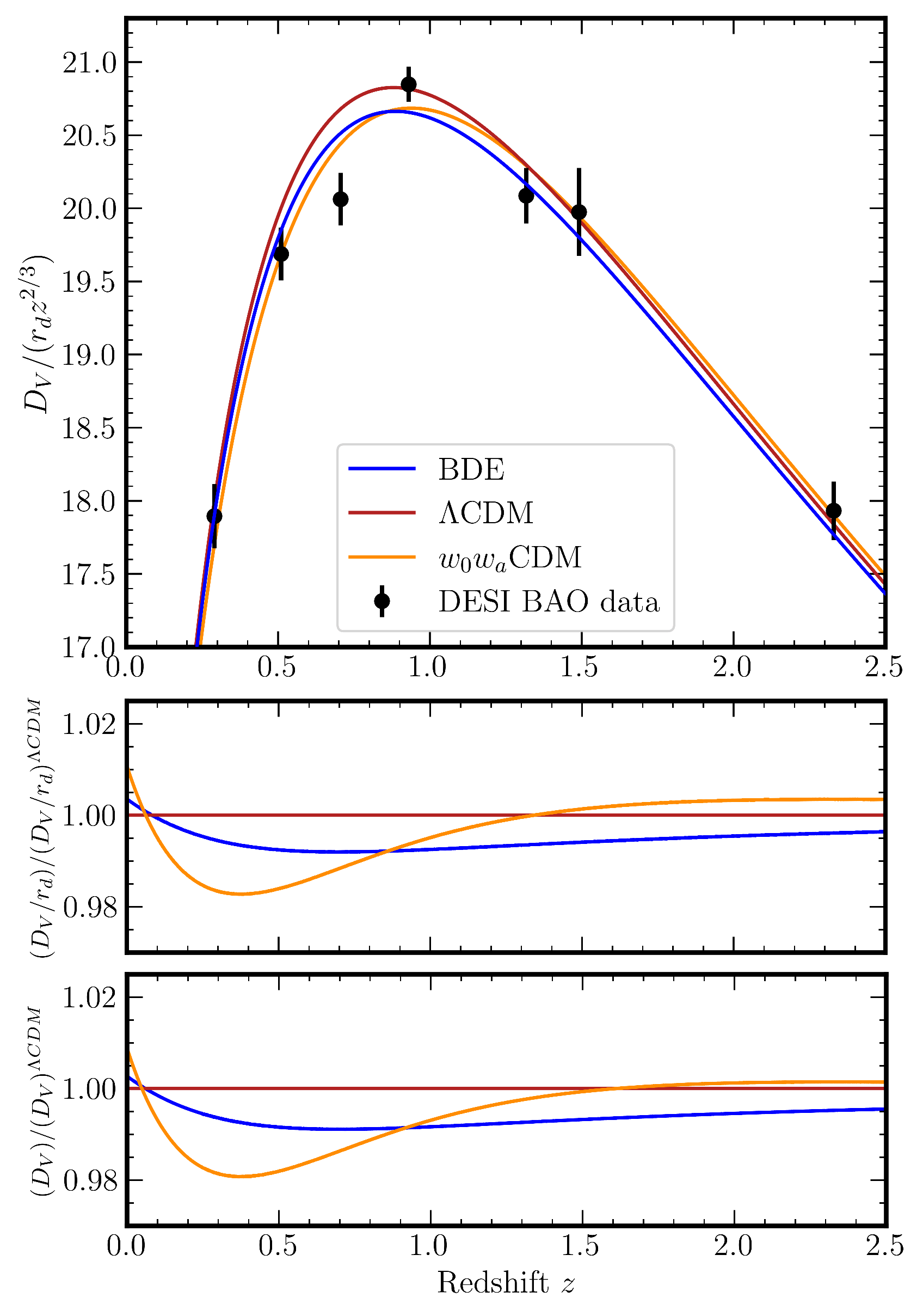}
      \includegraphics[width=24em,height=30em]{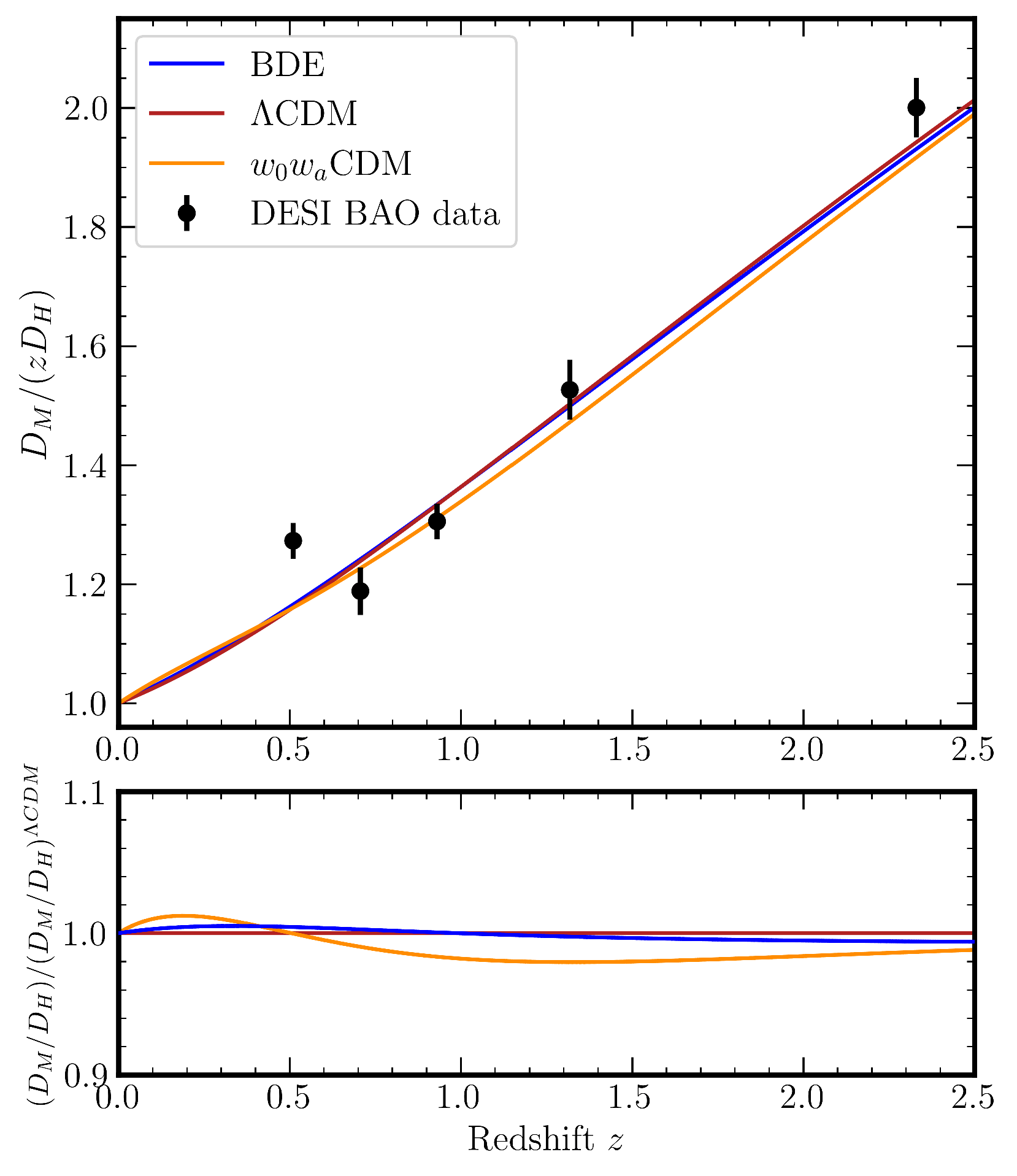}
  \end{tabular}
  \vspace*{8pt}
  \caption{\label{Fig:3} The DESI project measures the Baryon Acoustic Oscillation (BAO) distance scales at different redshifts. These measurements compare the angle-averaged distance, defined as $D_{V} \equiv (zD^{2}_M D_H)^{1/3}$ (for clarity, an arbitrary scaling of $z^{-2/3}$ is applied), to the sound horizon at the baryon drag epoch, $r_{d}$, shown in the left panel, while the right panel shows the ratio of transverse and line-of-sight comoving distances $D_{M}/D_{H}$. The data includes contributions from all tracers and redshift bins, as labeled. The solid lines represent predictions from different cosmological models: BDE (in blue), $\Lambda$CDM (in red), and the $w_0w_a$CDM model (in orange). These predictions are based on their best fits. This analysis combines data from DESI, CMB, and DESY5. The bottom panel shows how the best fits for BDE and $w_0w_a$CDM models compare to $\Lambda$CDM model.  }
\end{figure*}
\vspace{-0.4cm}
\subsection{Cosmological Constraints}
\vspace{-0.1cm}
 The marginalized constraints for various cosmological parameters at $68\%$ and $95\%$ confidence levels is presented for the BDE model, along with the $w_0w_a$CDM and $\Lambda$CDM models, using DESI BAO \cite{desicollaboration2024desi2024vicosmological}, in conjunction with CMB from the \textit{Planck} satellite \cite{refId0} and the DES Supernova (DESY5) dataset \cite{descollaboration2024darkenergysurveycosmology} are visually presented in Fig.\ref{Fig:2}, which illustrates the contours of the marginalized distributions. The mean values derived from the $68\%$ confidence level of these marginalized distributions, along with the best-fit parameters for the three cosmological models, are comprehensively laid out in table \ref{tab: Results of some parameters in the BDE, CPL, LCDM for the full dataset}. Notably, the constraints on the six base $\Lambda$CDM parameters—namely $\Omega_{b}h^{2}$ (baryon density), $\Omega_{c}h^{2}$ (cold dark matter density), $100\theta_{MC}$ (acoustic scale), $\tau$ (reionization optical depth), $\mathrm{ln}(10^{10}A_{s})$ (scalar amplitude), and $n_{s}$ (scalar spectral index)— and derived parameters ($H_0, \Omega_{m},\Omega_{\mathrm{DE}}, D_{\mathrm{M}}(z_{*}),r_{d}, \sigma_{8}$) show remarkable consistency across the $\Lambda$CDM, $w_0w_a$CDM, and BDE models at the 1$\sigma$ confidence level, as evidenced in table \ref{tab: Results of some parameters in the BDE, CPL, LCDM for the full dataset} and figure \ref{Fig:2}. Importantly, our computed value for $\Lambda_{c} = 43.806 \pm 0.190$ is consistent with the theoretical limit of $\Lambda_c^{\text{th}} = 34^{+16}_{-11} \, \text{eV}$ when considering the $68\%$ credible interval, which relies on datasets from high-energy physics. We find that the synergistic combination of DESI BAO, CMB, and DESY5 datasets, as detailed in table \ref{tab: Results of some parameters in the BDE, CPL, LCDM for the full dataset} and depicted in figure \ref{Fig:2}, enables cutting-edge constraints on fundamental cosmological parameters, including the composition of the universe, the expansion rate over time, and the initial conditions governing its evolution. 
\begin{figure*}[ht]
  \centering
  \begin{tabular}{c@{\hspace{1em}}c@{\hspace{1em}}c@{\hspace{1em}}c}
      \includegraphics[width=\textwidth]{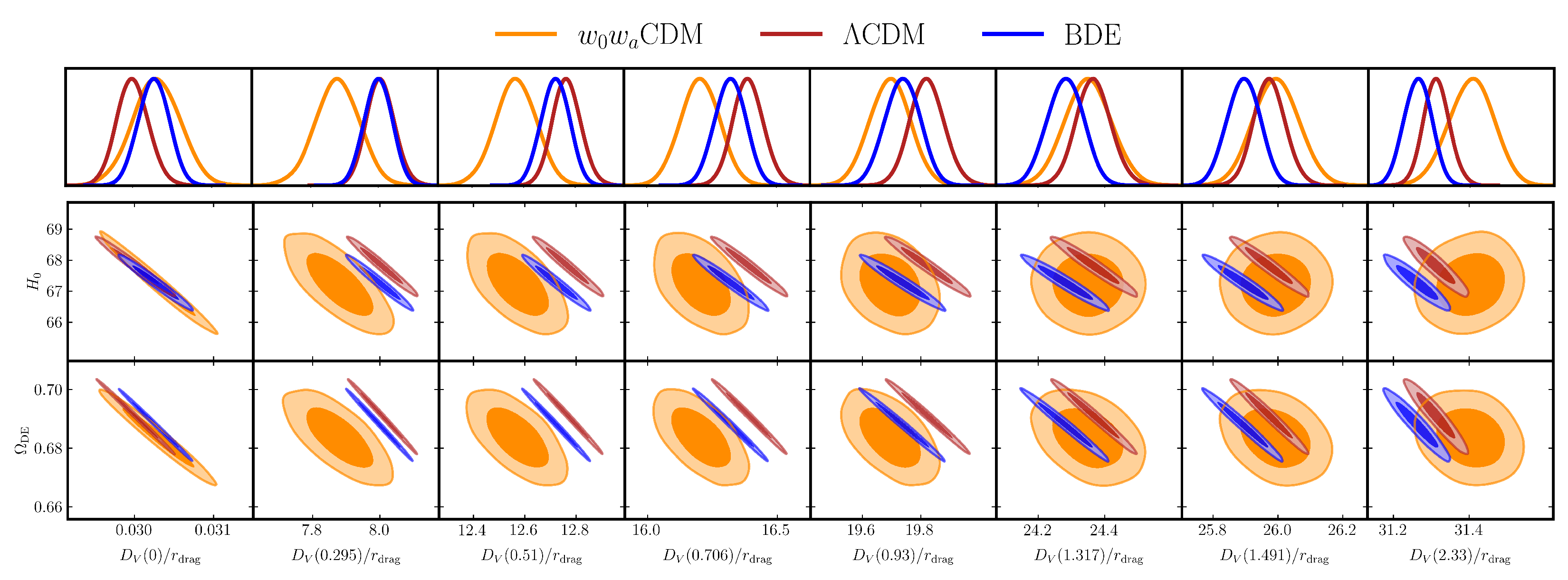}&\\
      \includegraphics[width=\textwidth]{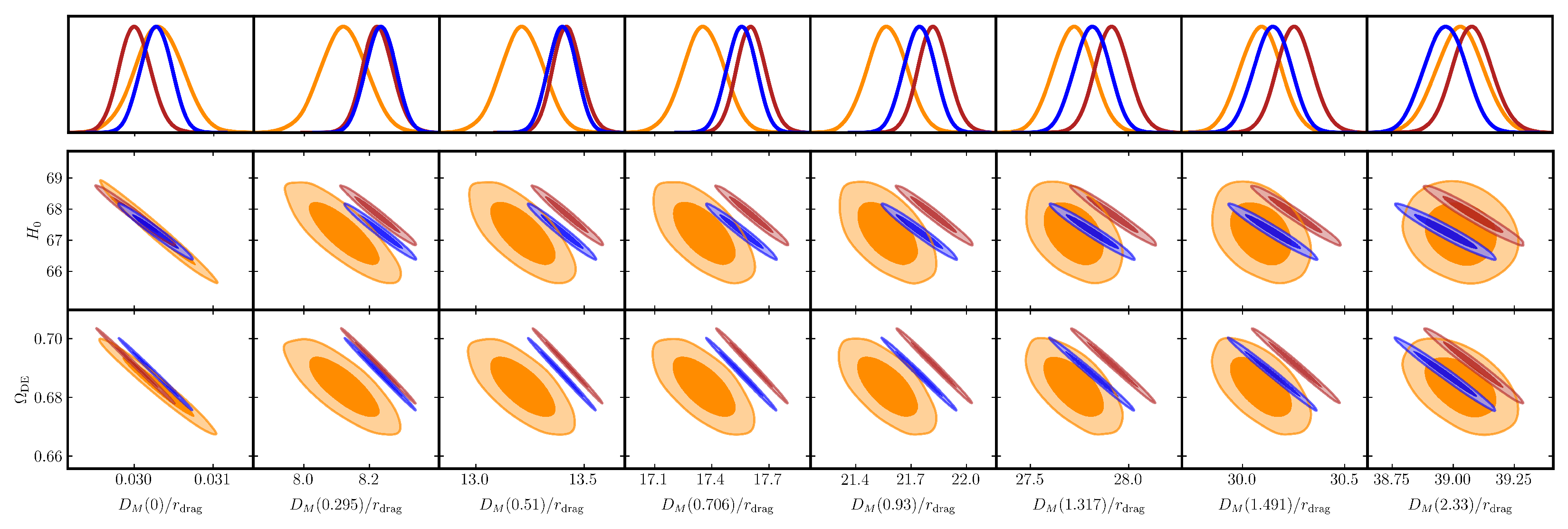}&\\
      \includegraphics[width=\textwidth]{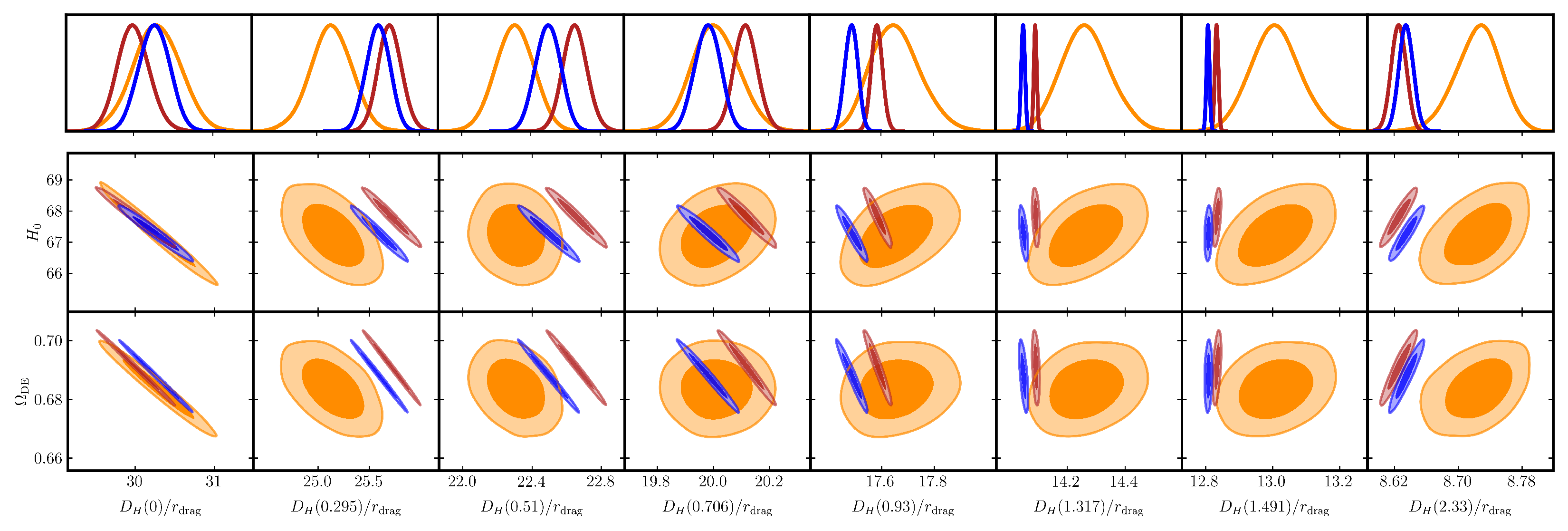} 
  \end{tabular}
  \vspace*{8pt}
  \caption{\label{Fig:4} The upper, middle, and lower panels present the $68\%$ and $95\%$ constraints for the angle-averaged distance, which is defined as $D_{V} \equiv (zD^{2}_M D_H)^{1/3}$, as well as for the comoving angular distance $D_{M}(z)$ and the Hubble distance $D_{H}$ to the sound horizon at the baryon drag epoch, $r_{d}$, with respect to Hubble parameter at present time $H_{0}$ and dark energy density parameter $\Omega_{\mathrm{DE}}$ at different effective redshifts according to BDE (in blue), $w_0w_a$CDM (in orange) and $\Lambda$CDM (in red). The confidence contours are derived from the joint analysis of DESI BAO, CMB, and DESY5 datasets. }
\end{figure*}
\begin{figure*}[ht]
  \centering
  \begin{tabular}{c@{\hspace{2em}}c@{\hspace{1em}}c@{\hspace{1em}}c}
      \includegraphics[width=26em,height=28.6em]{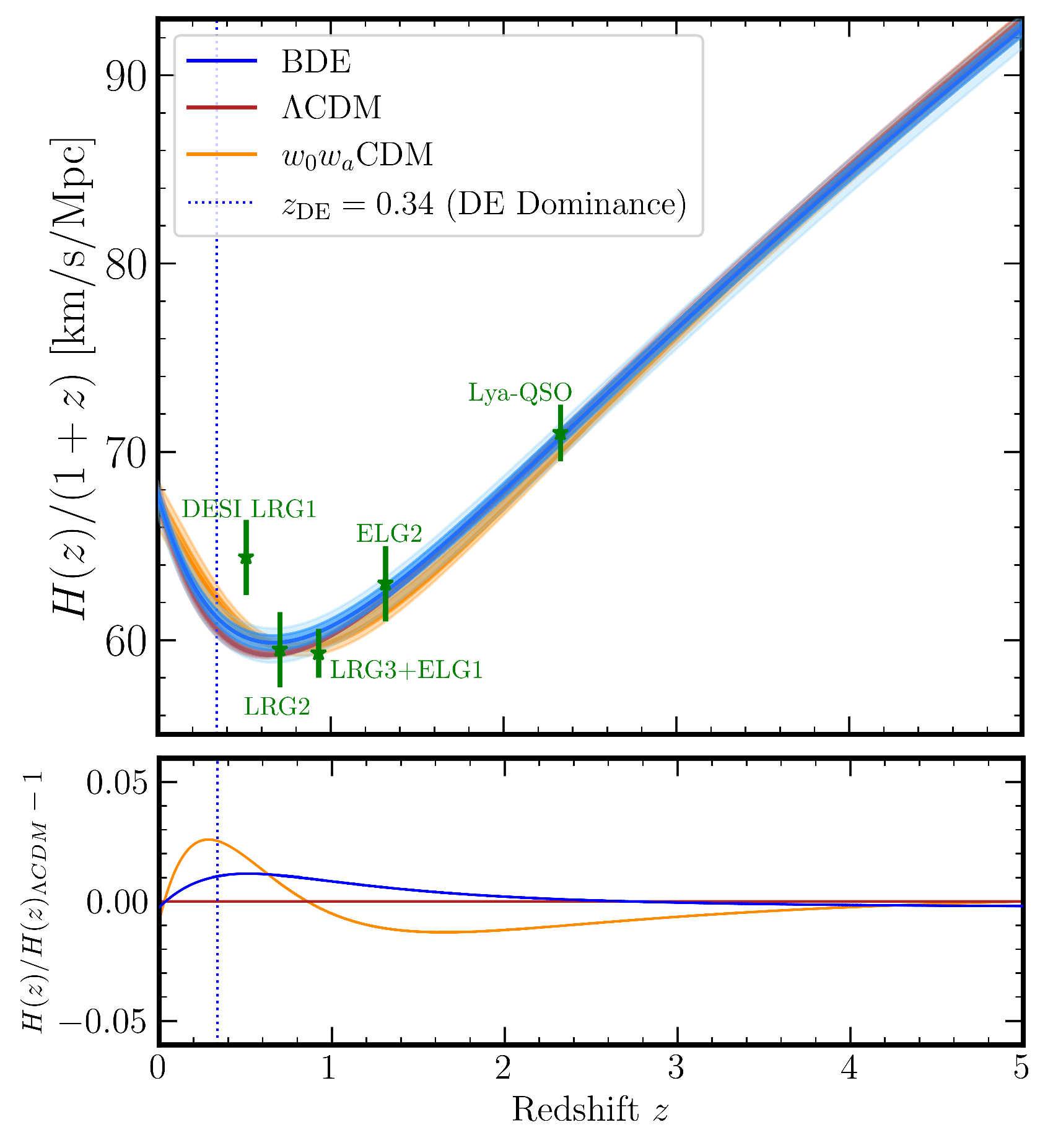}
      \includegraphics[width=26em]{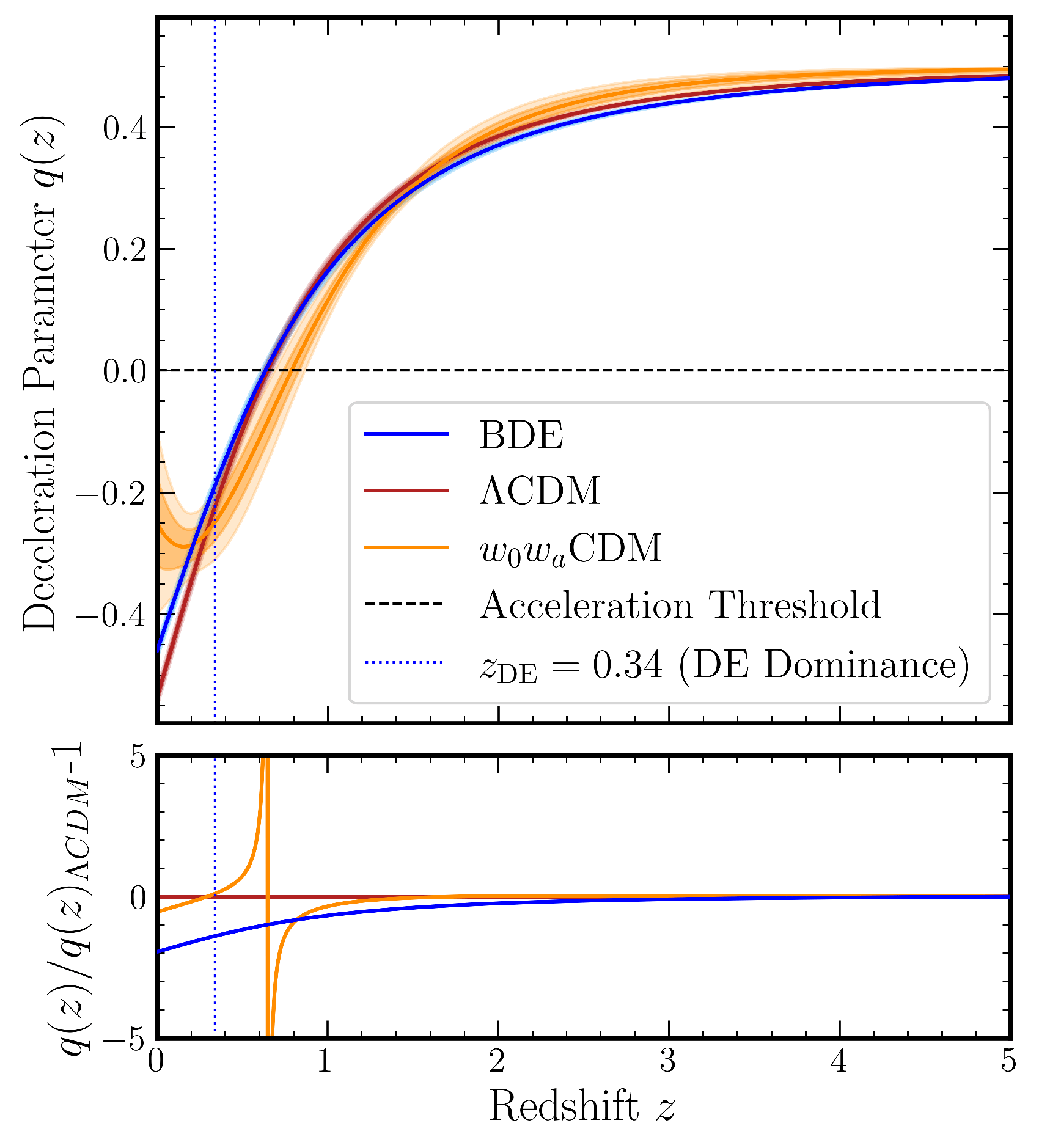} & \\
      \includegraphics[width=\textwidth]{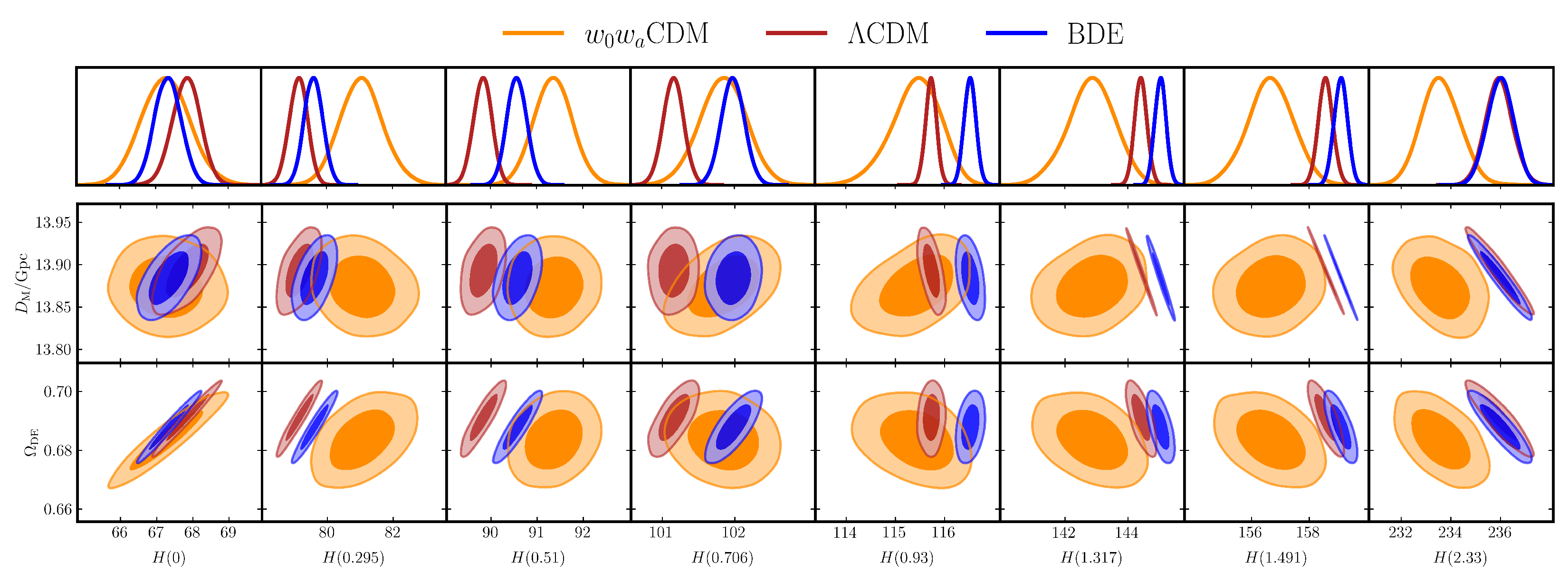}     
  \end{tabular}
  \vspace*{8pt}
  \caption{\label{Fig:5} Main aspects of the accelerated cosmic expansion of the Universe in BDE (blue), $w0w_a$CDM (orange), and $\Lambda$CDM (red) models. \textit{Upper-left}: Conformal expansion rate $H(z)/(1+z)$ as a function of redshift for the given models with shaded bands representing the $68\%$ and $95\%$ C.L. . \textit{Upper-right}: deceleration parameter $q(z)$ as a function of redshift $z$. \textit{Bottom}: $68\%$ and $95\%$ credible-interval contours of the Hubble parameter $H(z)$ with respect to $D_{M}(z_{*})$ and $\Omega_{DE}$. The vertical dashed line in the upper panels mark the redshift $z_{DE}=0.34$, corresponding to the onset of dark energy dominance. Green data points indicate measurements from DESI surveys.  }
\end{figure*}
\subsection{Distance-redshift and expansion rate measurements}
\begin{figure*}
  \centering
  \begin{tabular}{c@{\hspace{1em}}c}
      \includegraphics[width=25em,height=29em]{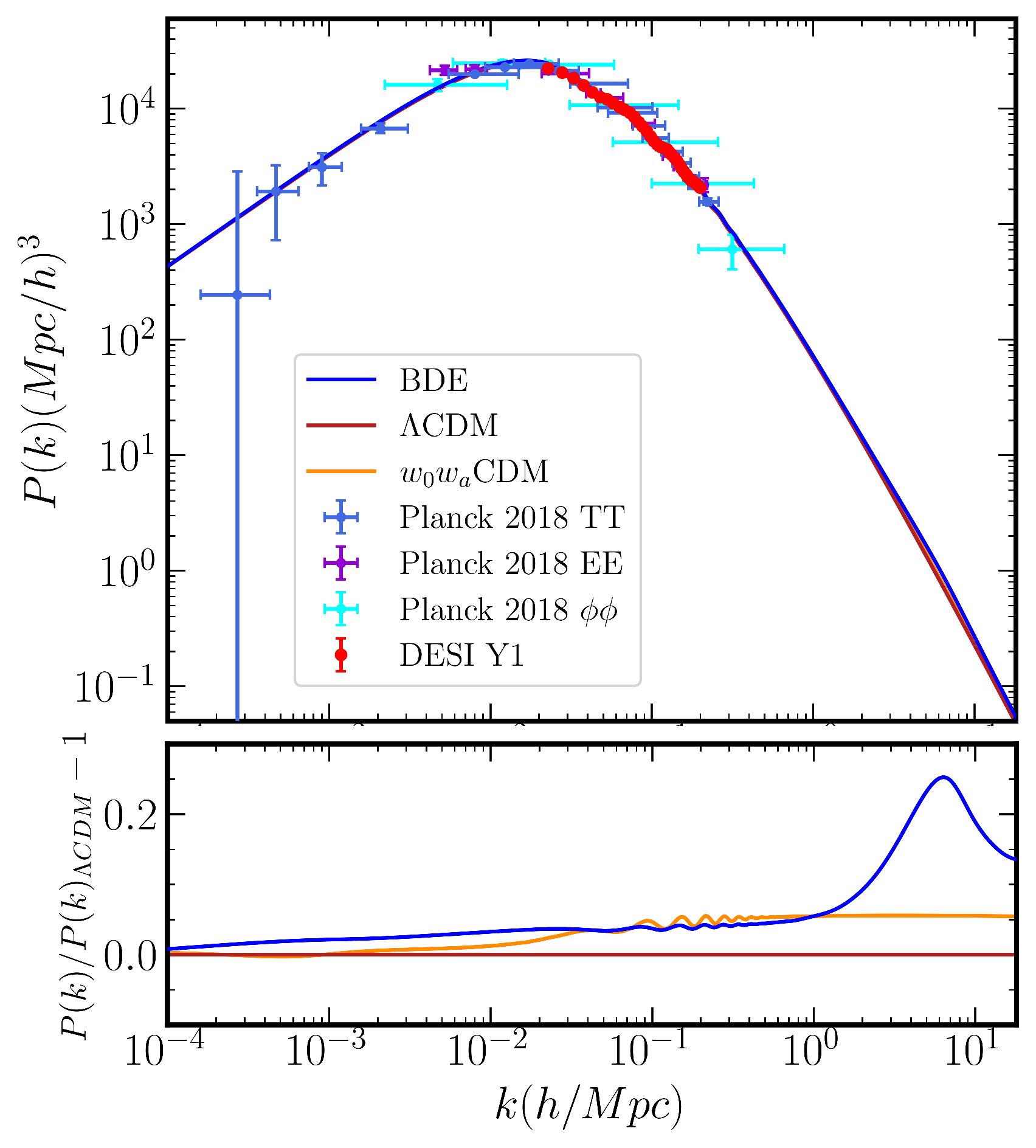}
      \includegraphics[width=25em,height=29em]{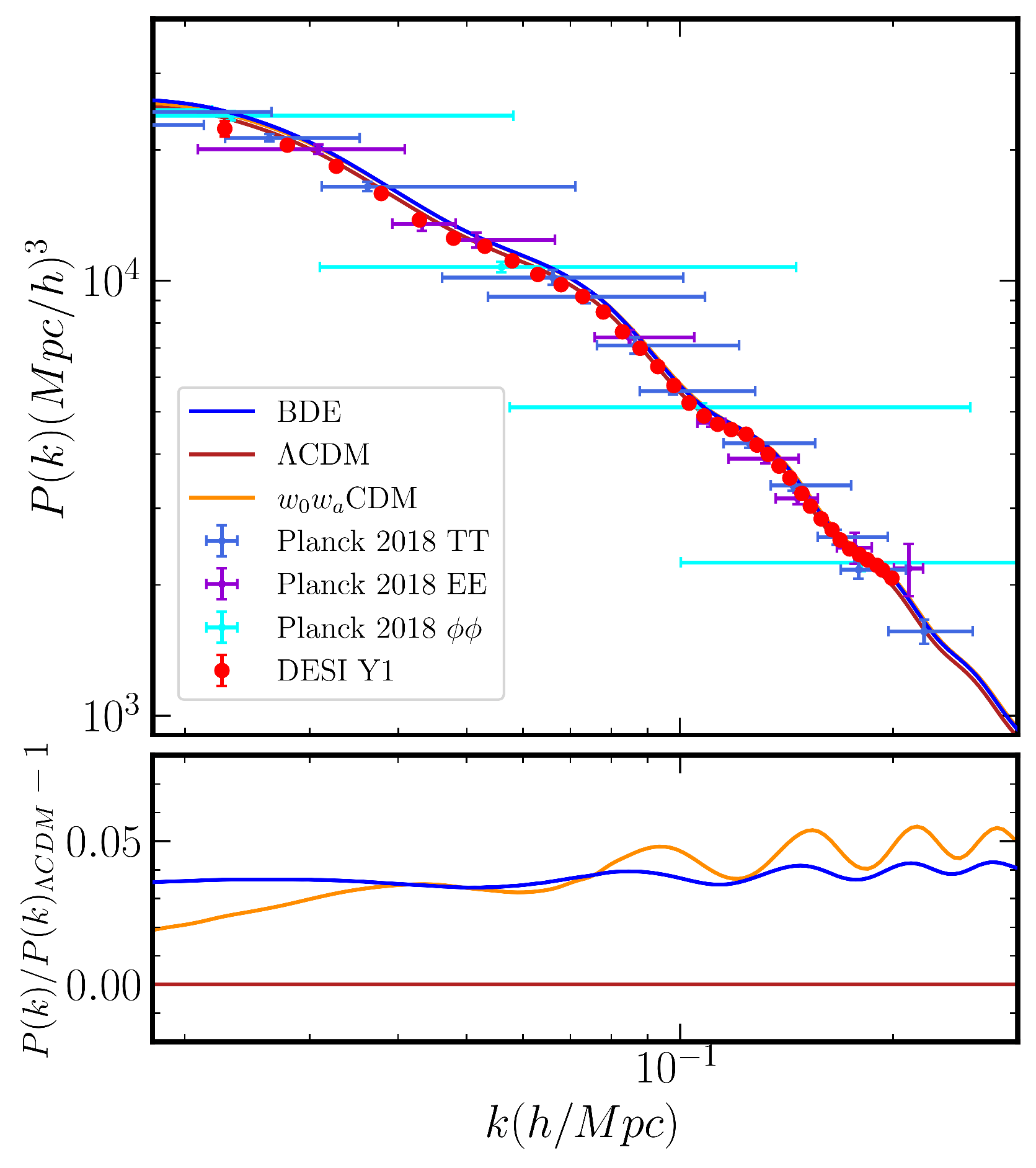}&\\
  \end{tabular}
  \vspace*{8pt}
  \caption{\label{Fig:6} \textit{Left panel}: The linear matter power spectrum $P(k)$ (top) at redshift $z=0$ and its fractional deviation from $\Lambda$CDM, $P(k)/P(k)_{\Lambda}CDM-1$ (bottom), for the BDE (blue), $w_0w_a$CDM (orange), and $\Lambda$CDM (red) models. The three models show consistency with \textit{Planck} 2018 CMB data (TT, EE, $\phi \phi$) at all scales. BDE exhibits a progressive enhancement of power (up to $ \sim 20\%$) at small scales ($k \gtrsim 1$), aligning with DESI DR1 measurements. The $w_0w_a$CDM model displays also deviations indicating parametric flexibility in its dark energy equation of state. \textit{Right panel}: A closer view on the BAO wiggle in the linear matter power spectrum.    }
\end{figure*}
 BAO measurements are systematically conducted across a diverse range of redshifts, which provides a rich dataset to analyze the fundamental cosmological parameters regulating the intricate relationship between distance and redshift. This analysis encompasses essential elements such as the curvature of the universe, which describes its geometric shape, the enigmatic properties of dark energy that drive its accelerated expansion, and the Hubble constant, a crucial value that defines the rate of expansion of the universe. Moreover, the precision of these measurements is significantly enhanced when additional external information regarding the absolute BAO scale is integrated, allowing for more robust constraints on these cosmic phenomena. The distance-redshift results of DESI BAO measurements are displayed in both plots given in Fig.\ref{Fig:3} in a Hubble diagram. The left plot from Fig.\ref{Fig:3} shows  $D_{V}(z)/r_{d}$ (scaled by an arbitrary factor of $z^{-2/3}$ for clarity), in which $D_{V}(z)$ is the angle-average distance that quantifies the average of the distances measured along, and perpendicular to, the line of sight to the observer. On the other hand, right plot shows $D_{M}/D_{H}$ (scaled by $z^{-1}$), defining $D_{M}(z)=D_{L}/(1+z_{\mathrm{obs}})$ and $ D_{H}(z)=c/H(z)$, where $H(z)$ is the Hubble parameter, $z$ is the redshift due to the expansion of the Universe, $z_{obs}$ is the observed redshift and $D_{L}(z)$ is the luminosity distance. The solid lines in each plot represent the model predictions for the best-fit in BDE, $w_0w_a$CDM, and $\Lambda$CDM models, which best align with the combined analysis of the astrophysical data compared to the DESI BAO data points. The lower panels in both figures show the relative difference of the $D_{V}(z)/r_d$ and $D_{M}/D_{H}$ values of BDE and $w_0w_a$CDM models with respect to $\Lambda$CDM model. Furthermore, in figure \ref{Fig:4} are displayed the contours corresponding to the $68\%$ and $95\%$ credible intervals for $D_{V}/r_{d}$, $D_{M}/r_{d}$, $D_{H}(z)/r_{d}$ as a function of the Hubble constant $H_{0}$ and the DE density parameter $\Omega_{\mathrm{DE}}$. This analysis spans seven distinct redshift bins where BAO signals using different cosmic tracers were identified by DESI collaboration \cite{desicollaboration2024desi2024vicosmological}  across the redshift range $0.1 <z < 4.2$ at effective redshifts $z_{\mathrm{eff}}$ from fits to the clustering measurements of DESI DR1 galaxies, quasars and the Ly$\alpha$ forest: 0.295, 0.510, 0.706, 0.930, 1.317, 1.491, and 2.330. Interestingly, we observe that the contours for BDE and $\Lambda$CDM in the joint constraints for $H_0$ and $\Omega_{\mathrm{DE}}$ begin to split across all distances as the effective redshift increases. However, in most cases, these contours remain consistent with the constraints of the contours of $w_0w_a$CDM at the 1$\sigma$ level for $D_{V}/r_{d}$, $D_{M}/r_{d}$ and mildly for $D_{H}/r_{d}$. The upper plots of figure \ref{Fig:5} display a general description of the dynamical evolution of accelerated cosmic expansion for the cosmological scenarios of BDE, $w_0w_a$CDM, and $\Lambda$CDM, intricately illustrating key components such as the conformal expansion rate in terms of $H(z)/(1+z)$ alongside the deceleration parameter $q(z)$ given by $q(z)\equiv -a\ddot{a}/\dot{a}^{2}=\mathrm{d}\text{ } \mathrm{ln} (H)/\mathrm{d} \text{ }\mathrm{ln}(1+z)-1 $. Furthermore, the lower plot of figure \ref{Fig:5} delineates the  $68\%$ and $95\%$ marginalized posterior constraints on the Hubble parameter $H(z)$  as a function of comoving angular distance $D_{M}$ (in Gpc) and DE density parameter $\Omega_{\mathrm{DE}}$. Observational measurements from DESI surveys (e.g., LRGs, ELGs, and Ly$\alpha$-QSO) are overlaid as green data points with their error bars in the upper-left plot of figure \ref{Fig:5}, while the vertical dashed line indicates the redshift $z_{\mathrm{DE}}=0.34$, marking the transition to dark energy dominance in same figure \ref{Fig:5}. As it is observed in the first column of the lower plot from Fig.\ref{Fig:5}, the dark energy content at present time is roughly similar across the three models, so the expansion rate. In the low redshift regime at $z\sim0.3$ seen in the upper-left and lower plot of figure \ref{Fig:5}, the $w_0w_a$CDM model exhibits a conformal expansion rate higher than that of $\Lambda$CDM and BDE models. This behavior can be due to the EoS of $w_0w_a$CDM begins to become less negative than $w=-1$ at redshift ($z\le0.5)$, where the dark energy contribution is higher and leading to a more rapid expansion. The expansion rate in BDE model is similar to $\Lambda$CDM at high-redshift regime ($z\ge1)$, with a small  deviation around $0.8\%$ at redshift $z\sim0.6$. This trend of the expansion rate can also be seen in the upper-right plot of figure \ref{Fig:5} in terms of $q(z)$. The reconstructed deceleration parameters $q(z)$ indicates that the cosmic acceleration$(q<0)$--according to $w_0w_a$CDM model--started to take off earlier at $z\sim0.8$ than those predicted by BDE and $\Lambda$CDM models, at $z\sim0.6$, with a slowing down of cosmic acceleration at late times. However, this trend in $q(z)$ has been also observed in recent studies performed by DESI collaboration with DESI DR2 data \cite{lodha2025extendeddarkenergyanalysis}.
\begin{figure}
  \centering
      \includegraphics[width=25em,height=27em]{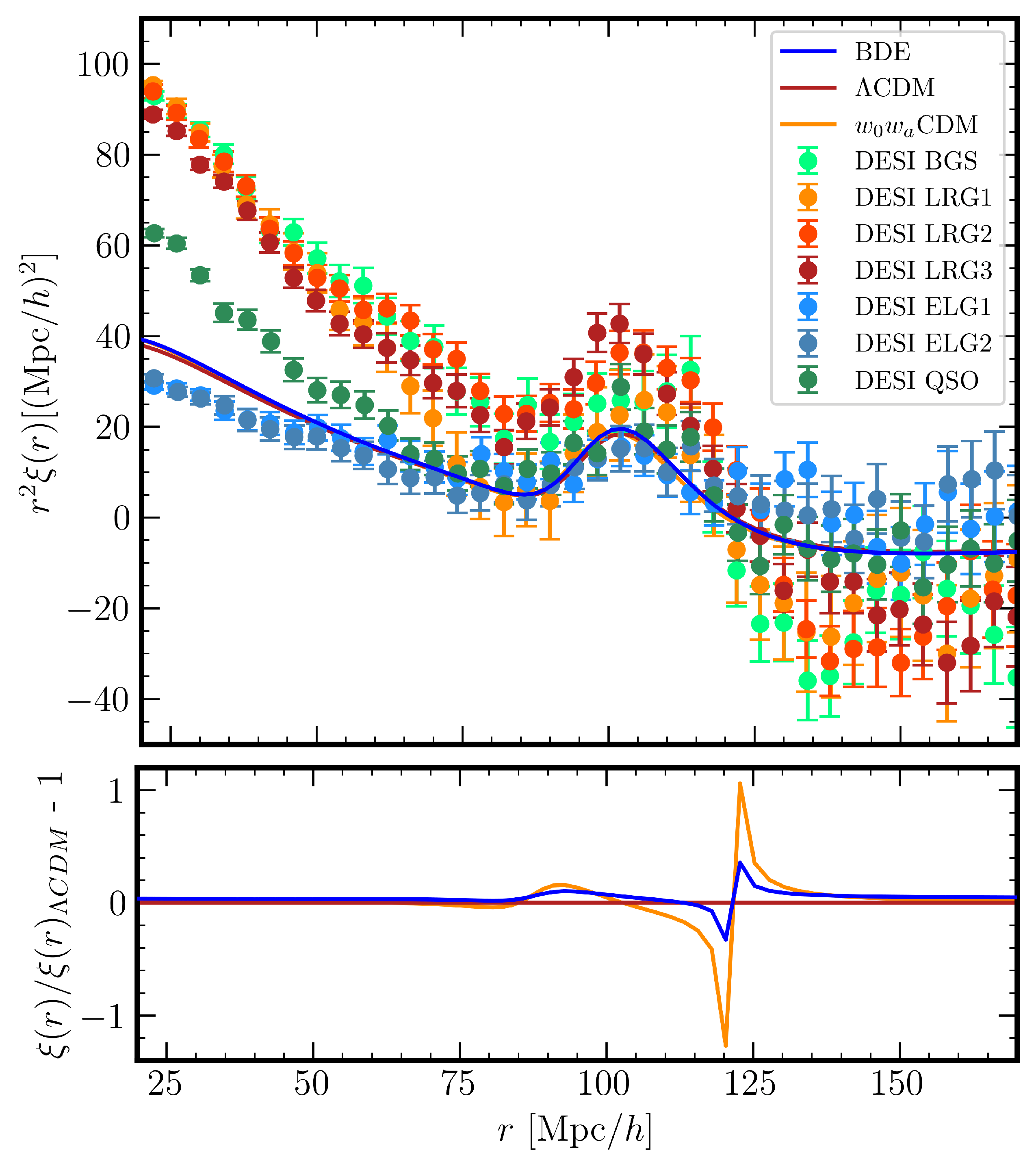}
  \caption{\label{Fig:7} The BAO peak in the cosmological matter correlation function at a distance corresponding to the sound horizon, $r_d \sim 100 h^{-1}$ Mpc in the cosmological scenarios of BDE, $w_0w_a$CDM and $\Lambda$CDM. The colored data points correspond to DESI BAO signal data for different tracers.}
\end{figure}
\begin{figure}
  \centering
      \includegraphics[width=25em,height=27em]{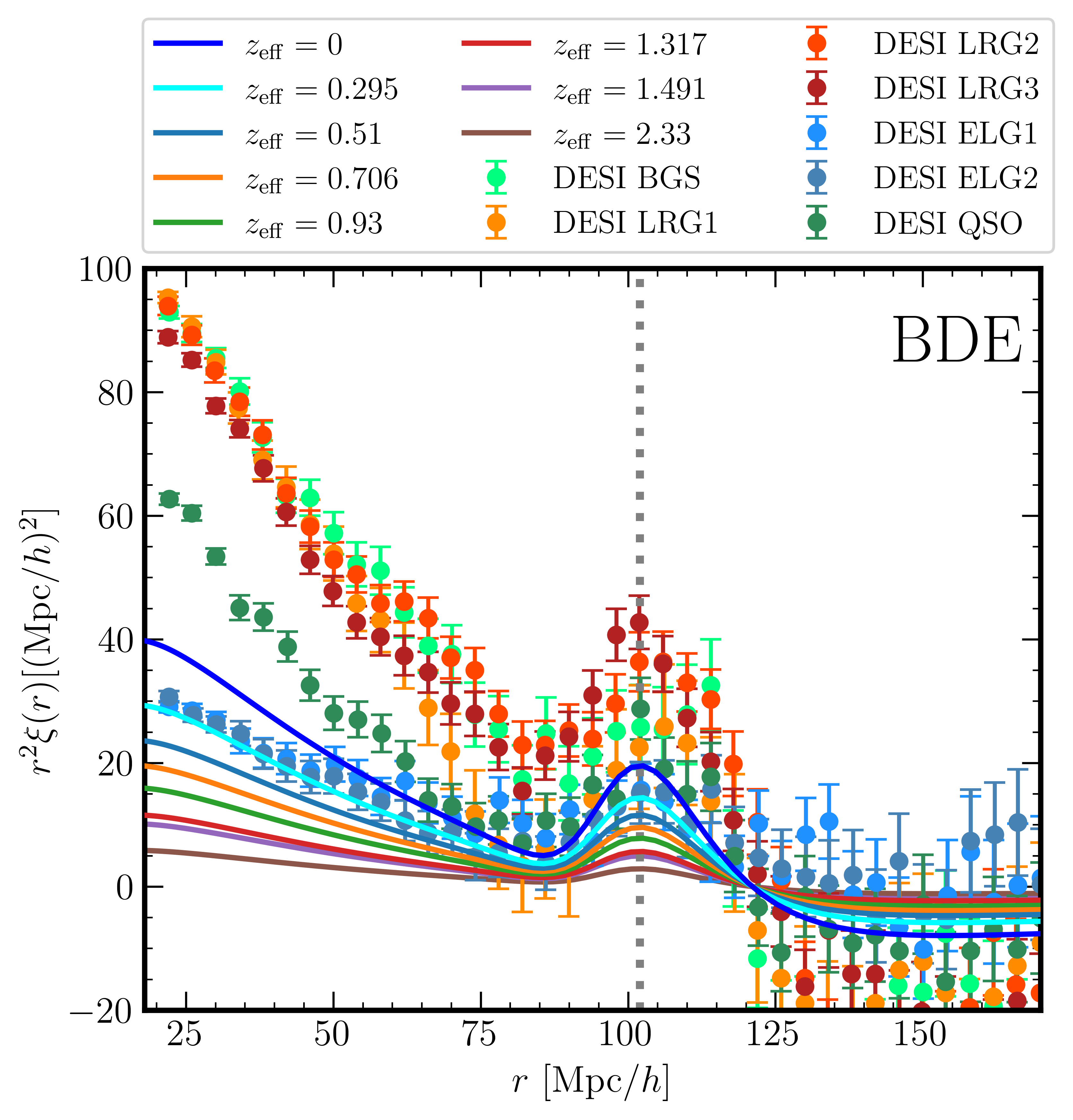}
  \caption{\label{Fig:8} Evolution of the baryon acoustic oscillation (BAO) peak given by the two-point correlation function at a distance corresponding to the sound horizon $r_d$ at several redshifts in the framework of BDE model. The dotted line marks the location of the BAO peak  $\sim 102 h^{-1}$ Mpc.   }
\end{figure}
\begin{figure}
  \centering
      \includegraphics[width=27em]{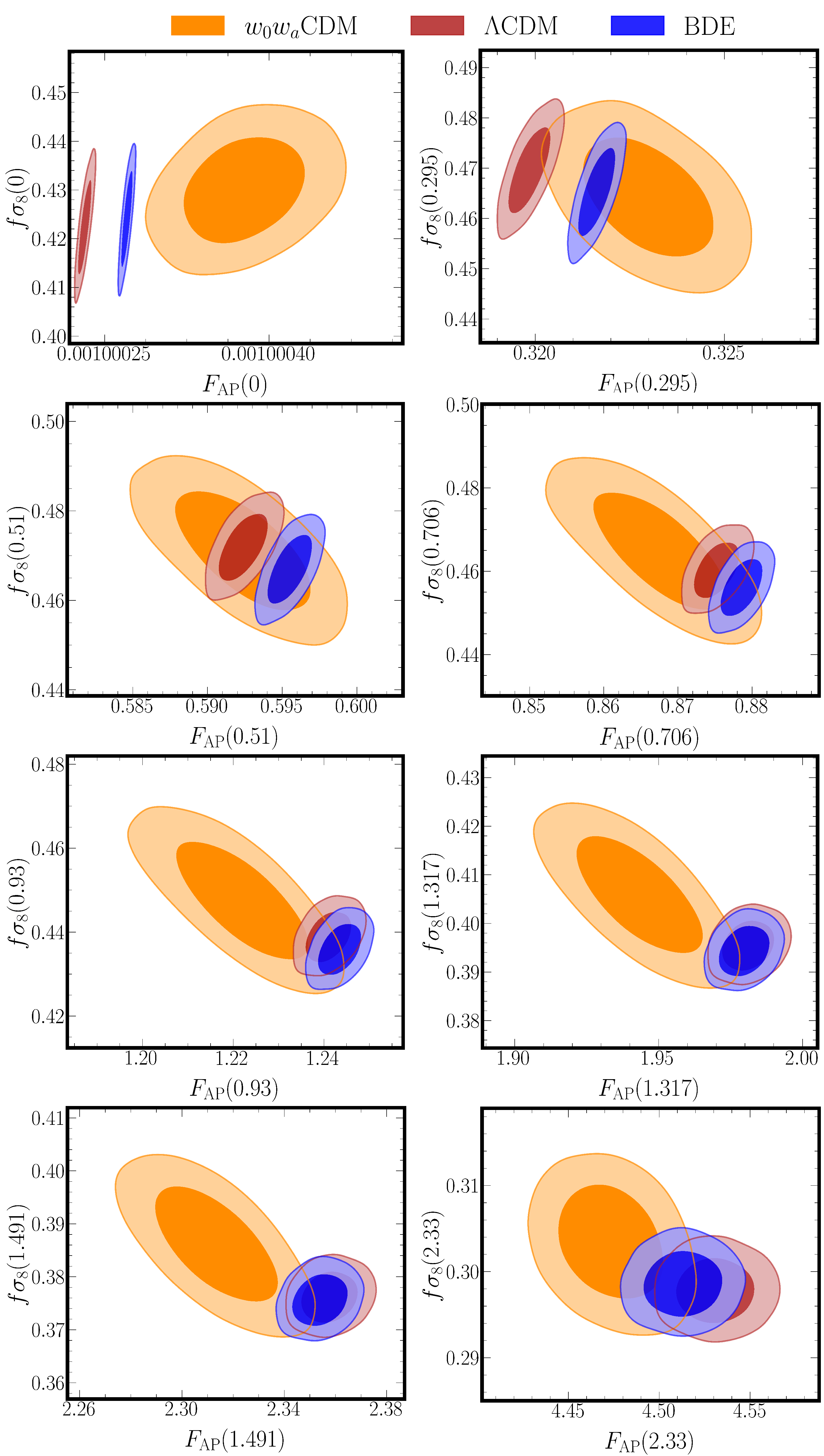}
  \caption{\label{Fig:9} Constraints on $f\sigma_8$ and $F_{AP}$ from analysis of redshift-space distortions. The contours show the $68\%$ and $95\%$ confidence ranges on $(f\sigma_8, F_{AP})$ from BDE (in blue), $w_0w_a$CDM (in orange), and $\Lambda$CDM (in red) at different effective redshifts.   }
\end{figure}
\begin{figure*}
  \centering
  \begin{tabular}{c@{\hspace{1em}}c@{\hspace{1em}}c@{\hspace{1em}}c}
      \includegraphics[width=\textwidth]{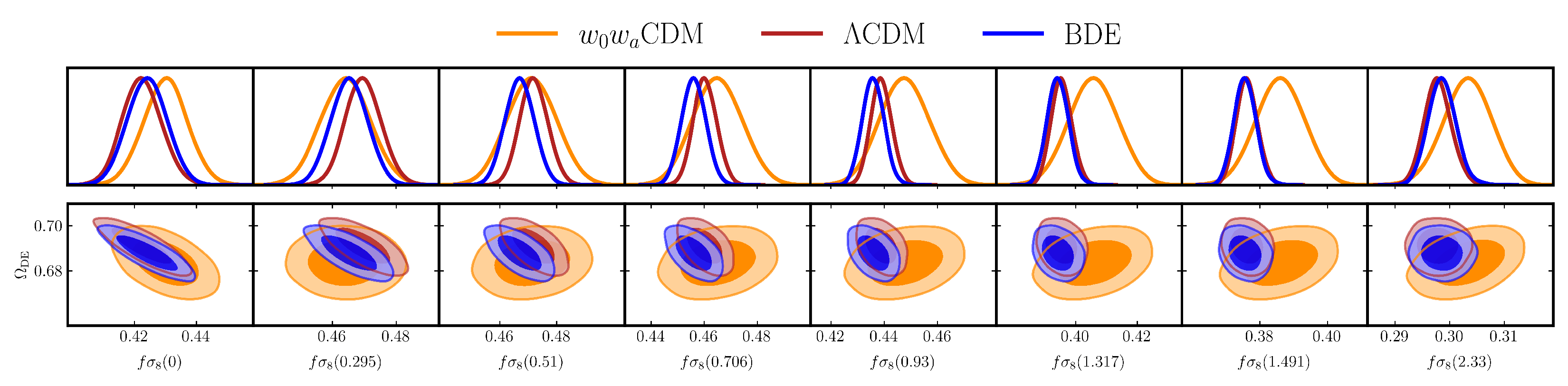}&\\
      \includegraphics[width=\textwidth]{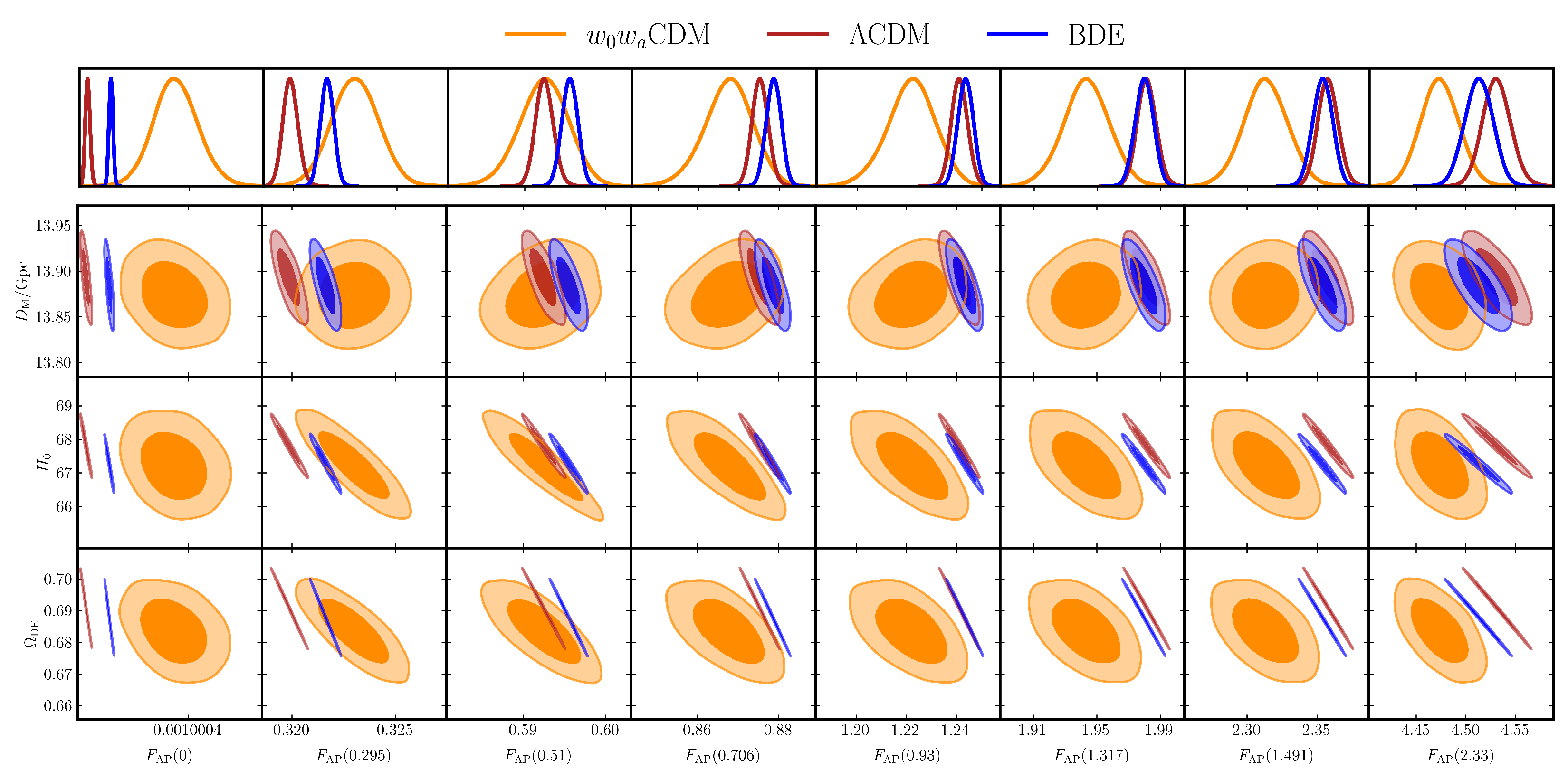} 
  \end{tabular}
  \vspace*{8pt}
  \caption{\label{Fig:10} \textit{Upper plot}: Marginalized distributions in BDE (blue), $w_0w_a$CDM (in dark orange), and $\Lambda$CDM (in red)  of the combination $f\sigma_{8}$ at different effective redshifts where BAO signals are detected by DESI. It is also displayed the joint constraints on $\Omega_{DE}$ and $f\sigma_8$ for each case. \textit{lower plot}: Marginalized distributions of the combination of Alcock-Paczinski parameter at different effective redshifts of DESI BAO signals. It is also displayed the joint constraints on $\Omega_{DE}$, $H_0$, and $D_{M}$ with $F_{AP}$ for each case.  }
\end{figure*}
\subsection{Matter power spectrum and structure formation}

\noindent The linear matter power spectrum, denoted as $P(k)$, serves as a crucial tool in cosmology, quantifying the distribution of matter density fluctuations across different wavenumbers $k$. This relationship is described by $k \propto 1/\lambda$, indicating that larger values of $k$ correspond to smaller physical scales, or more localized structures. Figure \ref{Fig:6} displays the linear matter power spectrum in BDE (in blue), $w_0w_a$CDM (in orange), and $\Lambda$CDM (in red) models. Accompanying these plots are the fractional deviations of BDE and $w_0w_a$CDM from the baseline $\Lambda$CDM model displayed in the lower panels, along with observational data sourced from the \textit{Planck} 2018 mission \cite{refId0} and DESI Year 1 survey \cite{cereskaite2025inferencematterpowerspectrum} . Focusing on the characteristics of $P(k)$ in relation to BDE as compared to $\Lambda$CDM reveals intriguing insights. At large scales  $k < 0.001$ $h/\mathrm{Mpc}$ , we observe a  minor deviation of $\Delta P(k)/P(k)_{\Lambda\mathrm{CDM}}$, which can be attributed to the distinct primordial power spectrum inherent to the BDE model. This outcome is expected since these long-wavelength modes traverse the horizon at a later epoch, thereby affording them less time to evolve. 
As we delve into intermediate scales ($k \sim 0.1 - 1$ $h/\mathrm{Mpc}$), a significant escalation in the growth of density fluctuations is noted, with $\Delta P(k) > 0$. This trend continues until we reach a peak enhancement at $k \approx 4.3$ $\mathrm{Mpc}^{-1}$, where the BDE exhibits $\sim 20\%$ increase in power. The growth rate of matter overdensities is notably influenced at small scales. Initially, the amplitude of the modes that cross the horizon before $ a_c $ are suppressed compared to $\Lambda$CDM and $w_0w_a$CDM due to the free streaming particles from DGG, i.e., $a<a_{c}$. Subsequently, these modes experience an enhancement caused by the rapid dilution of the BDE after the condensation scale \cite{PhysRevLett.121.161303,PhysRevD.72.043508}. As a result, the matter power spectrum in BDE is enhanced at all small scales, with respect to $\Lambda$CDM and $w_0w_a$CDM increasing by $16\%-20\%$ at the peak of the bump $k \approx 4.3 \, \mathrm{Mpc}^{-1} $ \cite{PhysRevLett.121.161303,PhysRevD.72.043508}. This enhancement is a hallmark of its distinctive expansion history and the dynamics of dark energy, illustrating how the BDE model modifies the growth of structure in our universe. On the other hand, comparing $w_0w_a$CDM model with respect to $\Lambda$CDM, we find that it yields a net positive deviation  $\sim6\%$ in comparison to $\Lambda$CDM at small scales. This enhancement is primarily due to the amplification of late-time growth effects, which supersede the suppression experienced at earlier epochs. At smaller scales ($k > 0.1$ $h/\mathrm{Mpc}$), deviations in the matter power spectrum become more stable, as nonlinear effects gain dominance leaving this analysis for future work. This contrast underscores the complexity and richness of cosmic structure formation within different dark energy frameworks. 
\begin{figure}
  \centering
      \includegraphics[width=26em,height=29em]{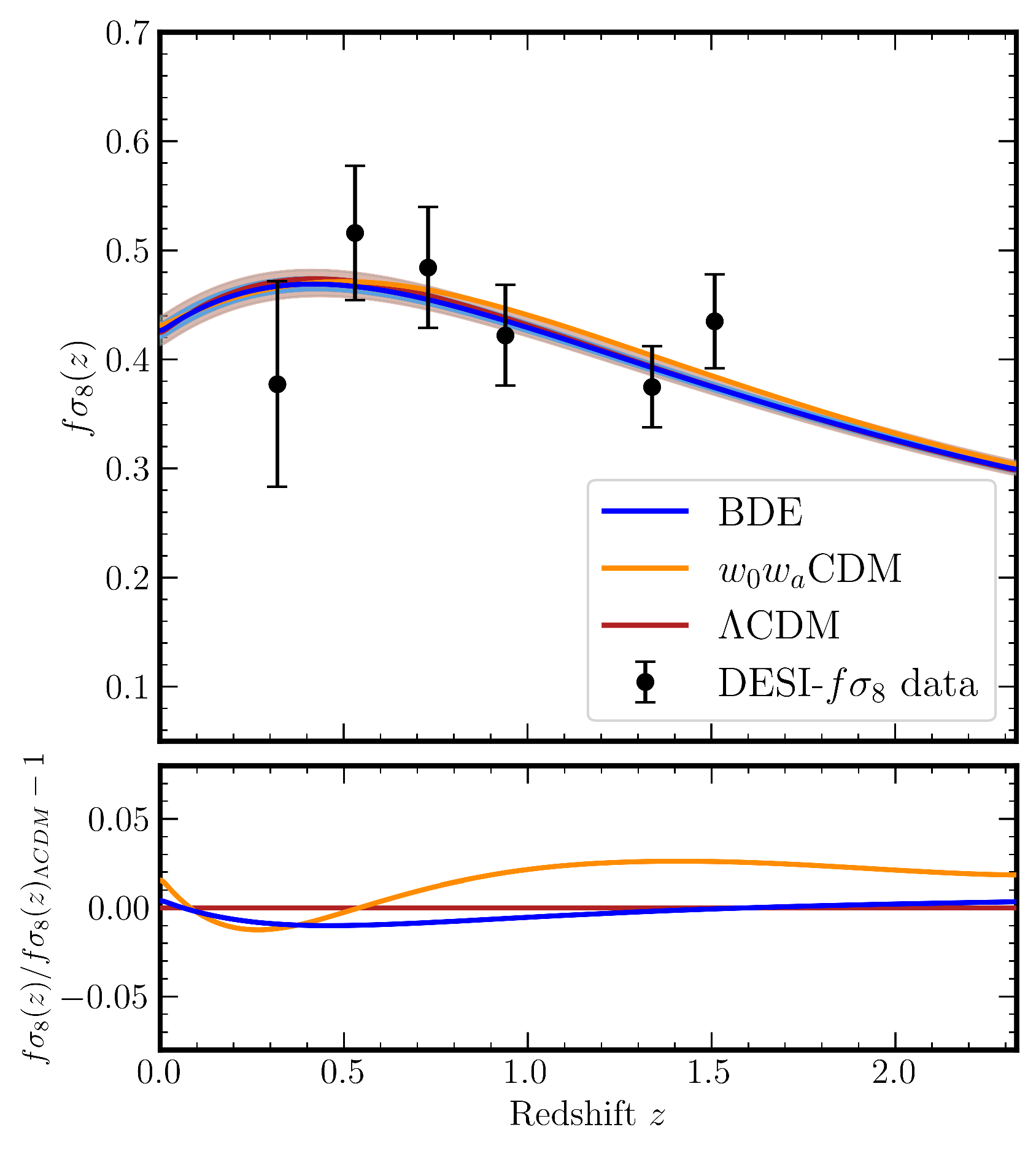}
  \caption{\label{Fig:11} The growth of structure measurement parameter as a function of redshift, $f\sigma_{8}(z)$, for BDE (in blue), $w_0w_a$CDM (in dark orange), and $\Lambda$CDM (in red) plotted along with the shaded bands correspond to $68\%$ and $95\%$ confidence regions for each model. In the bottom panel, it is shown the relative difference of BDE and $w_0w_a$CDM with respect to $\Lambda$CDM.   }
\end{figure}
\noindent We will now expand our analysis of the matter power spectrum by examining its Fourier transform: the two-point correlation function (2PCF), $ \xi(r) $. The function $ \xi(r) $ serves as a vital statistical metric for large-scale structure (LSS), encapsulating information about matter clustering and its evolution due to gravitational collapse. A key feature of $ \xi(r) $ is the baryon acoustic oscillation (BAO) peak, which is a remnant of sound waves propagating through the primordial plasma. This peak appears as a localized enhancement in clustering at the sound horizon scale, approximately $ r_{d} \sim 100 h^{-1} \mathrm{Mpc} $. This characteristic has proven to be a reliable standard ruler for constraining cosmological models. Figure \ref{Fig:7} presents the best-fit function $\xi(r)$ from BDE, $w_0w_a$CDM, and $\Lambda$CDM models, alongside observational data derived from DESI tracers. The bottom panel provides a measurement of deviations from $\Lambda$CDM characterized by the ratio $\xi(r)/\xi(r)_{\Lambda\textrm{CDM}}-1$, where $\xi(r)$ refers to the correlation function of the BDE and $w_0w_a$CDM models. This metric isolates fractional differences in clustering amplitude relative to the $\Lambda$CDM reference framework: values greater than zero signify enhanced clustering, whereas negative values indicate a suppression of clustering. The discrepancies observed for the BDE and $w_0w_a$CDM models stem from their distinctive modifications to the cosmological structure as discussed previously, each reflecting unique theoretical implications. The observational DESI tracers, including BGS, LRGs, ELGs, and QSOs, present deviations largely attributed to variations in galaxy bias, redshift distributions, and survey systematics. The BAO peak position remains firmly anchored to the sound horizon $r_d$, governed by pre-recombination physics. For BDE, the sound horizon aligns with that of $\Lambda$CDM when calibrated against CMB data, thereby ensuring that the BAO peak's position is consistent with observational evidence. This consistency arises because $r_d$ is predominantly determined by early-universe physics, which remains unaffected by the late-time dynamics inherent to BDE. Examining figure \ref{Fig:7}, we observe that the amplitude of the BAO peak in BDE mildly changes from that predicted by $\Lambda$CDM due to altered behavior of dark energy. Following the condensation epoch $a_c$, the EoS for BDE transitions from a stiff phase ($w_{\mathrm{BDE}}=1$) to a near-cosmological constant ($w_{\mathrm{BDE}} \approx -0.93$). This swift dilution temporarily lowers the dark energy density, consequently affecting the structure growth rate and the clustering amplitude at late times. Conversely, deviations in the bottom panel for the $w_0w_a$CDM model may occur due to the null physical constraints on $w_0$ and $w_a$, as this model does not engage with early-universe dynamics or introduce scale-dependent interactions. However, the deviations attributed to BDE may exhibit both amplitude and scale-dependent characteristics due to the introduction of the DGG, which increases the total radiation density. This modification influences the sound speed of the baryon-photon plasma, resulting in an amplification of primordial sound wave amplitudes; thus, positive deviations observed at $r \sim r_d$ imply that BDE's correlation function mildly surpasses that of $\Lambda$CDM at the BAO scale, directly indicating the enhanced peak amplitude. Figure \ref{Fig:8} illustrates the evolution of the BAO peak in BDE model across different redshifts, in which DESI collaboration measured the BAO signal from different tracers \cite{desicollaboration2024desi2024vicosmological}. In figure \ref{Fig:9}, we present a marginalized contours for the $ f\sigma_{8}$ (with $f$ representing the growth rate of the matter density perturbations and $\sigma_{8}$ is the matter variance on an $8h^{-1}$ Mpc scale) and the Alcock-Paczynski $F_{AP}$ parameters, across seven distinct redshift bins (redshift bins analyzed by DESI, culminate in a measurement at the current epoch $(z=0)$). The $F_{AP}$ parameter is formulated as $F_{AP}(z) = D_M(H(z)/c) $, where $ D_M$ is the comoving angular diameter distance, $H(z)$ is the Hubble parameter at redshift $z$, and $c$ denotes the speed of light. This parameter is essential for correcting any geometric distortions due to the expansion of the universe, enabling accurate comparisons between the observed and theoretical distributions of galaxies. The upper plot in figure \ref{Fig:10} displays the joint constraints of $f\sigma_{8}$ with respect to $\Omega_{DE}$ at the same effective redshifts as shown in figure \ref{Fig:9}. While the confidence regions align within the $1\sigma$ level, we note that the relation follows: $f\sigma_8(\mathrm{BDE}) < f\sigma_8(\Lambda\mathrm{CDM}) < f\sigma_8(w_0w_a\mathrm{CDM})$. However, the observed differences among these models are quite subtle, indicating no significant tension present in the results. The lower plot in figure \ref{Fig:10} displays the joint constraints of the Alcock-Paczynski $F_{AP}$ with respect to $\Omega_{DE}$, $H_0$ and $D_{M}$. In figure \ref{Fig:11}, we present the best-fit values along with the $68\%$ and $95\%$ confidence contours for the evolution of the structure growth parameter, $f\sigma_{8}(z)$, according to BDE, the $w_0w_a$CDM, and $\Lambda$CDM models. The contours illustrate the uncertainties associated with the growth measurements as a function of redshift. Additionally, we include the observational results from the six distinct redshift bins derived from DESI DR1 data, which are indicated in black. These measurements provide critical insights into cosmic structure formation and help constrain the parameters governing dark energy evolution and the expansion history of the universe.

\begin{table}[ht!]
\caption{\label{tab:cosmological_fit} Cosmological fitting results for BDE, $w_0w_a$CDM and $\Lambda$CDM models. The number of DE free parameters $N_{\mathrm{DE}}$ are: for BDE ($N_{\mathrm{DE}}=0$), $w_0w_a$CDM ($N_{\mathrm{DE}}=3$) and $\Lambda$CDM ($N_{\mathrm{DE}}=1$). $\Delta\mathrm{X}=\mathrm{X}_{\mathrm{y}}-\mathrm{X}_{\Lambda\mathrm{CDM}}$,  with $\mathrm{X}=\mathrm{AIC,BIC}$ and $\mathrm{y=BDE},w_0w_a\mathrm{CDM}$.}
\begin{ruledtabular}
\begin{tabular}{llll}
\textrm{Models}& \textrm{BDE}& \multicolumn{1}{c}{$w_0w_a$CDM}& \textrm{$\Lambda$CDM}\\ 
Parameters &  5   & 8 & 6 \\ 
\hline 
$\chi^{2,\mathrm{BAO}}_{\mathrm{reduced}}$ & 1.73 & 3 & 2.76\\ \vspace{0.05cm}
$\chi^{2,\mathrm{CMB}}_{\mathrm{reduced}}$ & 0.908 & 0.904 & 0.904\\ \vspace{0.05cm}
$\chi^{2,\mathrm{DESY5}}_{\mathrm{reduced}}$ & 0.602 & 0.598 & 0.599\\ 
AIC$_{\mathrm{BAO}}$ & 22.11 & 28.00 & 28.55 \\
BIC$_{\mathrm{BAO}}$& 24.53  & 31.87 & 31.45\\
AIC$_{\mathrm{DESY5}}$& 1658.63  & 1662.76 & 1668.39  \\
BIC$_{\mathrm{DESY5}}$& 1686.18  & 1706.85 & 1701.45 \\
AIC$_{\mathrm{CMB}}$ & 2790.06 & 2780.98 & 2778.80 \\
BIC$_{\mathrm{CMB}}$& 2823.89  & 2835.10 & 2819.39 \\
\hline
$\Delta$AIC$_{\mathrm{BAO}}$ & -6.44  & -0.55  & 0  \\
$\Delta$BIC$_{\mathrm{BAO}}$ & -6.92  & 0.42    &  0 \\
$\Delta$AIC$_{\mathrm{DESY5}}$ & -9.76 & -5.63   & 0  \\
$\Delta$BIC$_{\mathrm{DESY5}}$ & -15.27  & 5.4    &  0 \\
$\Delta$AIC$_{\mathrm{CMB}}$ & 11.26  & 2.18   & 0  \\
$\Delta$BIC$_{\mathrm{CMB}}$ & 4.5    & 15.71    &  0 \\
 \end{tabular}
\end{ruledtabular}
\end{table}
In order to compare the different cosmological models, $\Lambda\mathrm{CDM}$, $w_0w_a$CDM, and BDE, we compute the values of the Akaike information criterion (AIC) expressed as $\mathrm{AIC}=\chi^{2}+2p$ \cite{1100705} , the Bayesian information criterion (BIC) expressed as $\mathrm{BIC}=\chi^{2} + p\text{ }\mathrm{ln}(N)$ \cite{10.1214/aos/1176344136} and reduced chi-square $\chi^{2}_{\mathrm{reduced}}=\chi^{2}/(N-p)$,  where $p$ is the number of free parameters in a model and $N$ the number of data points given by the data set. As we discussed earlier, $\Lambda\mathrm{CDM}$ has one DE free parameter, namely ($\Lambda$), $w_0w_a$CDM has three DE free parameters $(w_0,w_a,\Lambda)$ and BDE has null DE free parameters. The values of $\Delta$AIC and $\Delta$BIC, which represent the difference between AIC and BIC with respect to the reference model, taking $\Lambda$CDM as our reference model, are summarized in table \ref{tab:cosmological_fit}. For the values of $\Delta$AIC and $\Delta$BIC, $0<\Delta\mathrm{AIC}(\Delta\mathrm{BIC})<2$ implies indistinguishable in preferring a given model, $2<\Delta\mathrm{AIC}(\Delta\mathrm{BIC})<6$ indicates average evidence against the given model, and $\Delta\mathrm{AIC}(\Delta\mathrm{BIC})>6$ indicates strong evidence against the model, while a negative value of $\Delta$AIC($\Delta$BIC) means otherwise \cite{10.1093/mnras/stad838}. We find that the results of $\Delta$AIC and $\Delta$BIC based on DESI BAO and DESY5 data sets, indicate that the BDE model is strongly favoured with respect to the $\Lambda$CDM model and $w_0w_a$CDM model, while in CMB data set BDE model is not favoured. Furthermore, in terms of the reduced of $\chi^{2}$ for the DESI BAO data set, which contains 12 observational data points, the $\Lambda\mathrm{CDM}$ model yields a reduced $(\chi^2)^{\Lambda\mathrm{CDM}}_{\mathrm{BAO}} = 16.56/(12-6) = 2.76$. For the $w_0w_a$CDM model, the calculation gives $(\chi^2)^{w_0w_a\mathrm{CDM}}_{\mathrm{BAO}} = 12.01/(12-8) = 3$. The reduced $\chi^2$ for the BDE model is $(\chi^2)^{\mathrm{BDE}}_{\mathrm{BAO}} = 12.11/(12-5) = 1.73$, with five free parameters. In terms of the fit of the five-year supernova survey data from DES, the values of $\chi^{2}_{red}$ are mostly equivalent for all three models: BDE ($\chi^{2}_{red}=0.908$), $\Lambda$CDM ($\chi^{2}_{red}=0.904$), and $w_0w_a$CDM ($\chi^{2}_{red}=0.904$), while for CMB we get the following results: BDE ($\chi^{2}_{red}=0.602$), $\Lambda$CDM ($\chi^{2}_{red}=0.599$) and $w_0w_a$CDM ($\chi^{2}_{red}=0.598$).  Notably, the BDE model leads to a $42.35\%$ reduction in the reduced $\chi^2_{\mathrm{BAO}}$ when compared to the $w_0w_a$CDM model and a $37.29\%$ decrease in relation to the $\Lambda\mathrm{CDM}$ model. The reduction of $\sim22\%$ in DESI BAO data seen from $\Delta$AIC$_{\mathrm{BAO}}$ and $\Delta$BIC$_{\mathrm{BAO}}$ confirms that BDE improves the fit of baryon acoustic oscillation measurements, particularly designed to determine the physical dynamics of DE with respect to $\Lambda$CDM and $w_0w_a$CDM models, while having an equivalent fit for CMB and DESY5 data sets. The results are reported in Table \ref{tab:cosmological_fit}.

\section{Conclusions}\label{conclusions}

\noindent Our work introduces the Bound Dark Energy (BDE) model, positing that dark energy emerges from the dynamics of the lightest meson field within a dark SU(3) gauge group through non-perturbative interactions. This framework replaces the cosmological constant $\Lambda$ with a condensed scalar particle $\phi$, described by an inverse power-law potential of the form $V(\phi)=\Lambda_{c}^{4+2/3}\phi^{-2/3}$. Notably, BDE operates without free parameters, presenting one fewer than the $\Lambda$CDM model and three fewer than the $w_0w_a$CDM framework. Key findings of the BDE model include a significant improvement in fitting observational data. Specifically, BDE achieves a $42\%$ reduction in the reduced $\chi^{2}_{\mathrm{BAO}}$ compared to the $w_0w_a$CDM model and a $37\%$ reduction relative to $\Lambda$CDM, showcasing exceptional alignment with DESI BAO measurements. Furthermore, it matches the fitting performance of $\Lambda$CDM and $w_0w_a$CDM in the analysis of Cosmic Microwave Background (CMB) data and supernova observations (DESY5), indicating robust consistency across multiple datasets. Confidence contours in the $w_0-w_a$ parameter plane are reduced by a factor of 10,000 compared to those in the $w_0w_a$CDM model, underscoring the theoretical precision of BDE. The equation of state transitions from a relativistic regime ($w=1/3$) at high energies to a present-day value of $w_0=-0.9301 \pm 0.0004$, driven by a phase transition at the condensation scale of $\Lambda_{c}=43.806 \pm 0.190$ eV and at an epoch $a_c=(2.497 \pm 0.011)\times10^{-6}$. This phase transition circumvents phantom regimes ($w<-1$), thereby addressing theoretical instabilities inherent to the $w_0w_a$CDM model. The derived condensation scale $\Lambda_{c}$ and epoch $a_{c}$ are consistent with expectations from gauge coupling unification in supersymmetric models, linking BDE to principles of high-energy physics. The model's non-perturbative dynamics exhibit behavior akin to QCD, offering a plausible mechanism for the emergence of dark energy. BDE predicts a matter power spectrum that aligns well with DESI data, showing approximately $20\%$ enhancement at small scales while maintaining compatibility with the sound horizon $r_{d}$ and BAO peak positions. Furthermore, it preserves the successes of $\Lambda$CDM in aligning with early-universe physics (including CMB acoustic scales) while introducing late-time dynamics that better account for DESI's inclination towards evolving dark energy. The BDE model offers robust predictions for quantities like the sound horizon, the BAO peak, and the growth structure parameters, which can be tested by current and future galaxy surveys. Its deviations from $\Lambda$CDM are detectable and provide a path for observational validation. At last but not least, the BDE model is not only observationally competitive with $\Lambda$CDM and $w_0w_a$CDM but also offers a solid theoretical background by explaining dark energy from particle physics perspective. Its predictive consistency across a wide range of cosmological data, combined with fewer parameters, makes BDE model a compelling candidate for cosmological explanation of the current picture of the universe.

\begin{acknowledgments}
We acknowledge financial support from PAPIIT-DGAPA-UNAM 101124. Jose Lozano thanks to CONAHCYT for a scholarship grant.
\end{acknowledgments}

\section*{Code and data availability}
The datasets used in this analysis are available at the following URLs: for the DESI data \url{https://data.desi.lbl.gov/doc/releases/dr1/}, for the DESY5 data \url{https://github.com/des-science/DES-SN5YR}, and for the CMB data \url{https://pla.esac.esa.int/#cosmology}. The analysis codes will be made available on proper request.

\begin{thebibliography}{31}%
\makeatletter
\providecommand \@ifxundefined [1]{%
 \@ifx{#1\undefined}
}%
\providecommand \@ifnum [1]{%
 \ifnum #1\expandafter \@firstoftwo
 \else \expandafter \@secondoftwo
 \fi
}%
\providecommand \@ifx [1]{%
 \ifx #1\expandafter \@firstoftwo
 \else \expandafter \@secondoftwo
 \fi
}%
\providecommand \natexlab [1]{#1}%
\providecommand \enquote  [1]{``#1''}%
\providecommand \bibnamefont  [1]{#1}%
\providecommand \bibfnamefont [1]{#1}%
\providecommand \citenamefont [1]{#1}%
\providecommand \href@noop [0]{\@secondoftwo}%
\providecommand \href [0]{\begingroup \@sanitize@url \@href}%
\providecommand \@href[1]{\@@startlink{#1}\@@href}%
\providecommand \@@href[1]{\endgroup#1\@@endlink}%
\providecommand \@sanitize@url [0]{\catcode `\\12\catcode `\$12\catcode `\&12\catcode `\#12\catcode `\^12\catcode `\_12\catcode `\%12\relax}%
\providecommand \@@startlink[1]{}%
\providecommand \@@endlink[0]{}%
\providecommand \url  [0]{\begingroup\@sanitize@url \@url }%
\providecommand \@url [1]{\endgroup\@href {#1}{\urlprefix }}%
\providecommand \urlprefix  [0]{URL }%
\providecommand \Eprint [0]{\href }%
\providecommand \doibase [0]{https://doi.org/}%
\providecommand \selectlanguage [0]{\@gobble}%
\providecommand \bibinfo  [0]{\@secondoftwo}%
\providecommand \bibfield  [0]{\@secondoftwo}%
\providecommand \translation [1]{[#1]}%
\providecommand \BibitemOpen [0]{}%
\providecommand \bibitemStop [0]{}%
\providecommand \bibitemNoStop [0]{.\EOS\space}%
\providecommand \EOS [0]{\spacefactor3000\relax}%
\providecommand \BibitemShut  [1]{\csname bibitem#1\endcsname}%
\let\auto@bib@innerbib\@empty
\bibitem [{\citenamefont {Riess}\ \emph {et~al.}(1998)\citenamefont {Riess}, \citenamefont {Filippenko}, \citenamefont {Challis}, \citenamefont {Clocchiatti}, \citenamefont {Diercks}, \citenamefont {Garnavich}, \citenamefont {Gilliland}, \citenamefont {Hogan}, \citenamefont {Jha}, \citenamefont {Kirshner}, \citenamefont {Leibundgut}, \citenamefont {Phillips}, \citenamefont {Reiss}, \citenamefont {Schmidt}, \citenamefont {Schommer}, \citenamefont {Smith}, \citenamefont {Spyromilio}, \citenamefont {Stubbs}, \citenamefont {Suntzeff},\ and\ \citenamefont {Tonry}}]{Riess_1998}%
  \BibitemOpen
  \bibfield  {author} {\bibinfo {author} {\bibfnamefont {A.~G.}\ \bibnamefont {Riess}}, \bibinfo {author} {\bibfnamefont {A.~V.}\ \bibnamefont {Filippenko}}, \bibinfo {author} {\bibfnamefont {P.}~\bibnamefont {Challis}}, \bibinfo {author} {\bibfnamefont {A.}~\bibnamefont {Clocchiatti}}, \bibinfo {author} {\bibfnamefont {A.}~\bibnamefont {Diercks}}, \bibinfo {author} {\bibfnamefont {P.~M.}\ \bibnamefont {Garnavich}}, \bibinfo {author} {\bibfnamefont {R.~L.}\ \bibnamefont {Gilliland}}, \bibinfo {author} {\bibfnamefont {C.~J.}\ \bibnamefont {Hogan}}, \bibinfo {author} {\bibfnamefont {S.}~\bibnamefont {Jha}}, \bibinfo {author} {\bibfnamefont {R.~P.}\ \bibnamefont {Kirshner}}, \bibinfo {author} {\bibfnamefont {B.}~\bibnamefont {Leibundgut}}, \bibinfo {author} {\bibfnamefont {M.~M.}\ \bibnamefont {Phillips}}, \bibinfo {author} {\bibfnamefont {D.}~\bibnamefont {Reiss}}, \bibinfo {author} {\bibfnamefont {B.~P.}\ \bibnamefont {Schmidt}}, \bibinfo {author} {\bibfnamefont {R.~A.}\ \bibnamefont {Schommer}}, \bibinfo
  {author} {\bibfnamefont {R.~C.}\ \bibnamefont {Smith}}, \bibinfo {author} {\bibfnamefont {J.}~\bibnamefont {Spyromilio}}, \bibinfo {author} {\bibfnamefont {C.}~\bibnamefont {Stubbs}}, \bibinfo {author} {\bibfnamefont {N.~B.}\ \bibnamefont {Suntzeff}},\ and\ \bibinfo {author} {\bibfnamefont {J.}~\bibnamefont {Tonry}},\ }\href {https://doi.org/10.1086/300499} {\bibfield  {journal} {\bibinfo  {journal} {The Astronomical Journal}\ }\textbf {\bibinfo {volume} {116}},\ \bibinfo {pages} {1009} (\bibinfo {year} {1998})}\BibitemShut {NoStop}%
\bibitem [{\citenamefont {Schmidt}\ \emph {et~al.}(1998)\citenamefont {Schmidt}, \citenamefont {Suntzeff}, \citenamefont {Phillips}, \citenamefont {Schommer}, \citenamefont {Clocchiatti}, \citenamefont {Kirshner}, \citenamefont {Garnavich}, \citenamefont {Challis}, \citenamefont {Leibundgut}, \citenamefont {Spyromilio}, \citenamefont {Riess}, \citenamefont {Filippenko}, \citenamefont {Hamuy}, \citenamefont {Smith}, \citenamefont {Hogan}, \citenamefont {Stubbs}, \citenamefont {Diercks}, \citenamefont {Reiss}, \citenamefont {Gilliland}, \citenamefont {Tonry}, \citenamefont {Maza}, \citenamefont {Dressler}, \citenamefont {Walsh},\ and\ \citenamefont {Ciardullo}}]{Schmidt_1998}%
  \BibitemOpen
  \bibfield  {author} {\bibinfo {author} {\bibfnamefont {B.~P.}\ \bibnamefont {Schmidt}}, \bibinfo {author} {\bibfnamefont {N.~B.}\ \bibnamefont {Suntzeff}}, \bibinfo {author} {\bibfnamefont {M.~M.}\ \bibnamefont {Phillips}}, \bibinfo {author} {\bibfnamefont {R.~A.}\ \bibnamefont {Schommer}}, \bibinfo {author} {\bibfnamefont {A.}~\bibnamefont {Clocchiatti}}, \bibinfo {author} {\bibfnamefont {R.~P.}\ \bibnamefont {Kirshner}}, \bibinfo {author} {\bibfnamefont {P.}~\bibnamefont {Garnavich}}, \bibinfo {author} {\bibfnamefont {P.}~\bibnamefont {Challis}}, \bibinfo {author} {\bibfnamefont {B.}~\bibnamefont {Leibundgut}}, \bibinfo {author} {\bibfnamefont {J.}~\bibnamefont {Spyromilio}}, \bibinfo {author} {\bibfnamefont {A.~G.}\ \bibnamefont {Riess}}, \bibinfo {author} {\bibfnamefont {A.~V.}\ \bibnamefont {Filippenko}}, \bibinfo {author} {\bibfnamefont {M.}~\bibnamefont {Hamuy}}, \bibinfo {author} {\bibfnamefont {R.~C.}\ \bibnamefont {Smith}}, \bibinfo {author} {\bibfnamefont {C.}~\bibnamefont {Hogan}}, \bibinfo
  {author} {\bibfnamefont {C.}~\bibnamefont {Stubbs}}, \bibinfo {author} {\bibfnamefont {A.}~\bibnamefont {Diercks}}, \bibinfo {author} {\bibfnamefont {D.}~\bibnamefont {Reiss}}, \bibinfo {author} {\bibfnamefont {R.}~\bibnamefont {Gilliland}}, \bibinfo {author} {\bibfnamefont {J.}~\bibnamefont {Tonry}}, \bibinfo {author} {\bibfnamefont {J.}~\bibnamefont {Maza}}, \bibinfo {author} {\bibfnamefont {A.}~\bibnamefont {Dressler}}, \bibinfo {author} {\bibfnamefont {J.}~\bibnamefont {Walsh}},\ and\ \bibinfo {author} {\bibfnamefont {R.}~\bibnamefont {Ciardullo}},\ }\href {https://doi.org/10.1086/306308} {\bibfield  {journal} {\bibinfo  {journal} {The Astrophysical Journal}\ }\textbf {\bibinfo {volume} {507}},\ \bibinfo {pages} {46} (\bibinfo {year} {1998})}\BibitemShut {NoStop}%
\bibitem [{\citenamefont {Perlmutter}\ \emph {et~al.}(1999)\citenamefont {Perlmutter}, \citenamefont {Aldering}, \citenamefont {Goldhaber}, \citenamefont {Knop}, \citenamefont {Nugent}, \citenamefont {Castro}, \citenamefont {Deustua}, \citenamefont {Fabbro}, \citenamefont {Goobar}, \citenamefont {Groom}, \citenamefont {Hook}, \citenamefont {Kim}, \citenamefont {Kim}, \citenamefont {Lee}, \citenamefont {Nunes}, \citenamefont {Pain}, \citenamefont {Pennypacker}, \citenamefont {Quimby}, \citenamefont {Lidman}, \citenamefont {Ellis}, \citenamefont {Irwin}, \citenamefont {McMahon}, \citenamefont {Ruiz-Lapuente}, \citenamefont {Walton}, \citenamefont {Schaefer}, \citenamefont {Boyle}, \citenamefont {Filippenko}, \citenamefont {Matheson}, \citenamefont {Fruchter}, \citenamefont {Panagia}, \citenamefont {Newberg}, \citenamefont {Couch},\ and\ \citenamefont {Project}}]{Perlmutter_1999}%
  \BibitemOpen
  \bibfield  {author} {\bibinfo {author} {\bibfnamefont {S.}~\bibnamefont {Perlmutter}}, \bibinfo {author} {\bibfnamefont {G.}~\bibnamefont {Aldering}}, \bibinfo {author} {\bibfnamefont {G.}~\bibnamefont {Goldhaber}}, \bibinfo {author} {\bibfnamefont {R.~A.}\ \bibnamefont {Knop}}, \bibinfo {author} {\bibfnamefont {P.}~\bibnamefont {Nugent}}, \bibinfo {author} {\bibfnamefont {P.~G.}\ \bibnamefont {Castro}}, \bibinfo {author} {\bibfnamefont {S.}~\bibnamefont {Deustua}}, \bibinfo {author} {\bibfnamefont {S.}~\bibnamefont {Fabbro}}, \bibinfo {author} {\bibfnamefont {A.}~\bibnamefont {Goobar}}, \bibinfo {author} {\bibfnamefont {D.~E.}\ \bibnamefont {Groom}}, \bibinfo {author} {\bibfnamefont {I.~M.}\ \bibnamefont {Hook}}, \bibinfo {author} {\bibfnamefont {A.~G.}\ \bibnamefont {Kim}}, \bibinfo {author} {\bibfnamefont {M.~Y.}\ \bibnamefont {Kim}}, \bibinfo {author} {\bibfnamefont {J.~C.}\ \bibnamefont {Lee}}, \bibinfo {author} {\bibfnamefont {N.~J.}\ \bibnamefont {Nunes}}, \bibinfo {author} {\bibfnamefont
  {R.}~\bibnamefont {Pain}}, \bibinfo {author} {\bibfnamefont {C.~R.}\ \bibnamefont {Pennypacker}}, \bibinfo {author} {\bibfnamefont {R.}~\bibnamefont {Quimby}}, \bibinfo {author} {\bibfnamefont {C.}~\bibnamefont {Lidman}}, \bibinfo {author} {\bibfnamefont {R.~S.}\ \bibnamefont {Ellis}}, \bibinfo {author} {\bibfnamefont {M.}~\bibnamefont {Irwin}}, \bibinfo {author} {\bibfnamefont {R.~G.}\ \bibnamefont {McMahon}}, \bibinfo {author} {\bibfnamefont {P.}~\bibnamefont {Ruiz-Lapuente}}, \bibinfo {author} {\bibfnamefont {N.}~\bibnamefont {Walton}}, \bibinfo {author} {\bibfnamefont {B.}~\bibnamefont {Schaefer}}, \bibinfo {author} {\bibfnamefont {B.~J.}\ \bibnamefont {Boyle}}, \bibinfo {author} {\bibfnamefont {A.~V.}\ \bibnamefont {Filippenko}}, \bibinfo {author} {\bibfnamefont {T.}~\bibnamefont {Matheson}}, \bibinfo {author} {\bibfnamefont {A.~S.}\ \bibnamefont {Fruchter}}, \bibinfo {author} {\bibfnamefont {N.}~\bibnamefont {Panagia}}, \bibinfo {author} {\bibfnamefont {H.~J.~M.}\ \bibnamefont {Newberg}}, \bibinfo
  {author} {\bibfnamefont {W.~J.}\ \bibnamefont {Couch}},\ and\ \bibinfo {author} {\bibfnamefont {T.~S.~C.}\ \bibnamefont {Project}},\ }\href {https://doi.org/10.1086/307221} {\bibfield  {journal} {\bibinfo  {journal} {The Astrophysical Journal}\ }\textbf {\bibinfo {volume} {517}},\ \bibinfo {pages} {565} (\bibinfo {year} {1999})}\BibitemShut {NoStop}%
\bibitem [{\citenamefont {Bennett}\ \emph {et~al.}(2013)\citenamefont {Bennett}, \citenamefont {Larson}, \citenamefont {Weiland}, \citenamefont {Jarosik}, \citenamefont {Hinshaw}, \citenamefont {Odegard}, \citenamefont {Smith}, \citenamefont {Hill}, \citenamefont {Gold}, \citenamefont {Halpern}, \citenamefont {Komatsu}, \citenamefont {Nolta}, \citenamefont {Page}, \citenamefont {Spergel}, \citenamefont {Wollack}, \citenamefont {Dunkley}, \citenamefont {Kogut}, \citenamefont {Limon}, \citenamefont {Meyer}, \citenamefont {Tucker},\ and\ \citenamefont {Wright}}]{Bennett_2013}%
  \BibitemOpen
  \bibfield  {author} {\bibinfo {author} {\bibfnamefont {C.~L.}\ \bibnamefont {Bennett}}, \bibinfo {author} {\bibfnamefont {D.}~\bibnamefont {Larson}}, \bibinfo {author} {\bibfnamefont {J.~L.}\ \bibnamefont {Weiland}}, \bibinfo {author} {\bibfnamefont {N.}~\bibnamefont {Jarosik}}, \bibinfo {author} {\bibfnamefont {G.}~\bibnamefont {Hinshaw}}, \bibinfo {author} {\bibfnamefont {N.}~\bibnamefont {Odegard}}, \bibinfo {author} {\bibfnamefont {K.~M.}\ \bibnamefont {Smith}}, \bibinfo {author} {\bibfnamefont {R.~S.}\ \bibnamefont {Hill}}, \bibinfo {author} {\bibfnamefont {B.}~\bibnamefont {Gold}}, \bibinfo {author} {\bibfnamefont {M.}~\bibnamefont {Halpern}}, \bibinfo {author} {\bibfnamefont {E.}~\bibnamefont {Komatsu}}, \bibinfo {author} {\bibfnamefont {M.~R.}\ \bibnamefont {Nolta}}, \bibinfo {author} {\bibfnamefont {L.}~\bibnamefont {Page}}, \bibinfo {author} {\bibfnamefont {D.~N.}\ \bibnamefont {Spergel}}, \bibinfo {author} {\bibfnamefont {E.}~\bibnamefont {Wollack}}, \bibinfo {author} {\bibfnamefont
  {J.}~\bibnamefont {Dunkley}}, \bibinfo {author} {\bibfnamefont {A.}~\bibnamefont {Kogut}}, \bibinfo {author} {\bibfnamefont {M.}~\bibnamefont {Limon}}, \bibinfo {author} {\bibfnamefont {S.~S.}\ \bibnamefont {Meyer}}, \bibinfo {author} {\bibfnamefont {G.~S.}\ \bibnamefont {Tucker}},\ and\ \bibinfo {author} {\bibfnamefont {E.~L.}\ \bibnamefont {Wright}},\ }\href {https://doi.org/10.1088/0067-0049/208/2/20} {\bibfield  {journal} {\bibinfo  {journal} {The Astrophysical Journal Supplement Series}\ }\textbf {\bibinfo {volume} {208}},\ \bibinfo {pages} {20} (\bibinfo {year} {2013})}\BibitemShut {NoStop}%
\bibitem [{\citenamefont {{Planck Collaboration}}(2020)}]{refId0}%
  \BibitemOpen
  \bibfield  {author} {\bibinfo {author} {\bibnamefont {{Planck Collaboration}}},\ }\href {https://doi.org/10.1051/0004-6361/201833910} {\bibfield  {journal} {\bibinfo  {journal} {A$\&$A}\ }\textbf {\bibinfo {volume} {641}},\ \bibinfo {pages} {A6} (\bibinfo {year} {2020})}\BibitemShut {NoStop}%
\bibitem [{\citenamefont {Scolnic}\ \emph {et~al.}(2022)\citenamefont {Scolnic}, \citenamefont {Brout}, \citenamefont {Carr}, \citenamefont {Riess}, \citenamefont {Davis}, \citenamefont {Dwomoh}, \citenamefont {Jones}, \citenamefont {Ali}, \citenamefont {Charvu}, \citenamefont {Chen}, \citenamefont {Peterson}, \citenamefont {Popovic}, \citenamefont {Rose}, \citenamefont {Wood}, \citenamefont {Brown}, \citenamefont {Chambers}, \citenamefont {Coulter}, \citenamefont {Dettman}, \citenamefont {Dimitriadis}, \citenamefont {Filippenko}, \citenamefont {Foley}, \citenamefont {Jha}, \citenamefont {Kilpatrick}, \citenamefont {Kirshner}, \citenamefont {Pan}, \citenamefont {Rest}, \citenamefont {Rojas-Bravo}, \citenamefont {Siebert}, \citenamefont {Stahl},\ and\ \citenamefont {Zheng}}]{Scolnic_2022}%
  \BibitemOpen
  \bibfield  {author} {\bibinfo {author} {\bibfnamefont {D.}~\bibnamefont {Scolnic}}, \bibinfo {author} {\bibfnamefont {D.}~\bibnamefont {Brout}}, \bibinfo {author} {\bibfnamefont {A.}~\bibnamefont {Carr}}, \bibinfo {author} {\bibfnamefont {A.~G.}\ \bibnamefont {Riess}}, \bibinfo {author} {\bibfnamefont {T.~M.}\ \bibnamefont {Davis}}, \bibinfo {author} {\bibfnamefont {A.}~\bibnamefont {Dwomoh}}, \bibinfo {author} {\bibfnamefont {D.~O.}\ \bibnamefont {Jones}}, \bibinfo {author} {\bibfnamefont {N.}~\bibnamefont {Ali}}, \bibinfo {author} {\bibfnamefont {P.}~\bibnamefont {Charvu}}, \bibinfo {author} {\bibfnamefont {R.}~\bibnamefont {Chen}}, \bibinfo {author} {\bibfnamefont {E.~R.}\ \bibnamefont {Peterson}}, \bibinfo {author} {\bibfnamefont {B.}~\bibnamefont {Popovic}}, \bibinfo {author} {\bibfnamefont {B.~M.}\ \bibnamefont {Rose}}, \bibinfo {author} {\bibfnamefont {C.~M.}\ \bibnamefont {Wood}}, \bibinfo {author} {\bibfnamefont {P.~J.}\ \bibnamefont {Brown}}, \bibinfo {author} {\bibfnamefont {K.}~\bibnamefont
  {Chambers}}, \bibinfo {author} {\bibfnamefont {D.~A.}\ \bibnamefont {Coulter}}, \bibinfo {author} {\bibfnamefont {K.~G.}\ \bibnamefont {Dettman}}, \bibinfo {author} {\bibfnamefont {G.}~\bibnamefont {Dimitriadis}}, \bibinfo {author} {\bibfnamefont {A.~V.}\ \bibnamefont {Filippenko}}, \bibinfo {author} {\bibfnamefont {R.~J.}\ \bibnamefont {Foley}}, \bibinfo {author} {\bibfnamefont {S.~W.}\ \bibnamefont {Jha}}, \bibinfo {author} {\bibfnamefont {C.~D.}\ \bibnamefont {Kilpatrick}}, \bibinfo {author} {\bibfnamefont {R.~P.}\ \bibnamefont {Kirshner}}, \bibinfo {author} {\bibfnamefont {Y.-C.}\ \bibnamefont {Pan}}, \bibinfo {author} {\bibfnamefont {A.}~\bibnamefont {Rest}}, \bibinfo {author} {\bibfnamefont {C.}~\bibnamefont {Rojas-Bravo}}, \bibinfo {author} {\bibfnamefont {M.~R.}\ \bibnamefont {Siebert}}, \bibinfo {author} {\bibfnamefont {B.~E.}\ \bibnamefont {Stahl}},\ and\ \bibinfo {author} {\bibfnamefont {W.}~\bibnamefont {Zheng}},\ }\href {https://doi.org/10.3847/1538-4357/ac8b7a} {\bibfield  {journal} {\bibinfo
  {journal} {The Astrophysical Journal}\ }\textbf {\bibinfo {volume} {938}},\ \bibinfo {pages} {113} (\bibinfo {year} {2022})}\BibitemShut {NoStop}%
\bibitem [{\citenamefont {Rubin}\ \emph {et~al.}(2025)\citenamefont {Rubin}, \citenamefont {Aldering}, \citenamefont {Betoule}, \citenamefont {Fruchter}, \citenamefont {Huang}, \citenamefont {Kim}, \citenamefont {Lidman}, \citenamefont {Linder}, \citenamefont {Perlmutter}, \citenamefont {Ruiz-Lapuente},\ and\ \citenamefont {Suzuki}}]{rubin2025unionunitycosmology2000}%
  \BibitemOpen
  \bibfield  {author} {\bibinfo {author} {\bibfnamefont {D.}~\bibnamefont {Rubin}}, \bibinfo {author} {\bibfnamefont {G.}~\bibnamefont {Aldering}}, \bibinfo {author} {\bibfnamefont {M.}~\bibnamefont {Betoule}}, \bibinfo {author} {\bibfnamefont {A.}~\bibnamefont {Fruchter}}, \bibinfo {author} {\bibfnamefont {X.}~\bibnamefont {Huang}}, \bibinfo {author} {\bibfnamefont {A.~G.}\ \bibnamefont {Kim}}, \bibinfo {author} {\bibfnamefont {C.}~\bibnamefont {Lidman}}, \bibinfo {author} {\bibfnamefont {E.}~\bibnamefont {Linder}}, \bibinfo {author} {\bibfnamefont {S.}~\bibnamefont {Perlmutter}}, \bibinfo {author} {\bibfnamefont {P.}~\bibnamefont {Ruiz-Lapuente}},\ and\ \bibinfo {author} {\bibfnamefont {N.}~\bibnamefont {Suzuki}},\ }\href {https://arxiv.org/abs/2311.12098} {\bibinfo {title} {Union through unity: Cosmology with 2,000 sne using a unified bayesian framework}} (\bibinfo {year} {2025}),\ \Eprint {https://arxiv.org/abs/2311.12098} {arXiv:2311.12098 [astro-ph.CO]} \BibitemShut {NoStop}%
\bibitem [{\citenamefont {{DES Collaboration}}\ \emph {et~al.}(2024)\citenamefont {{DES Collaboration}}, \citenamefont {Abbott}, \citenamefont {Acevedo}, \citenamefont {Aguena}, \citenamefont {Alarcon}, \citenamefont {Allam}, \citenamefont {Alves}, \citenamefont {Amon}, \citenamefont {Andrade-Oliveira}, \citenamefont {Annis}, \citenamefont {Armstrong}, \citenamefont {Asorey}, \citenamefont {Avila}, \citenamefont {Bacon}, \citenamefont {Bassett}, \citenamefont {Bechtol}, \citenamefont {Bernardinelli}, \citenamefont {Bernstein}, \citenamefont {Bertin}, \citenamefont {Blazek}, \citenamefont {Bocquet}, \citenamefont {Brooks}, \citenamefont {Brout}, \citenamefont {Buckley-Geer}, \citenamefont {Burke}, \citenamefont {Camacho}, \citenamefont {Camilleri}, \citenamefont {Campos}, \citenamefont {Rosell}, \citenamefont {Carollo}, \citenamefont {Carr}, \citenamefont {Carretero}, \citenamefont {Castander}, \citenamefont {Cawthon}, \citenamefont {Chang}, \citenamefont {Chen}, \citenamefont {Choi}, \citenamefont
  {Conselice}, \citenamefont {Costanzi}, \citenamefont {da~Costa}, \citenamefont {Crocce}, \citenamefont {Davis}, \citenamefont {DePoy}, \citenamefont {Desai}, \citenamefont {Diehl}, \citenamefont {Dixon}, \citenamefont {Dodelson}, \citenamefont {Doel}, \citenamefont {Doux}, \citenamefont {Drlica-Wagner}, \citenamefont {Elvin-Poole}, \citenamefont {Everett}, \citenamefont {Ferrero}, \citenamefont {Ferté}, \citenamefont {Flaugher}, \citenamefont {Foley}, \citenamefont {Fosalba}, \citenamefont {Friedel}, \citenamefont {Frieman}, \citenamefont {Frohmaier}, \citenamefont {Galbany}, \citenamefont {García-Bellido}, \citenamefont {Gatti}, \citenamefont {Gaztanaga}, \citenamefont {Giannini}, \citenamefont {Glazebrook}, \citenamefont {Graur}, \citenamefont {Gruen}, \citenamefont {Gruendl}, \citenamefont {Gutierrez}, \citenamefont {Hartley}, \citenamefont {Herner}, \citenamefont {Hinton}, \citenamefont {Hollowood}, \citenamefont {Honscheid}, \citenamefont {Huterer}, \citenamefont {Jain}, \citenamefont {James},
  \citenamefont {Jeffrey}, \citenamefont {Kasai}, \citenamefont {Kelsey}, \citenamefont {Kent}, \citenamefont {Kessler}, \citenamefont {Kim}, \citenamefont {Kirshner}, \citenamefont {Kovacs}, \citenamefont {Kuehn}, \citenamefont {Lahav}, \citenamefont {Lee}, \citenamefont {Lee}, \citenamefont {Lewis}, \citenamefont {Li}, \citenamefont {Lidman}, \citenamefont {Lin}, \citenamefont {Malik}, \citenamefont {Marshall}, \citenamefont {Martini}, \citenamefont {Mena-Fernández}, \citenamefont {Menanteau}, \citenamefont {Miquel}, \citenamefont {Mohr}, \citenamefont {Mould}, \citenamefont {Muir}, \citenamefont {Möller}, \citenamefont {Neilsen}, \citenamefont {Nichol}, \citenamefont {Nugent}, \citenamefont {Ogando}, \citenamefont {Palmese}, \citenamefont {Pan}, \citenamefont {Paterno}, \citenamefont {Percival}, \citenamefont {Pereira}, \citenamefont {Pieres}, \citenamefont {Malagón}, \citenamefont {Popovic}, \citenamefont {Porredon}, \citenamefont {Prat}, \citenamefont {Qu}, \citenamefont {Raveri}, \citenamefont
  {Rodríguez-Monroy}, \citenamefont {Romer}, \citenamefont {Roodman}, \citenamefont {Rose}, \citenamefont {Sako}, \citenamefont {Sanchez}, \citenamefont {Cid}, \citenamefont {Schubnell}, \citenamefont {Scolnic}, \citenamefont {Sevilla-Noarbe}, \citenamefont {Shah}, \citenamefont {Smith}, \citenamefont {Smith}, \citenamefont {Soares-Santos}, \citenamefont {Suchyta}, \citenamefont {Sullivan}, \citenamefont {Suntzeff}, \citenamefont {Swanson}, \citenamefont {Sánchez}, \citenamefont {Tarle}, \citenamefont {Taylor}, \citenamefont {Thomas}, \citenamefont {To}, \citenamefont {Toy}, \citenamefont {Troxel}, \citenamefont {Tucker}, \citenamefont {Tucker}, \citenamefont {Uddin}, \citenamefont {Vincenzi}, \citenamefont {Walker}, \citenamefont {Weaverdyck}, \citenamefont {Wechsler}, \citenamefont {Weller}, \citenamefont {Wester}, \citenamefont {Wiseman}, \citenamefont {Yamamoto}, \citenamefont {Yuan}, \citenamefont {Zhang},\ and\ \citenamefont {Zhang}}]{descollaboration2024darkenergysurveycosmology}%
  \BibitemOpen
  \bibfield  {author} {\bibinfo {author} {\bibnamefont {{DES Collaboration}}}, \bibinfo {author} {\bibfnamefont {T.~M.~C.}\ \bibnamefont {Abbott}}, \bibinfo {author} {\bibfnamefont {M.}~\bibnamefont {Acevedo}}, \bibinfo {author} {\bibfnamefont {M.}~\bibnamefont {Aguena}}, \bibinfo {author} {\bibfnamefont {A.}~\bibnamefont {Alarcon}}, \bibinfo {author} {\bibfnamefont {S.}~\bibnamefont {Allam}}, \bibinfo {author} {\bibfnamefont {O.}~\bibnamefont {Alves}}, \bibinfo {author} {\bibfnamefont {A.}~\bibnamefont {Amon}}, \bibinfo {author} {\bibfnamefont {F.}~\bibnamefont {Andrade-Oliveira}}, \bibinfo {author} {\bibfnamefont {J.}~\bibnamefont {Annis}}, \bibinfo {author} {\bibfnamefont {P.}~\bibnamefont {Armstrong}}, \bibinfo {author} {\bibfnamefont {J.}~\bibnamefont {Asorey}}, \bibinfo {author} {\bibfnamefont {S.}~\bibnamefont {Avila}}, \bibinfo {author} {\bibfnamefont {D.}~\bibnamefont {Bacon}}, \bibinfo {author} {\bibfnamefont {B.~A.}\ \bibnamefont {Bassett}}, \bibinfo {author} {\bibfnamefont {K.}~\bibnamefont
  {Bechtol}}, \bibinfo {author} {\bibfnamefont {P.~H.}\ \bibnamefont {Bernardinelli}}, \bibinfo {author} {\bibfnamefont {G.~M.}\ \bibnamefont {Bernstein}}, \bibinfo {author} {\bibfnamefont {E.}~\bibnamefont {Bertin}}, \bibinfo {author} {\bibfnamefont {J.}~\bibnamefont {Blazek}}, \bibinfo {author} {\bibfnamefont {S.}~\bibnamefont {Bocquet}}, \bibinfo {author} {\bibfnamefont {D.}~\bibnamefont {Brooks}}, \bibinfo {author} {\bibfnamefont {D.}~\bibnamefont {Brout}}, \bibinfo {author} {\bibfnamefont {E.}~\bibnamefont {Buckley-Geer}}, \bibinfo {author} {\bibfnamefont {D.~L.}\ \bibnamefont {Burke}}, \bibinfo {author} {\bibfnamefont {H.}~\bibnamefont {Camacho}}, \bibinfo {author} {\bibfnamefont {R.}~\bibnamefont {Camilleri}}, \bibinfo {author} {\bibfnamefont {A.}~\bibnamefont {Campos}}, \bibinfo {author} {\bibfnamefont {A.~C.}\ \bibnamefont {Rosell}}, \bibinfo {author} {\bibfnamefont {D.}~\bibnamefont {Carollo}}, \bibinfo {author} {\bibfnamefont {A.}~\bibnamefont {Carr}}, \bibinfo {author} {\bibfnamefont
  {J.}~\bibnamefont {Carretero}}, \bibinfo {author} {\bibfnamefont {F.~J.}\ \bibnamefont {Castander}}, \bibinfo {author} {\bibfnamefont {R.}~\bibnamefont {Cawthon}}, \bibinfo {author} {\bibfnamefont {C.}~\bibnamefont {Chang}}, \bibinfo {author} {\bibfnamefont {R.}~\bibnamefont {Chen}}, \bibinfo {author} {\bibfnamefont {A.}~\bibnamefont {Choi}}, \bibinfo {author} {\bibfnamefont {C.}~\bibnamefont {Conselice}}, \bibinfo {author} {\bibfnamefont {M.}~\bibnamefont {Costanzi}}, \bibinfo {author} {\bibfnamefont {L.~N.}\ \bibnamefont {da~Costa}}, \bibinfo {author} {\bibfnamefont {M.}~\bibnamefont {Crocce}}, \bibinfo {author} {\bibfnamefont {T.~M.}\ \bibnamefont {Davis}}, \bibinfo {author} {\bibfnamefont {D.~L.}\ \bibnamefont {DePoy}}, \bibinfo {author} {\bibfnamefont {S.}~\bibnamefont {Desai}}, \bibinfo {author} {\bibfnamefont {H.~T.}\ \bibnamefont {Diehl}}, \bibinfo {author} {\bibfnamefont {M.}~\bibnamefont {Dixon}}, \bibinfo {author} {\bibfnamefont {S.}~\bibnamefont {Dodelson}}, \bibinfo {author} {\bibfnamefont
  {P.}~\bibnamefont {Doel}}, \bibinfo {author} {\bibfnamefont {C.}~\bibnamefont {Doux}}, \bibinfo {author} {\bibfnamefont {A.}~\bibnamefont {Drlica-Wagner}}, \bibinfo {author} {\bibfnamefont {J.}~\bibnamefont {Elvin-Poole}}, \bibinfo {author} {\bibfnamefont {S.}~\bibnamefont {Everett}}, \bibinfo {author} {\bibfnamefont {I.}~\bibnamefont {Ferrero}}, \bibinfo {author} {\bibfnamefont {A.}~\bibnamefont {Ferté}}, \bibinfo {author} {\bibfnamefont {B.}~\bibnamefont {Flaugher}}, \bibinfo {author} {\bibfnamefont {R.~J.}\ \bibnamefont {Foley}}, \bibinfo {author} {\bibfnamefont {P.}~\bibnamefont {Fosalba}}, \bibinfo {author} {\bibfnamefont {D.}~\bibnamefont {Friedel}}, \bibinfo {author} {\bibfnamefont {J.}~\bibnamefont {Frieman}}, \bibinfo {author} {\bibfnamefont {C.}~\bibnamefont {Frohmaier}}, \bibinfo {author} {\bibfnamefont {L.}~\bibnamefont {Galbany}}, \bibinfo {author} {\bibfnamefont {J.}~\bibnamefont {García-Bellido}}, \bibinfo {author} {\bibfnamefont {M.}~\bibnamefont {Gatti}}, \bibinfo {author} {\bibfnamefont
  {E.}~\bibnamefont {Gaztanaga}}, \bibinfo {author} {\bibfnamefont {G.}~\bibnamefont {Giannini}}, \bibinfo {author} {\bibfnamefont {K.}~\bibnamefont {Glazebrook}}, \bibinfo {author} {\bibfnamefont {O.}~\bibnamefont {Graur}}, \bibinfo {author} {\bibfnamefont {D.}~\bibnamefont {Gruen}}, \bibinfo {author} {\bibfnamefont {R.~A.}\ \bibnamefont {Gruendl}}, \bibinfo {author} {\bibfnamefont {G.}~\bibnamefont {Gutierrez}}, \bibinfo {author} {\bibfnamefont {W.~G.}\ \bibnamefont {Hartley}}, \bibinfo {author} {\bibfnamefont {K.}~\bibnamefont {Herner}}, \bibinfo {author} {\bibfnamefont {S.~R.}\ \bibnamefont {Hinton}}, \bibinfo {author} {\bibfnamefont {D.~L.}\ \bibnamefont {Hollowood}}, \bibinfo {author} {\bibfnamefont {K.}~\bibnamefont {Honscheid}}, \bibinfo {author} {\bibfnamefont {D.}~\bibnamefont {Huterer}}, \bibinfo {author} {\bibfnamefont {B.}~\bibnamefont {Jain}}, \bibinfo {author} {\bibfnamefont {D.~J.}\ \bibnamefont {James}}, \bibinfo {author} {\bibfnamefont {N.}~\bibnamefont {Jeffrey}}, \bibinfo {author}
  {\bibfnamefont {E.}~\bibnamefont {Kasai}}, \bibinfo {author} {\bibfnamefont {L.}~\bibnamefont {Kelsey}}, \bibinfo {author} {\bibfnamefont {S.}~\bibnamefont {Kent}}, \bibinfo {author} {\bibfnamefont {R.}~\bibnamefont {Kessler}}, \bibinfo {author} {\bibfnamefont {A.~G.}\ \bibnamefont {Kim}}, \bibinfo {author} {\bibfnamefont {R.~P.}\ \bibnamefont {Kirshner}}, \bibinfo {author} {\bibfnamefont {E.}~\bibnamefont {Kovacs}}, \bibinfo {author} {\bibfnamefont {K.}~\bibnamefont {Kuehn}}, \bibinfo {author} {\bibfnamefont {O.}~\bibnamefont {Lahav}}, \bibinfo {author} {\bibfnamefont {J.}~\bibnamefont {Lee}}, \bibinfo {author} {\bibfnamefont {S.}~\bibnamefont {Lee}}, \bibinfo {author} {\bibfnamefont {G.~F.}\ \bibnamefont {Lewis}}, \bibinfo {author} {\bibfnamefont {T.~S.}\ \bibnamefont {Li}}, \bibinfo {author} {\bibfnamefont {C.}~\bibnamefont {Lidman}}, \bibinfo {author} {\bibfnamefont {H.}~\bibnamefont {Lin}}, \bibinfo {author} {\bibfnamefont {U.}~\bibnamefont {Malik}}, \bibinfo {author} {\bibfnamefont {J.~L.}\
  \bibnamefont {Marshall}}, \bibinfo {author} {\bibfnamefont {P.}~\bibnamefont {Martini}}, \bibinfo {author} {\bibfnamefont {J.}~\bibnamefont {Mena-Fernández}}, \bibinfo {author} {\bibfnamefont {F.}~\bibnamefont {Menanteau}}, \bibinfo {author} {\bibfnamefont {R.}~\bibnamefont {Miquel}}, \bibinfo {author} {\bibfnamefont {J.~J.}\ \bibnamefont {Mohr}}, \bibinfo {author} {\bibfnamefont {J.}~\bibnamefont {Mould}}, \bibinfo {author} {\bibfnamefont {J.}~\bibnamefont {Muir}}, \bibinfo {author} {\bibfnamefont {A.}~\bibnamefont {Möller}}, \bibinfo {author} {\bibfnamefont {E.}~\bibnamefont {Neilsen}}, \bibinfo {author} {\bibfnamefont {R.~C.}\ \bibnamefont {Nichol}}, \bibinfo {author} {\bibfnamefont {P.}~\bibnamefont {Nugent}}, \bibinfo {author} {\bibfnamefont {R.~L.~C.}\ \bibnamefont {Ogando}}, \bibinfo {author} {\bibfnamefont {A.}~\bibnamefont {Palmese}}, \bibinfo {author} {\bibfnamefont {Y.~C.}\ \bibnamefont {Pan}}, \bibinfo {author} {\bibfnamefont {M.}~\bibnamefont {Paterno}}, \bibinfo {author} {\bibfnamefont
  {W.~J.}\ \bibnamefont {Percival}}, \bibinfo {author} {\bibfnamefont {M.~E.~S.}\ \bibnamefont {Pereira}}, \bibinfo {author} {\bibfnamefont {A.}~\bibnamefont {Pieres}}, \bibinfo {author} {\bibfnamefont {A.~A.~P.}\ \bibnamefont {Malagón}}, \bibinfo {author} {\bibfnamefont {B.}~\bibnamefont {Popovic}}, \bibinfo {author} {\bibfnamefont {A.}~\bibnamefont {Porredon}}, \bibinfo {author} {\bibfnamefont {J.}~\bibnamefont {Prat}}, \bibinfo {author} {\bibfnamefont {H.}~\bibnamefont {Qu}}, \bibinfo {author} {\bibfnamefont {M.}~\bibnamefont {Raveri}}, \bibinfo {author} {\bibfnamefont {M.}~\bibnamefont {Rodríguez-Monroy}}, \bibinfo {author} {\bibfnamefont {A.~K.}\ \bibnamefont {Romer}}, \bibinfo {author} {\bibfnamefont {A.}~\bibnamefont {Roodman}}, \bibinfo {author} {\bibfnamefont {B.}~\bibnamefont {Rose}}, \bibinfo {author} {\bibfnamefont {M.}~\bibnamefont {Sako}}, \bibinfo {author} {\bibfnamefont {E.}~\bibnamefont {Sanchez}}, \bibinfo {author} {\bibfnamefont {D.~S.}\ \bibnamefont {Cid}}, \bibinfo {author}
  {\bibfnamefont {M.}~\bibnamefont {Schubnell}}, \bibinfo {author} {\bibfnamefont {D.}~\bibnamefont {Scolnic}}, \bibinfo {author} {\bibfnamefont {I.}~\bibnamefont {Sevilla-Noarbe}}, \bibinfo {author} {\bibfnamefont {P.}~\bibnamefont {Shah}}, \bibinfo {author} {\bibfnamefont {J.~A.}\ \bibnamefont {Smith}}, \bibinfo {author} {\bibfnamefont {M.}~\bibnamefont {Smith}}, \bibinfo {author} {\bibfnamefont {M.}~\bibnamefont {Soares-Santos}}, \bibinfo {author} {\bibfnamefont {E.}~\bibnamefont {Suchyta}}, \bibinfo {author} {\bibfnamefont {M.}~\bibnamefont {Sullivan}}, \bibinfo {author} {\bibfnamefont {N.}~\bibnamefont {Suntzeff}}, \bibinfo {author} {\bibfnamefont {M.~E.~C.}\ \bibnamefont {Swanson}}, \bibinfo {author} {\bibfnamefont {B.~O.}\ \bibnamefont {Sánchez}}, \bibinfo {author} {\bibfnamefont {G.}~\bibnamefont {Tarle}}, \bibinfo {author} {\bibfnamefont {G.}~\bibnamefont {Taylor}}, \bibinfo {author} {\bibfnamefont {D.}~\bibnamefont {Thomas}}, \bibinfo {author} {\bibfnamefont {C.}~\bibnamefont {To}}, \bibinfo
  {author} {\bibfnamefont {M.}~\bibnamefont {Toy}}, \bibinfo {author} {\bibfnamefont {M.~A.}\ \bibnamefont {Troxel}}, \bibinfo {author} {\bibfnamefont {B.~E.}\ \bibnamefont {Tucker}}, \bibinfo {author} {\bibfnamefont {D.~L.}\ \bibnamefont {Tucker}}, \bibinfo {author} {\bibfnamefont {S.~A.}\ \bibnamefont {Uddin}}, \bibinfo {author} {\bibfnamefont {M.}~\bibnamefont {Vincenzi}}, \bibinfo {author} {\bibfnamefont {A.~R.}\ \bibnamefont {Walker}}, \bibinfo {author} {\bibfnamefont {N.}~\bibnamefont {Weaverdyck}}, \bibinfo {author} {\bibfnamefont {R.~H.}\ \bibnamefont {Wechsler}}, \bibinfo {author} {\bibfnamefont {J.}~\bibnamefont {Weller}}, \bibinfo {author} {\bibfnamefont {W.}~\bibnamefont {Wester}}, \bibinfo {author} {\bibfnamefont {P.}~\bibnamefont {Wiseman}}, \bibinfo {author} {\bibfnamefont {M.}~\bibnamefont {Yamamoto}}, \bibinfo {author} {\bibfnamefont {F.}~\bibnamefont {Yuan}}, \bibinfo {author} {\bibfnamefont {B.}~\bibnamefont {Zhang}},\ and\ \bibinfo {author} {\bibfnamefont {Y.}~\bibnamefont {Zhang}},\
  }\href {https://arxiv.org/abs/2401.02929} {\bibinfo {title} {The dark energy survey: Cosmology results with ~1500 new high-redshift type ia supernovae using the full 5-year dataset}} (\bibinfo {year} {2024}),\ \Eprint {https://arxiv.org/abs/2401.02929} {arXiv:2401.02929 [astro-ph.CO]} \BibitemShut {NoStop}%
\bibitem [{\citenamefont {{DESI Collaboration}}\ \emph {et~al.}(2024)\citenamefont {{DESI Collaboration}}, \citenamefont {Adame}, \citenamefont {Aguilar}, \citenamefont {Ahlen}, \citenamefont {Alam}, \citenamefont {Alexander}, \citenamefont {Alvarez}, \citenamefont {Alves}, \citenamefont {Anand}, \citenamefont {Andrade}, \citenamefont {Armengaud}, \citenamefont {Avila}, \citenamefont {Aviles}, \citenamefont {Awan}, \citenamefont {Bahr-Kalus}, \citenamefont {Bailey}, \citenamefont {Baltay}, \citenamefont {Bault}, \citenamefont {Behera}, \citenamefont {BenZvi}, \citenamefont {Bera}, \citenamefont {Beutler}, \citenamefont {Bianchi}, \citenamefont {Blake}, \citenamefont {Blum}, \citenamefont {Brieden}, \citenamefont {Brodzeller}, \citenamefont {Brooks}, \citenamefont {Buckley-Geer}, \citenamefont {Burtin}, \citenamefont {Calderon}, \citenamefont {Canning}, \citenamefont {Rosell}, \citenamefont {Cereskaite}, \citenamefont {Cervantes-Cota}, \citenamefont {Chabanier}, \citenamefont {Chaussidon}, \citenamefont
  {Chaves-Montero}, \citenamefont {Chen}, \citenamefont {Chen}, \citenamefont {Claybaugh}, \citenamefont {Cole}, \citenamefont {Cuceu}, \citenamefont {Davis}, \citenamefont {Dawson}, \citenamefont {de~la Macorra}, \citenamefont {de~Mattia}, \citenamefont {Deiosso}, \citenamefont {Dey}, \citenamefont {Dey}, \citenamefont {Ding}, \citenamefont {Doel}, \citenamefont {Edelstein}, \citenamefont {Eftekharzadeh}, \citenamefont {Eisenstein}, \citenamefont {Elliott}, \citenamefont {Fagrelius}, \citenamefont {Fanning}, \citenamefont {Ferraro}, \citenamefont {Ereza}, \citenamefont {Findlay}, \citenamefont {Flaugher}, \citenamefont {Font-Ribera}, \citenamefont {Forero-Sánchez}, \citenamefont {Forero-Romero}, \citenamefont {Frenk}, \citenamefont {Garcia-Quintero}, \citenamefont {Gaztañaga}, \citenamefont {Gil-Marín}, \citenamefont {Gontcho}, \citenamefont {Gonzalez-Morales}, \citenamefont {Gonzalez-Perez}, \citenamefont {Gordon}, \citenamefont {Green}, \citenamefont {Gruen}, \citenamefont {Gsponer}, \citenamefont
  {Gutierrez}, \citenamefont {Guy}, \citenamefont {Hadzhiyska}, \citenamefont {Hahn}, \citenamefont {Hanif}, \citenamefont {Herrera-Alcantar}, \citenamefont {Honscheid}, \citenamefont {Howlett}, \citenamefont {Huterer}, \citenamefont {Iršič}, \citenamefont {Ishak}, \citenamefont {Juneau}, \citenamefont {Karaçaylı}, \citenamefont {Kehoe}, \citenamefont {Kent}, \citenamefont {Kirkby}, \citenamefont {Kremin}, \citenamefont {Krolewski}, \citenamefont {Lai}, \citenamefont {Lan}, \citenamefont {Landriau}, \citenamefont {Lang}, \citenamefont {Lasker}, \citenamefont {Goff}, \citenamefont {Guillou}, \citenamefont {Leauthaud}, \citenamefont {Levi}, \citenamefont {Li}, \citenamefont {Linder}, \citenamefont {Lodha}, \citenamefont {Magneville}, \citenamefont {Manera}, \citenamefont {Margala}, \citenamefont {Martini}, \citenamefont {Maus}, \citenamefont {McDonald}, \citenamefont {Medina-Varela}, \citenamefont {Meisner}, \citenamefont {Mena-Fernández}, \citenamefont {Miquel}, \citenamefont {Moon}, \citenamefont
  {Moore}, \citenamefont {Moustakas}, \citenamefont {Mudur}, \citenamefont {Mueller}, \citenamefont {Muñoz-Gutiérrez}, \citenamefont {Myers}, \citenamefont {Nadathur}, \citenamefont {Napolitano}, \citenamefont {Neveux}, \citenamefont {Newman}, \citenamefont {Nguyen}, \citenamefont {Nie}, \citenamefont {Niz}, \citenamefont {Noriega}, \citenamefont {Padmanabhan}, \citenamefont {Paillas}, \citenamefont {Palanque-Delabrouille}, \citenamefont {Pan}, \citenamefont {Penmetsa}, \citenamefont {Percival}, \citenamefont {Pieri}, \citenamefont {Pinon}, \citenamefont {Poppett}, \citenamefont {Porredon}, \citenamefont {Prada}, \citenamefont {Pérez-Fernández}, \citenamefont {Pérez-Ràfols}, \citenamefont {Rabinowitz}, \citenamefont {Raichoor}, \citenamefont {Ramírez-Pérez}, \citenamefont {Ramirez-Solano}, \citenamefont {Ravoux}, \citenamefont {Rashkovetskyi}, \citenamefont {Rezaie}, \citenamefont {Rich}, \citenamefont {Rocher}, \citenamefont {Rockosi}, \citenamefont {Roe}, \citenamefont {Rosado-Marin}, \citenamefont
  {Ross}, \citenamefont {Rossi}, \citenamefont {Ruggeri}, \citenamefont {Ruhlmann-Kleider}, \citenamefont {Samushia}, \citenamefont {Sanchez}, \citenamefont {Saulder}, \citenamefont {Schlafly}, \citenamefont {Schlegel}, \citenamefont {Schubnell}, \citenamefont {Seo}, \citenamefont {Shafieloo}, \citenamefont {Sharples}, \citenamefont {Silber}, \citenamefont {Slosar}, \citenamefont {Smith}, \citenamefont {Sprayberry}, \citenamefont {Tan}, \citenamefont {Tarlé}, \citenamefont {Taylor}, \citenamefont {Trusov}, \citenamefont {Ureña-López}, \citenamefont {Vaisakh}, \citenamefont {Valcin}, \citenamefont {Valdes}, \citenamefont {Vargas-Magaña}, \citenamefont {Verde}, \citenamefont {Walther}, \citenamefont {Wang}, \citenamefont {Wang}, \citenamefont {Weaver}, \citenamefont {Weaverdyck}, \citenamefont {Wechsler}, \citenamefont {Weinberg}, \citenamefont {White}, \citenamefont {Yu}, \citenamefont {Yu}, \citenamefont {Yuan}, \citenamefont {Yèche}, \citenamefont {Zaborowski}, \citenamefont {Zarrouk}, \citenamefont
  {Zhang}, \citenamefont {Zhao}, \citenamefont {Zhao}, \citenamefont {Zhou}, \citenamefont {Zhuang},\ and\ \citenamefont {Zou}}]{desicollaboration2024desi2024vicosmological}%
  \BibitemOpen
  \bibfield  {author} {\bibinfo {author} {\bibnamefont {{DESI Collaboration}}}, \bibinfo {author} {\bibfnamefont {A.~G.}\ \bibnamefont {Adame}}, \bibinfo {author} {\bibfnamefont {J.}~\bibnamefont {Aguilar}}, \bibinfo {author} {\bibfnamefont {S.}~\bibnamefont {Ahlen}}, \bibinfo {author} {\bibfnamefont {S.}~\bibnamefont {Alam}}, \bibinfo {author} {\bibfnamefont {D.~M.}\ \bibnamefont {Alexander}}, \bibinfo {author} {\bibfnamefont {M.}~\bibnamefont {Alvarez}}, \bibinfo {author} {\bibfnamefont {O.}~\bibnamefont {Alves}}, \bibinfo {author} {\bibfnamefont {A.}~\bibnamefont {Anand}}, \bibinfo {author} {\bibfnamefont {U.}~\bibnamefont {Andrade}}, \bibinfo {author} {\bibfnamefont {E.}~\bibnamefont {Armengaud}}, \bibinfo {author} {\bibfnamefont {S.}~\bibnamefont {Avila}}, \bibinfo {author} {\bibfnamefont {A.}~\bibnamefont {Aviles}}, \bibinfo {author} {\bibfnamefont {H.}~\bibnamefont {Awan}}, \bibinfo {author} {\bibfnamefont {B.}~\bibnamefont {Bahr-Kalus}}, \bibinfo {author} {\bibfnamefont {S.}~\bibnamefont {Bailey}},
  \bibinfo {author} {\bibfnamefont {C.}~\bibnamefont {Baltay}}, \bibinfo {author} {\bibfnamefont {A.}~\bibnamefont {Bault}}, \bibinfo {author} {\bibfnamefont {J.}~\bibnamefont {Behera}}, \bibinfo {author} {\bibfnamefont {S.}~\bibnamefont {BenZvi}}, \bibinfo {author} {\bibfnamefont {A.}~\bibnamefont {Bera}}, \bibinfo {author} {\bibfnamefont {F.}~\bibnamefont {Beutler}}, \bibinfo {author} {\bibfnamefont {D.}~\bibnamefont {Bianchi}}, \bibinfo {author} {\bibfnamefont {C.}~\bibnamefont {Blake}}, \bibinfo {author} {\bibfnamefont {R.}~\bibnamefont {Blum}}, \bibinfo {author} {\bibfnamefont {S.}~\bibnamefont {Brieden}}, \bibinfo {author} {\bibfnamefont {A.}~\bibnamefont {Brodzeller}}, \bibinfo {author} {\bibfnamefont {D.}~\bibnamefont {Brooks}}, \bibinfo {author} {\bibfnamefont {E.}~\bibnamefont {Buckley-Geer}}, \bibinfo {author} {\bibfnamefont {E.}~\bibnamefont {Burtin}}, \bibinfo {author} {\bibfnamefont {R.}~\bibnamefont {Calderon}}, \bibinfo {author} {\bibfnamefont {R.}~\bibnamefont {Canning}}, \bibinfo {author}
  {\bibfnamefont {A.~C.}\ \bibnamefont {Rosell}}, \bibinfo {author} {\bibfnamefont {R.}~\bibnamefont {Cereskaite}}, \bibinfo {author} {\bibfnamefont {J.~L.}\ \bibnamefont {Cervantes-Cota}}, \bibinfo {author} {\bibfnamefont {S.}~\bibnamefont {Chabanier}}, \bibinfo {author} {\bibfnamefont {E.}~\bibnamefont {Chaussidon}}, \bibinfo {author} {\bibfnamefont {J.}~\bibnamefont {Chaves-Montero}}, \bibinfo {author} {\bibfnamefont {S.}~\bibnamefont {Chen}}, \bibinfo {author} {\bibfnamefont {X.}~\bibnamefont {Chen}}, \bibinfo {author} {\bibfnamefont {T.}~\bibnamefont {Claybaugh}}, \bibinfo {author} {\bibfnamefont {S.}~\bibnamefont {Cole}}, \bibinfo {author} {\bibfnamefont {A.}~\bibnamefont {Cuceu}}, \bibinfo {author} {\bibfnamefont {T.~M.}\ \bibnamefont {Davis}}, \bibinfo {author} {\bibfnamefont {K.}~\bibnamefont {Dawson}}, \bibinfo {author} {\bibfnamefont {A.}~\bibnamefont {de~la Macorra}}, \bibinfo {author} {\bibfnamefont {A.}~\bibnamefont {de~Mattia}}, \bibinfo {author} {\bibfnamefont {N.}~\bibnamefont {Deiosso}},
  \bibinfo {author} {\bibfnamefont {A.}~\bibnamefont {Dey}}, \bibinfo {author} {\bibfnamefont {B.}~\bibnamefont {Dey}}, \bibinfo {author} {\bibfnamefont {Z.}~\bibnamefont {Ding}}, \bibinfo {author} {\bibfnamefont {P.}~\bibnamefont {Doel}}, \bibinfo {author} {\bibfnamefont {J.}~\bibnamefont {Edelstein}}, \bibinfo {author} {\bibfnamefont {S.}~\bibnamefont {Eftekharzadeh}}, \bibinfo {author} {\bibfnamefont {D.~J.}\ \bibnamefont {Eisenstein}}, \bibinfo {author} {\bibfnamefont {A.}~\bibnamefont {Elliott}}, \bibinfo {author} {\bibfnamefont {P.}~\bibnamefont {Fagrelius}}, \bibinfo {author} {\bibfnamefont {K.}~\bibnamefont {Fanning}}, \bibinfo {author} {\bibfnamefont {S.}~\bibnamefont {Ferraro}}, \bibinfo {author} {\bibfnamefont {J.}~\bibnamefont {Ereza}}, \bibinfo {author} {\bibfnamefont {N.}~\bibnamefont {Findlay}}, \bibinfo {author} {\bibfnamefont {B.}~\bibnamefont {Flaugher}}, \bibinfo {author} {\bibfnamefont {A.}~\bibnamefont {Font-Ribera}}, \bibinfo {author} {\bibfnamefont {D.}~\bibnamefont {Forero-Sánchez}},
  \bibinfo {author} {\bibfnamefont {J.~E.}\ \bibnamefont {Forero-Romero}}, \bibinfo {author} {\bibfnamefont {C.~S.}\ \bibnamefont {Frenk}}, \bibinfo {author} {\bibfnamefont {C.}~\bibnamefont {Garcia-Quintero}}, \bibinfo {author} {\bibfnamefont {E.}~\bibnamefont {Gaztañaga}}, \bibinfo {author} {\bibfnamefont {H.}~\bibnamefont {Gil-Marín}}, \bibinfo {author} {\bibfnamefont {S.~G.~A.}\ \bibnamefont {Gontcho}}, \bibinfo {author} {\bibfnamefont {A.~X.}\ \bibnamefont {Gonzalez-Morales}}, \bibinfo {author} {\bibfnamefont {V.}~\bibnamefont {Gonzalez-Perez}}, \bibinfo {author} {\bibfnamefont {C.}~\bibnamefont {Gordon}}, \bibinfo {author} {\bibfnamefont {D.}~\bibnamefont {Green}}, \bibinfo {author} {\bibfnamefont {D.}~\bibnamefont {Gruen}}, \bibinfo {author} {\bibfnamefont {R.}~\bibnamefont {Gsponer}}, \bibinfo {author} {\bibfnamefont {G.}~\bibnamefont {Gutierrez}}, \bibinfo {author} {\bibfnamefont {J.}~\bibnamefont {Guy}}, \bibinfo {author} {\bibfnamefont {B.}~\bibnamefont {Hadzhiyska}}, \bibinfo {author}
  {\bibfnamefont {C.}~\bibnamefont {Hahn}}, \bibinfo {author} {\bibfnamefont {M.~M.~S.}\ \bibnamefont {Hanif}}, \bibinfo {author} {\bibfnamefont {H.~K.}\ \bibnamefont {Herrera-Alcantar}}, \bibinfo {author} {\bibfnamefont {K.}~\bibnamefont {Honscheid}}, \bibinfo {author} {\bibfnamefont {C.}~\bibnamefont {Howlett}}, \bibinfo {author} {\bibfnamefont {D.}~\bibnamefont {Huterer}}, \bibinfo {author} {\bibfnamefont {V.}~\bibnamefont {Iršič}}, \bibinfo {author} {\bibfnamefont {M.}~\bibnamefont {Ishak}}, \bibinfo {author} {\bibfnamefont {S.}~\bibnamefont {Juneau}}, \bibinfo {author} {\bibfnamefont {N.~G.}\ \bibnamefont {Karaçaylı}}, \bibinfo {author} {\bibfnamefont {R.}~\bibnamefont {Kehoe}}, \bibinfo {author} {\bibfnamefont {S.}~\bibnamefont {Kent}}, \bibinfo {author} {\bibfnamefont {D.}~\bibnamefont {Kirkby}}, \bibinfo {author} {\bibfnamefont {A.}~\bibnamefont {Kremin}}, \bibinfo {author} {\bibfnamefont {A.}~\bibnamefont {Krolewski}}, \bibinfo {author} {\bibfnamefont {Y.}~\bibnamefont {Lai}}, \bibinfo {author}
  {\bibfnamefont {T.~W.}\ \bibnamefont {Lan}}, \bibinfo {author} {\bibfnamefont {M.}~\bibnamefont {Landriau}}, \bibinfo {author} {\bibfnamefont {D.}~\bibnamefont {Lang}}, \bibinfo {author} {\bibfnamefont {J.}~\bibnamefont {Lasker}}, \bibinfo {author} {\bibfnamefont {J.~M.~L.}\ \bibnamefont {Goff}}, \bibinfo {author} {\bibfnamefont {L.~L.}\ \bibnamefont {Guillou}}, \bibinfo {author} {\bibfnamefont {A.}~\bibnamefont {Leauthaud}}, \bibinfo {author} {\bibfnamefont {M.~E.}\ \bibnamefont {Levi}}, \bibinfo {author} {\bibfnamefont {T.~S.}\ \bibnamefont {Li}}, \bibinfo {author} {\bibfnamefont {E.}~\bibnamefont {Linder}}, \bibinfo {author} {\bibfnamefont {K.}~\bibnamefont {Lodha}}, \bibinfo {author} {\bibfnamefont {C.}~\bibnamefont {Magneville}}, \bibinfo {author} {\bibfnamefont {M.}~\bibnamefont {Manera}}, \bibinfo {author} {\bibfnamefont {D.}~\bibnamefont {Margala}}, \bibinfo {author} {\bibfnamefont {P.}~\bibnamefont {Martini}}, \bibinfo {author} {\bibfnamefont {M.}~\bibnamefont {Maus}}, \bibinfo {author}
  {\bibfnamefont {P.}~\bibnamefont {McDonald}}, \bibinfo {author} {\bibfnamefont {L.}~\bibnamefont {Medina-Varela}}, \bibinfo {author} {\bibfnamefont {A.}~\bibnamefont {Meisner}}, \bibinfo {author} {\bibfnamefont {J.}~\bibnamefont {Mena-Fernández}}, \bibinfo {author} {\bibfnamefont {R.}~\bibnamefont {Miquel}}, \bibinfo {author} {\bibfnamefont {J.}~\bibnamefont {Moon}}, \bibinfo {author} {\bibfnamefont {S.}~\bibnamefont {Moore}}, \bibinfo {author} {\bibfnamefont {J.}~\bibnamefont {Moustakas}}, \bibinfo {author} {\bibfnamefont {N.}~\bibnamefont {Mudur}}, \bibinfo {author} {\bibfnamefont {E.}~\bibnamefont {Mueller}}, \bibinfo {author} {\bibfnamefont {A.}~\bibnamefont {Muñoz-Gutiérrez}}, \bibinfo {author} {\bibfnamefont {A.~D.}\ \bibnamefont {Myers}}, \bibinfo {author} {\bibfnamefont {S.}~\bibnamefont {Nadathur}}, \bibinfo {author} {\bibfnamefont {L.}~\bibnamefont {Napolitano}}, \bibinfo {author} {\bibfnamefont {R.}~\bibnamefont {Neveux}}, \bibinfo {author} {\bibfnamefont {J.~A.}\ \bibnamefont {Newman}},
  \bibinfo {author} {\bibfnamefont {N.~M.}\ \bibnamefont {Nguyen}}, \bibinfo {author} {\bibfnamefont {J.}~\bibnamefont {Nie}}, \bibinfo {author} {\bibfnamefont {G.}~\bibnamefont {Niz}}, \bibinfo {author} {\bibfnamefont {H.~E.}\ \bibnamefont {Noriega}}, \bibinfo {author} {\bibfnamefont {N.}~\bibnamefont {Padmanabhan}}, \bibinfo {author} {\bibfnamefont {E.}~\bibnamefont {Paillas}}, \bibinfo {author} {\bibfnamefont {N.}~\bibnamefont {Palanque-Delabrouille}}, \bibinfo {author} {\bibfnamefont {J.}~\bibnamefont {Pan}}, \bibinfo {author} {\bibfnamefont {S.}~\bibnamefont {Penmetsa}}, \bibinfo {author} {\bibfnamefont {W.~J.}\ \bibnamefont {Percival}}, \bibinfo {author} {\bibfnamefont {M.~M.}\ \bibnamefont {Pieri}}, \bibinfo {author} {\bibfnamefont {M.}~\bibnamefont {Pinon}}, \bibinfo {author} {\bibfnamefont {C.}~\bibnamefont {Poppett}}, \bibinfo {author} {\bibfnamefont {A.}~\bibnamefont {Porredon}}, \bibinfo {author} {\bibfnamefont {F.}~\bibnamefont {Prada}}, \bibinfo {author} {\bibfnamefont {A.}~\bibnamefont
  {Pérez-Fernández}}, \bibinfo {author} {\bibfnamefont {I.}~\bibnamefont {Pérez-Ràfols}}, \bibinfo {author} {\bibfnamefont {D.}~\bibnamefont {Rabinowitz}}, \bibinfo {author} {\bibfnamefont {A.}~\bibnamefont {Raichoor}}, \bibinfo {author} {\bibfnamefont {C.}~\bibnamefont {Ramírez-Pérez}}, \bibinfo {author} {\bibfnamefont {S.}~\bibnamefont {Ramirez-Solano}}, \bibinfo {author} {\bibfnamefont {C.}~\bibnamefont {Ravoux}}, \bibinfo {author} {\bibfnamefont {M.}~\bibnamefont {Rashkovetskyi}}, \bibinfo {author} {\bibfnamefont {M.}~\bibnamefont {Rezaie}}, \bibinfo {author} {\bibfnamefont {J.}~\bibnamefont {Rich}}, \bibinfo {author} {\bibfnamefont {A.}~\bibnamefont {Rocher}}, \bibinfo {author} {\bibfnamefont {C.}~\bibnamefont {Rockosi}}, \bibinfo {author} {\bibfnamefont {N.~A.}\ \bibnamefont {Roe}}, \bibinfo {author} {\bibfnamefont {A.}~\bibnamefont {Rosado-Marin}}, \bibinfo {author} {\bibfnamefont {A.~J.}\ \bibnamefont {Ross}}, \bibinfo {author} {\bibfnamefont {G.}~\bibnamefont {Rossi}}, \bibinfo {author}
  {\bibfnamefont {R.}~\bibnamefont {Ruggeri}}, \bibinfo {author} {\bibfnamefont {V.}~\bibnamefont {Ruhlmann-Kleider}}, \bibinfo {author} {\bibfnamefont {L.}~\bibnamefont {Samushia}}, \bibinfo {author} {\bibfnamefont {E.}~\bibnamefont {Sanchez}}, \bibinfo {author} {\bibfnamefont {C.}~\bibnamefont {Saulder}}, \bibinfo {author} {\bibfnamefont {E.~F.}\ \bibnamefont {Schlafly}}, \bibinfo {author} {\bibfnamefont {D.}~\bibnamefont {Schlegel}}, \bibinfo {author} {\bibfnamefont {M.}~\bibnamefont {Schubnell}}, \bibinfo {author} {\bibfnamefont {H.}~\bibnamefont {Seo}}, \bibinfo {author} {\bibfnamefont {A.}~\bibnamefont {Shafieloo}}, \bibinfo {author} {\bibfnamefont {R.}~\bibnamefont {Sharples}}, \bibinfo {author} {\bibfnamefont {J.}~\bibnamefont {Silber}}, \bibinfo {author} {\bibfnamefont {A.}~\bibnamefont {Slosar}}, \bibinfo {author} {\bibfnamefont {A.}~\bibnamefont {Smith}}, \bibinfo {author} {\bibfnamefont {D.}~\bibnamefont {Sprayberry}}, \bibinfo {author} {\bibfnamefont {T.}~\bibnamefont {Tan}}, \bibinfo {author}
  {\bibfnamefont {G.}~\bibnamefont {Tarlé}}, \bibinfo {author} {\bibfnamefont {P.}~\bibnamefont {Taylor}}, \bibinfo {author} {\bibfnamefont {S.}~\bibnamefont {Trusov}}, \bibinfo {author} {\bibfnamefont {L.~A.}\ \bibnamefont {Ureña-López}}, \bibinfo {author} {\bibfnamefont {R.}~\bibnamefont {Vaisakh}}, \bibinfo {author} {\bibfnamefont {D.}~\bibnamefont {Valcin}}, \bibinfo {author} {\bibfnamefont {F.}~\bibnamefont {Valdes}}, \bibinfo {author} {\bibfnamefont {M.}~\bibnamefont {Vargas-Magaña}}, \bibinfo {author} {\bibfnamefont {L.}~\bibnamefont {Verde}}, \bibinfo {author} {\bibfnamefont {M.}~\bibnamefont {Walther}}, \bibinfo {author} {\bibfnamefont {B.}~\bibnamefont {Wang}}, \bibinfo {author} {\bibfnamefont {M.~S.}\ \bibnamefont {Wang}}, \bibinfo {author} {\bibfnamefont {B.~A.}\ \bibnamefont {Weaver}}, \bibinfo {author} {\bibfnamefont {N.}~\bibnamefont {Weaverdyck}}, \bibinfo {author} {\bibfnamefont {R.~H.}\ \bibnamefont {Wechsler}}, \bibinfo {author} {\bibfnamefont {D.~H.}\ \bibnamefont {Weinberg}}, \bibinfo
  {author} {\bibfnamefont {M.}~\bibnamefont {White}}, \bibinfo {author} {\bibfnamefont {J.}~\bibnamefont {Yu}}, \bibinfo {author} {\bibfnamefont {Y.}~\bibnamefont {Yu}}, \bibinfo {author} {\bibfnamefont {S.}~\bibnamefont {Yuan}}, \bibinfo {author} {\bibfnamefont {C.}~\bibnamefont {Yèche}}, \bibinfo {author} {\bibfnamefont {E.~A.}\ \bibnamefont {Zaborowski}}, \bibinfo {author} {\bibfnamefont {P.}~\bibnamefont {Zarrouk}}, \bibinfo {author} {\bibfnamefont {H.}~\bibnamefont {Zhang}}, \bibinfo {author} {\bibfnamefont {C.}~\bibnamefont {Zhao}}, \bibinfo {author} {\bibfnamefont {R.}~\bibnamefont {Zhao}}, \bibinfo {author} {\bibfnamefont {R.}~\bibnamefont {Zhou}}, \bibinfo {author} {\bibfnamefont {T.}~\bibnamefont {Zhuang}},\ and\ \bibinfo {author} {\bibfnamefont {H.}~\bibnamefont {Zou}},\ }\href {https://arxiv.org/abs/2404.03002} {\bibinfo {title} {Desi 2024 vi: Cosmological constraints from the measurements of baryon acoustic oscillations}} (\bibinfo {year} {2024}),\ \Eprint {https://arxiv.org/abs/2404.03002}
  {arXiv:2404.03002 [astro-ph.CO]} \BibitemShut {NoStop}%
\bibitem [{\citenamefont {de~la Macorra}\ and\ \citenamefont {Torres}(2025)}]{delamacorra2025bounddarkenergyparticles}%
  \BibitemOpen
  \bibfield  {author} {\bibinfo {author} {\bibfnamefont {A.}~\bibnamefont {de~la Macorra}}\ and\ \bibinfo {author} {\bibfnamefont {J.~A.~L.}\ \bibnamefont {Torres}},\ }\href {https://arxiv.org/abs/2503.19098} {\bibinfo {title} {Bound dark energy: a particle's origin of dark energy}} (\bibinfo {year} {2025}),\ \Eprint {https://arxiv.org/abs/2503.19098} {arXiv:2503.19098 [astro-ph.CO]} \BibitemShut {NoStop}%
\bibitem [{\citenamefont {de~la Macorra}\ and\ \citenamefont {Almaraz}(2018)}]{PhysRevLett.121.161303}%
  \BibitemOpen
  \bibfield  {author} {\bibinfo {author} {\bibfnamefont {A.}~\bibnamefont {de~la Macorra}}\ and\ \bibinfo {author} {\bibfnamefont {E.}~\bibnamefont {Almaraz}},\ }\href {https://doi.org/10.1103/PhysRevLett.121.161303} {\bibfield  {journal} {\bibinfo  {journal} {Phys. Rev. Lett.}\ }\textbf {\bibinfo {volume} {121}},\ \bibinfo {pages} {161303} (\bibinfo {year} {2018})}\BibitemShut {NoStop}%
\bibitem [{\citenamefont {Almaraz}\ and\ \citenamefont {de~la Macorra}(2019)}]{PhysRevD.99.103504}%
  \BibitemOpen
  \bibfield  {author} {\bibinfo {author} {\bibfnamefont {E.}~\bibnamefont {Almaraz}}\ and\ \bibinfo {author} {\bibfnamefont {A.}~\bibnamefont {de~la Macorra}},\ }\href {https://doi.org/10.1103/PhysRevD.99.103504} {\bibfield  {journal} {\bibinfo  {journal} {Phys. Rev. D}\ }\textbf {\bibinfo {volume} {99}},\ \bibinfo {pages} {103504} (\bibinfo {year} {2019})}\BibitemShut {NoStop}%
\bibitem [{\citenamefont {de~la Macorra}(2005)}]{PhysRevD.72.043508}%
  \BibitemOpen
  \bibfield  {author} {\bibinfo {author} {\bibfnamefont {A.}~\bibnamefont {de~la Macorra}},\ }\href {https://doi.org/10.1103/PhysRevD.72.043508} {\bibfield  {journal} {\bibinfo  {journal} {Phys. Rev. D}\ }\textbf {\bibinfo {volume} {72}},\ \bibinfo {pages} {043508} (\bibinfo {year} {2005})}\BibitemShut {NoStop}%
\bibitem [{\citenamefont {Affleck}\ \emph {et~al.}(1984)\citenamefont {Affleck}, \citenamefont {Dine},\ and\ \citenamefont {Seiberg}}]{AFFLECK1984493}%
  \BibitemOpen
  \bibfield  {author} {\bibinfo {author} {\bibfnamefont {I.}~\bibnamefont {Affleck}}, \bibinfo {author} {\bibfnamefont {M.}~\bibnamefont {Dine}},\ and\ \bibinfo {author} {\bibfnamefont {N.}~\bibnamefont {Seiberg}},\ }\href {https://doi.org/https://doi.org/10.1016/0550-3213(84)90058-0} {\bibfield  {journal} {\bibinfo  {journal} {Nuclear Physics B}\ }\textbf {\bibinfo {volume} {241}},\ \bibinfo {pages} {493} (\bibinfo {year} {1984})}\BibitemShut {NoStop}%
\bibitem [{\citenamefont {de~la Macorra}(2003)}]{Axel_de_la_Macorra_2003}%
  \BibitemOpen
  \bibfield  {author} {\bibinfo {author} {\bibfnamefont {A.}~\bibnamefont {de~la Macorra}},\ }\href {https://doi.org/10.1088/1126-6708/2003/01/033} {\bibfield  {journal} {\bibinfo  {journal} {Journal of High Energy Physics}\ }\textbf {\bibinfo {volume} {2003}},\ \bibinfo {pages} {033} (\bibinfo {year} {2003})}\BibitemShut {NoStop}%
\bibitem [{\citenamefont {Burgess}\ \emph {et~al.}(1997)\citenamefont {Burgess}, \citenamefont {{de la Macorra}}, \citenamefont {Maksymyk},\ and\ \citenamefont {Quevedo}}]{BURGESS1997181}%
  \BibitemOpen
  \bibfield  {author} {\bibinfo {author} {\bibfnamefont {C.}~\bibnamefont {Burgess}}, \bibinfo {author} {\bibfnamefont {A.}~\bibnamefont {{de la Macorra}}}, \bibinfo {author} {\bibfnamefont {I.}~\bibnamefont {Maksymyk}},\ and\ \bibinfo {author} {\bibfnamefont {F.}~\bibnamefont {Quevedo}},\ }\href {https://doi.org/https://doi.org/10.1016/S0370-2693(97)00973-8} {\bibfield  {journal} {\bibinfo  {journal} {Physics Letters B}\ }\textbf {\bibinfo {volume} {410}},\ \bibinfo {pages} {181} (\bibinfo {year} {1997})}\BibitemShut {NoStop}%
\bibitem [{\citenamefont {{de la Macorra}}(2003)}]{2003JHEP...01..033D}%
  \BibitemOpen
  \bibfield  {author} {\bibinfo {author} {\bibfnamefont {A.}~\bibnamefont {{de la Macorra}}},\ }\href {https://doi.org/10.1088/1126-6708/2003/01/033} {\bibfield  {journal} {\bibinfo  {journal} {Journal of High Energy Physics}\ }\textbf {\bibinfo {volume} {2003}},\ \bibinfo {eid} {033} (\bibinfo {year} {2003})},\ \Eprint {https://arxiv.org/abs/hep-ph/0111292} {arXiv:hep-ph/0111292 [hep-ph]} \BibitemShut {NoStop}%
\bibitem [{\citenamefont {Bourilkov}(2015)}]{Bourilkov_2015}%
  \BibitemOpen
  \bibfield  {author} {\bibinfo {author} {\bibfnamefont {D.}~\bibnamefont {Bourilkov}},\ }\bibfield  {journal} {\bibinfo  {journal} {Journal of High Energy Physics}\ }\textbf {\bibinfo {volume} {2015}},\ \href {https://doi.org/10.1007/jhep11(2015)117} {10.1007/jhep11(2015)117} (\bibinfo {year} {2015})\BibitemShut {NoStop}%
\bibitem [{\citenamefont {Navas}\ \emph {et~al.}(2024{\natexlab{a}})\citenamefont {Navas} \emph {et~al.}}]{ParticleDataGroup:2024cfk}%
  \BibitemOpen
  \bibfield  {author} {\bibinfo {author} {\bibfnamefont {S.}~\bibnamefont {Navas}} \emph {et~al.} (\bibinfo {collaboration} {Particle Data Group}),\ }\href {https://doi.org/10.1103/PhysRevD.110.030001} {\bibfield  {journal} {\bibinfo  {journal} {Phys. Rev. D}\ }\textbf {\bibinfo {volume} {110}},\ \bibinfo {pages} {030001} (\bibinfo {year} {2024}{\natexlab{a}})}\BibitemShut {NoStop}%
\bibitem [{\citenamefont {Navas}\ \emph {et~al.}(2024{\natexlab{b}})\citenamefont {Navas}, \citenamefont {Amsler}, \citenamefont {Gutsche}, \citenamefont {Hanhart}, \citenamefont {Hern\'andez-Rey}, \citenamefont {Louren\ifmmode~\mbox{\c{c}}\else \c{c}\fi{}o}, \citenamefont {Masoni}, \citenamefont {Mikhasenko}, \citenamefont {Mitchell}, \citenamefont {Patrignani}, \citenamefont {Schwanda}, \citenamefont {Spanier}, \citenamefont {Venanzoni}, \citenamefont {Yuan}, \citenamefont {Agashe}, \citenamefont {Aielli}, \citenamefont {Allanach}, \citenamefont {Alvarez-Mu\~niz}, \citenamefont {Antonelli}, \citenamefont {Aschenauer}, \citenamefont {Asner}, \citenamefont {Assamagan}, \citenamefont {Baer}, \citenamefont {Banerjee}, \citenamefont {Barnett}, \citenamefont {Baudis}, \citenamefont {Bauer}, \citenamefont {Beatty}, \citenamefont {Beringer}, \citenamefont {Bettini}, \citenamefont {Biebel}, \citenamefont {Black}, \citenamefont {Blucher}, \citenamefont {Bonventre}, \citenamefont {Briere}, \citenamefont {Buckley},
  \citenamefont {Burkert}, \citenamefont {Bychkov}, \citenamefont {Cahn}, \citenamefont {Cao}, \citenamefont {Carena}, \citenamefont {Casarosa}, \citenamefont {Ceccucci}, \citenamefont {Cerri}, \citenamefont {Chivukula}, \citenamefont {Cowan}, \citenamefont {Cranmer}, \citenamefont {Crede}, \citenamefont {Cremonesi}, \citenamefont {D'Ambrosio}, \citenamefont {Damour}, \citenamefont {de~Florian}, \citenamefont {de~Gouv\^ea}, \citenamefont {DeGrand}, \citenamefont {Demers}, \citenamefont {Demiragli}, \citenamefont {Dobrescu}, \citenamefont {D'Onofrio}, \citenamefont {Doser}, \citenamefont {Dreiner}, \citenamefont {Eerola}, \citenamefont {Egede}, \citenamefont {Eidelman}, \citenamefont {El-Khadra}, \citenamefont {Ellis}, \citenamefont {Eno}, \citenamefont {Erler}, \citenamefont {Ezhela}, \citenamefont {Fava}, \citenamefont {Fetscher}, \citenamefont {Fields}, \citenamefont {Freitas}, \citenamefont {Gallagher}, \citenamefont {Gershon}, \citenamefont {Gershtein}, \citenamefont {Gherghetta}, \citenamefont
  {Gonzalez-Garcia}, \citenamefont {Goodman}, \citenamefont {Grab}, \citenamefont {Gritsan}, \citenamefont {Grojean}, \citenamefont {Groom}, \citenamefont {Gr\"unewald}, \citenamefont {Gurtu}, \citenamefont {Haber}, \citenamefont {Hamel}, \citenamefont {Hashimoto}, \citenamefont {Hayato}, \citenamefont {Hebecker}, \citenamefont {Heinemeyer}, \citenamefont {Hikasa}, \citenamefont {Hisano}, \citenamefont {H\"ocker}, \citenamefont {Holder}, \citenamefont {Hsu}, \citenamefont {Huston}, \citenamefont {Hyodo}, \citenamefont {Ianni}, \citenamefont {Kado}, \citenamefont {Karliner}, \citenamefont {Katz}, \citenamefont {Kenzie}, \citenamefont {Khoze}, \citenamefont {Klein}, \citenamefont {Krauss}, \citenamefont {Kreps}, \citenamefont {Kri\ifmmode~\check{z}\else \v{z}\fi{}an}, \citenamefont {Krusche}, \citenamefont {Kwon}, \citenamefont {Lahav}, \citenamefont {Lellouch}, \citenamefont {Lesgourgues}, \citenamefont {Liddle}, \citenamefont {Ligeti}, \citenamefont {Lin}, \citenamefont {Lippmann}, \citenamefont {Liss},
  \citenamefont {Lister}, \citenamefont {Littenberg}, \citenamefont {Lugovsky}, \citenamefont {Lugovsky}, \citenamefont {Lusiani}, \citenamefont {Makida}, \citenamefont {Maltoni}, \citenamefont {Manohar}, \citenamefont {Marciano}, \citenamefont {Matthews}, \citenamefont {Mei\ss{}ner}, \citenamefont {Melzer-Pellmann}, \citenamefont {Mertsch}, \citenamefont {Miller}, \citenamefont {Milstead}, \citenamefont {M\"onig}, \citenamefont {Molaro}, \citenamefont {Moortgat}, \citenamefont {Moskovic}, \citenamefont {Nagata}, \citenamefont {Nakamura}, \citenamefont {Narain}, \citenamefont {Nason}, \citenamefont {Nelles}, \citenamefont {Neubert}, \citenamefont {Nir}, \citenamefont {O'Connell}, \citenamefont {O'Hare}, \citenamefont {Olive}, \citenamefont {Peacock}, \citenamefont {Pianori}, \citenamefont {Pich}, \citenamefont {Piepke}, \citenamefont {Pietropaolo}, \citenamefont {Pomarol}, \citenamefont {Pordes}, \citenamefont {Profumo}, \citenamefont {Quadt}, \citenamefont {Rabbertz}, \citenamefont {Rademacker},
  \citenamefont {Raffelt}, \citenamefont {Ramsey-Musolf}, \citenamefont {Richardson}, \citenamefont {Ringwald}, \citenamefont {Robinson}, \citenamefont {Roesler}, \citenamefont {Rolli}, \citenamefont {Romaniouk}, \citenamefont {Rosenberg}, \citenamefont {Rosner}, \citenamefont {Rybka}, \citenamefont {Ryskin}, \citenamefont {Ryutin}, \citenamefont {Safdi}, \citenamefont {Sakai}, \citenamefont {Sarkar}, \citenamefont {Sauli}, \citenamefont {Schneider}, \citenamefont {Sch\"onert}, \citenamefont {Scholberg}, \citenamefont {Schwartz}, \citenamefont {Schwiening}, \citenamefont {Scott}, \citenamefont {Sefkow}, \citenamefont {Seljak}, \citenamefont {Sharma}, \citenamefont {Sharpe}, \citenamefont {Shiltsev}, \citenamefont {Signorelli}, \citenamefont {Silari}, \citenamefont {Simon}, \citenamefont {Sj\"ostrand}, \citenamefont {Skands}, \citenamefont {Skwarnicki}, \citenamefont {Smoot}, \citenamefont {Soffer}, \citenamefont {Sozzi}, \citenamefont {Spiering}, \citenamefont {Stahl}, \citenamefont {Sumino}, \citenamefont
  {Takahashi}, \citenamefont {Tanabashi}, \citenamefont {Tanaka}, \citenamefont {Ta\ifmmode~\check{s}\else \v{s}\fi{}evsk\'y}, \citenamefont {Terao}, \citenamefont {Terashi}, \citenamefont {Terning}, \citenamefont {Thoma}, \citenamefont {Thorne}, \citenamefont {Tiator}, \citenamefont {Titov}, \citenamefont {Tovey}, \citenamefont {Trabelsi}, \citenamefont {Urquijo}, \citenamefont {Valencia}, \citenamefont {Van~de Water}, \citenamefont {Varelas}, \citenamefont {Verde}, \citenamefont {Vivarelli}, \citenamefont {Vogel}, \citenamefont {Vogelsang}, \citenamefont {Vorobyev}, \citenamefont {Wakely}, \citenamefont {Walkowiak}, \citenamefont {Walter}, \citenamefont {Wands}, \citenamefont {Weinberg}, \citenamefont {Weinberg}, \citenamefont {Wermes}, \citenamefont {White}, \citenamefont {Wiencke}, \citenamefont {Willocq}, \citenamefont {Woody}, \citenamefont {Workman}, \citenamefont {Yao}, \citenamefont {Yokoyama}, \citenamefont {Yoshida}, \citenamefont {Zanderighi}, \citenamefont {Zeller}, \citenamefont {Zhu},
  \citenamefont {Zhu}, \citenamefont {Zimmermann}, \citenamefont {Zyla}, \citenamefont {Anderson}, \citenamefont {Kramer}, \citenamefont {Schaffner},\ and\ \citenamefont {Zheng}}]{PhysRevD.110.030001}%
  \BibitemOpen
  \bibfield  {author} {\bibinfo {author} {\bibfnamefont {S.}~\bibnamefont {Navas}}, \bibinfo {author} {\bibfnamefont {C.}~\bibnamefont {Amsler}}, \bibinfo {author} {\bibfnamefont {T.}~\bibnamefont {Gutsche}}, \bibinfo {author} {\bibfnamefont {C.}~\bibnamefont {Hanhart}}, \bibinfo {author} {\bibfnamefont {J.~J.}\ \bibnamefont {Hern\'andez-Rey}}, \bibinfo {author} {\bibfnamefont {C.}~\bibnamefont {Louren\ifmmode~\mbox{\c{c}}\else \c{c}\fi{}o}}, \bibinfo {author} {\bibfnamefont {A.}~\bibnamefont {Masoni}}, \bibinfo {author} {\bibfnamefont {M.}~\bibnamefont {Mikhasenko}}, \bibinfo {author} {\bibfnamefont {R.~E.}\ \bibnamefont {Mitchell}}, \bibinfo {author} {\bibfnamefont {C.}~\bibnamefont {Patrignani}}, \bibinfo {author} {\bibfnamefont {C.}~\bibnamefont {Schwanda}}, \bibinfo {author} {\bibfnamefont {S.}~\bibnamefont {Spanier}}, \bibinfo {author} {\bibfnamefont {G.}~\bibnamefont {Venanzoni}}, \bibinfo {author} {\bibfnamefont {C.~Z.}\ \bibnamefont {Yuan}}, \bibinfo {author} {\bibfnamefont {K.}~\bibnamefont
  {Agashe}}, \bibinfo {author} {\bibfnamefont {G.}~\bibnamefont {Aielli}}, \bibinfo {author} {\bibfnamefont {B.~C.}\ \bibnamefont {Allanach}}, \bibinfo {author} {\bibfnamefont {J.}~\bibnamefont {Alvarez-Mu\~niz}}, \bibinfo {author} {\bibfnamefont {M.}~\bibnamefont {Antonelli}}, \bibinfo {author} {\bibfnamefont {E.~C.}\ \bibnamefont {Aschenauer}}, \bibinfo {author} {\bibfnamefont {D.~M.}\ \bibnamefont {Asner}}, \bibinfo {author} {\bibfnamefont {K.}~\bibnamefont {Assamagan}}, \bibinfo {author} {\bibfnamefont {H.}~\bibnamefont {Baer}}, \bibinfo {author} {\bibfnamefont {S.}~\bibnamefont {Banerjee}}, \bibinfo {author} {\bibfnamefont {R.~M.}\ \bibnamefont {Barnett}}, \bibinfo {author} {\bibfnamefont {L.}~\bibnamefont {Baudis}}, \bibinfo {author} {\bibfnamefont {C.~W.}\ \bibnamefont {Bauer}}, \bibinfo {author} {\bibfnamefont {J.~J.}\ \bibnamefont {Beatty}}, \bibinfo {author} {\bibfnamefont {J.}~\bibnamefont {Beringer}}, \bibinfo {author} {\bibfnamefont {A.}~\bibnamefont {Bettini}}, \bibinfo {author} {\bibfnamefont
  {O.}~\bibnamefont {Biebel}}, \bibinfo {author} {\bibfnamefont {K.~M.}\ \bibnamefont {Black}}, \bibinfo {author} {\bibfnamefont {E.}~\bibnamefont {Blucher}}, \bibinfo {author} {\bibfnamefont {R.}~\bibnamefont {Bonventre}}, \bibinfo {author} {\bibfnamefont {R.~A.}\ \bibnamefont {Briere}}, \bibinfo {author} {\bibfnamefont {A.}~\bibnamefont {Buckley}}, \bibinfo {author} {\bibfnamefont {V.~D.}\ \bibnamefont {Burkert}}, \bibinfo {author} {\bibfnamefont {M.~A.}\ \bibnamefont {Bychkov}}, \bibinfo {author} {\bibfnamefont {R.~N.}\ \bibnamefont {Cahn}}, \bibinfo {author} {\bibfnamefont {Z.}~\bibnamefont {Cao}}, \bibinfo {author} {\bibfnamefont {M.}~\bibnamefont {Carena}}, \bibinfo {author} {\bibfnamefont {G.}~\bibnamefont {Casarosa}}, \bibinfo {author} {\bibfnamefont {A.}~\bibnamefont {Ceccucci}}, \bibinfo {author} {\bibfnamefont {A.}~\bibnamefont {Cerri}}, \bibinfo {author} {\bibfnamefont {R.~S.}\ \bibnamefont {Chivukula}}, \bibinfo {author} {\bibfnamefont {G.}~\bibnamefont {Cowan}}, \bibinfo {author} {\bibfnamefont
  {K.}~\bibnamefont {Cranmer}}, \bibinfo {author} {\bibfnamefont {V.}~\bibnamefont {Crede}}, \bibinfo {author} {\bibfnamefont {O.}~\bibnamefont {Cremonesi}}, \bibinfo {author} {\bibfnamefont {G.}~\bibnamefont {D'Ambrosio}}, \bibinfo {author} {\bibfnamefont {T.}~\bibnamefont {Damour}}, \bibinfo {author} {\bibfnamefont {D.}~\bibnamefont {de~Florian}}, \bibinfo {author} {\bibfnamefont {A.}~\bibnamefont {de~Gouv\^ea}}, \bibinfo {author} {\bibfnamefont {T.}~\bibnamefont {DeGrand}}, \bibinfo {author} {\bibfnamefont {S.}~\bibnamefont {Demers}}, \bibinfo {author} {\bibfnamefont {Z.}~\bibnamefont {Demiragli}}, \bibinfo {author} {\bibfnamefont {B.~A.}\ \bibnamefont {Dobrescu}}, \bibinfo {author} {\bibfnamefont {M.}~\bibnamefont {D'Onofrio}}, \bibinfo {author} {\bibfnamefont {M.}~\bibnamefont {Doser}}, \bibinfo {author} {\bibfnamefont {H.~K.}\ \bibnamefont {Dreiner}}, \bibinfo {author} {\bibfnamefont {P.}~\bibnamefont {Eerola}}, \bibinfo {author} {\bibfnamefont {U.}~\bibnamefont {Egede}}, \bibinfo {author}
  {\bibfnamefont {S.}~\bibnamefont {Eidelman}}, \bibinfo {author} {\bibfnamefont {A.~X.}\ \bibnamefont {El-Khadra}}, \bibinfo {author} {\bibfnamefont {J.}~\bibnamefont {Ellis}}, \bibinfo {author} {\bibfnamefont {S.~C.}\ \bibnamefont {Eno}}, \bibinfo {author} {\bibfnamefont {J.}~\bibnamefont {Erler}}, \bibinfo {author} {\bibfnamefont {V.~V.}\ \bibnamefont {Ezhela}}, \bibinfo {author} {\bibfnamefont {A.}~\bibnamefont {Fava}}, \bibinfo {author} {\bibfnamefont {W.}~\bibnamefont {Fetscher}}, \bibinfo {author} {\bibfnamefont {B.~D.}\ \bibnamefont {Fields}}, \bibinfo {author} {\bibfnamefont {A.}~\bibnamefont {Freitas}}, \bibinfo {author} {\bibfnamefont {H.}~\bibnamefont {Gallagher}}, \bibinfo {author} {\bibfnamefont {T.}~\bibnamefont {Gershon}}, \bibinfo {author} {\bibfnamefont {Y.}~\bibnamefont {Gershtein}}, \bibinfo {author} {\bibfnamefont {T.}~\bibnamefont {Gherghetta}}, \bibinfo {author} {\bibfnamefont {M.~C.}\ \bibnamefont {Gonzalez-Garcia}}, \bibinfo {author} {\bibfnamefont {M.}~\bibnamefont {Goodman}},
  \bibinfo {author} {\bibfnamefont {C.}~\bibnamefont {Grab}}, \bibinfo {author} {\bibfnamefont {A.~V.}\ \bibnamefont {Gritsan}}, \bibinfo {author} {\bibfnamefont {C.}~\bibnamefont {Grojean}}, \bibinfo {author} {\bibfnamefont {D.~E.}\ \bibnamefont {Groom}}, \bibinfo {author} {\bibfnamefont {M.}~\bibnamefont {Gr\"unewald}}, \bibinfo {author} {\bibfnamefont {A.}~\bibnamefont {Gurtu}}, \bibinfo {author} {\bibfnamefont {H.~E.}\ \bibnamefont {Haber}}, \bibinfo {author} {\bibfnamefont {M.}~\bibnamefont {Hamel}}, \bibinfo {author} {\bibfnamefont {S.}~\bibnamefont {Hashimoto}}, \bibinfo {author} {\bibfnamefont {Y.}~\bibnamefont {Hayato}}, \bibinfo {author} {\bibfnamefont {A.}~\bibnamefont {Hebecker}}, \bibinfo {author} {\bibfnamefont {S.}~\bibnamefont {Heinemeyer}}, \bibinfo {author} {\bibfnamefont {K.}~\bibnamefont {Hikasa}}, \bibinfo {author} {\bibfnamefont {J.}~\bibnamefont {Hisano}}, \bibinfo {author} {\bibfnamefont {A.}~\bibnamefont {H\"ocker}}, \bibinfo {author} {\bibfnamefont {J.}~\bibnamefont {Holder}},
  \bibinfo {author} {\bibfnamefont {L.}~\bibnamefont {Hsu}}, \bibinfo {author} {\bibfnamefont {J.}~\bibnamefont {Huston}}, \bibinfo {author} {\bibfnamefont {T.}~\bibnamefont {Hyodo}}, \bibinfo {author} {\bibfnamefont {A.}~\bibnamefont {Ianni}}, \bibinfo {author} {\bibfnamefont {M.}~\bibnamefont {Kado}}, \bibinfo {author} {\bibfnamefont {M.}~\bibnamefont {Karliner}}, \bibinfo {author} {\bibfnamefont {U.~F.}\ \bibnamefont {Katz}}, \bibinfo {author} {\bibfnamefont {M.}~\bibnamefont {Kenzie}}, \bibinfo {author} {\bibfnamefont {V.~A.}\ \bibnamefont {Khoze}}, \bibinfo {author} {\bibfnamefont {S.~R.}\ \bibnamefont {Klein}}, \bibinfo {author} {\bibfnamefont {F.}~\bibnamefont {Krauss}}, \bibinfo {author} {\bibfnamefont {M.}~\bibnamefont {Kreps}}, \bibinfo {author} {\bibfnamefont {P.}~\bibnamefont {Kri\ifmmode~\check{z}\else \v{z}\fi{}an}}, \bibinfo {author} {\bibfnamefont {B.}~\bibnamefont {Krusche}}, \bibinfo {author} {\bibfnamefont {Y.}~\bibnamefont {Kwon}}, \bibinfo {author} {\bibfnamefont {O.}~\bibnamefont
  {Lahav}}, \bibinfo {author} {\bibfnamefont {L.~P.}\ \bibnamefont {Lellouch}}, \bibinfo {author} {\bibfnamefont {J.}~\bibnamefont {Lesgourgues}}, \bibinfo {author} {\bibfnamefont {A.~R.}\ \bibnamefont {Liddle}}, \bibinfo {author} {\bibfnamefont {Z.}~\bibnamefont {Ligeti}}, \bibinfo {author} {\bibfnamefont {C.-J.}\ \bibnamefont {Lin}}, \bibinfo {author} {\bibfnamefont {C.}~\bibnamefont {Lippmann}}, \bibinfo {author} {\bibfnamefont {T.~M.}\ \bibnamefont {Liss}}, \bibinfo {author} {\bibfnamefont {A.}~\bibnamefont {Lister}}, \bibinfo {author} {\bibfnamefont {L.}~\bibnamefont {Littenberg}}, \bibinfo {author} {\bibfnamefont {K.~S.}\ \bibnamefont {Lugovsky}}, \bibinfo {author} {\bibfnamefont {S.~B.}\ \bibnamefont {Lugovsky}}, \bibinfo {author} {\bibfnamefont {A.}~\bibnamefont {Lusiani}}, \bibinfo {author} {\bibfnamefont {Y.}~\bibnamefont {Makida}}, \bibinfo {author} {\bibfnamefont {F.}~\bibnamefont {Maltoni}}, \bibinfo {author} {\bibfnamefont {A.~V.}\ \bibnamefont {Manohar}}, \bibinfo {author} {\bibfnamefont
  {W.~J.}\ \bibnamefont {Marciano}}, \bibinfo {author} {\bibfnamefont {J.}~\bibnamefont {Matthews}}, \bibinfo {author} {\bibfnamefont {U.-G.}\ \bibnamefont {Mei\ss{}ner}}, \bibinfo {author} {\bibfnamefont {I.-A.}\ \bibnamefont {Melzer-Pellmann}}, \bibinfo {author} {\bibfnamefont {P.}~\bibnamefont {Mertsch}}, \bibinfo {author} {\bibfnamefont {D.~J.}\ \bibnamefont {Miller}}, \bibinfo {author} {\bibfnamefont {D.}~\bibnamefont {Milstead}}, \bibinfo {author} {\bibfnamefont {K.}~\bibnamefont {M\"onig}}, \bibinfo {author} {\bibfnamefont {P.}~\bibnamefont {Molaro}}, \bibinfo {author} {\bibfnamefont {F.}~\bibnamefont {Moortgat}}, \bibinfo {author} {\bibfnamefont {M.}~\bibnamefont {Moskovic}}, \bibinfo {author} {\bibfnamefont {N.}~\bibnamefont {Nagata}}, \bibinfo {author} {\bibfnamefont {K.}~\bibnamefont {Nakamura}}, \bibinfo {author} {\bibfnamefont {M.}~\bibnamefont {Narain}}, \bibinfo {author} {\bibfnamefont {P.}~\bibnamefont {Nason}}, \bibinfo {author} {\bibfnamefont {A.}~\bibnamefont {Nelles}}, \bibinfo {author}
  {\bibfnamefont {M.}~\bibnamefont {Neubert}}, \bibinfo {author} {\bibfnamefont {Y.}~\bibnamefont {Nir}}, \bibinfo {author} {\bibfnamefont {H.~B.}\ \bibnamefont {O'Connell}}, \bibinfo {author} {\bibfnamefont {C.~A.~J.}\ \bibnamefont {O'Hare}}, \bibinfo {author} {\bibfnamefont {K.~A.}\ \bibnamefont {Olive}}, \bibinfo {author} {\bibfnamefont {J.~A.}\ \bibnamefont {Peacock}}, \bibinfo {author} {\bibfnamefont {E.}~\bibnamefont {Pianori}}, \bibinfo {author} {\bibfnamefont {A.}~\bibnamefont {Pich}}, \bibinfo {author} {\bibfnamefont {A.}~\bibnamefont {Piepke}}, \bibinfo {author} {\bibfnamefont {F.}~\bibnamefont {Pietropaolo}}, \bibinfo {author} {\bibfnamefont {A.}~\bibnamefont {Pomarol}}, \bibinfo {author} {\bibfnamefont {S.}~\bibnamefont {Pordes}}, \bibinfo {author} {\bibfnamefont {S.}~\bibnamefont {Profumo}}, \bibinfo {author} {\bibfnamefont {A.}~\bibnamefont {Quadt}}, \bibinfo {author} {\bibfnamefont {K.}~\bibnamefont {Rabbertz}}, \bibinfo {author} {\bibfnamefont {J.}~\bibnamefont {Rademacker}}, \bibinfo {author}
  {\bibfnamefont {G.}~\bibnamefont {Raffelt}}, \bibinfo {author} {\bibfnamefont {M.}~\bibnamefont {Ramsey-Musolf}}, \bibinfo {author} {\bibfnamefont {P.}~\bibnamefont {Richardson}}, \bibinfo {author} {\bibfnamefont {A.}~\bibnamefont {Ringwald}}, \bibinfo {author} {\bibfnamefont {D.~J.}\ \bibnamefont {Robinson}}, \bibinfo {author} {\bibfnamefont {S.}~\bibnamefont {Roesler}}, \bibinfo {author} {\bibfnamefont {S.}~\bibnamefont {Rolli}}, \bibinfo {author} {\bibfnamefont {A.}~\bibnamefont {Romaniouk}}, \bibinfo {author} {\bibfnamefont {L.~J.}\ \bibnamefont {Rosenberg}}, \bibinfo {author} {\bibfnamefont {J.~L.}\ \bibnamefont {Rosner}}, \bibinfo {author} {\bibfnamefont {G.}~\bibnamefont {Rybka}}, \bibinfo {author} {\bibfnamefont {M.~G.}\ \bibnamefont {Ryskin}}, \bibinfo {author} {\bibfnamefont {R.~A.}\ \bibnamefont {Ryutin}}, \bibinfo {author} {\bibfnamefont {B.}~\bibnamefont {Safdi}}, \bibinfo {author} {\bibfnamefont {Y.}~\bibnamefont {Sakai}}, \bibinfo {author} {\bibfnamefont {S.}~\bibnamefont {Sarkar}}, \bibinfo
  {author} {\bibfnamefont {F.}~\bibnamefont {Sauli}}, \bibinfo {author} {\bibfnamefont {O.}~\bibnamefont {Schneider}}, \bibinfo {author} {\bibfnamefont {S.}~\bibnamefont {Sch\"onert}}, \bibinfo {author} {\bibfnamefont {K.}~\bibnamefont {Scholberg}}, \bibinfo {author} {\bibfnamefont {A.~J.}\ \bibnamefont {Schwartz}}, \bibinfo {author} {\bibfnamefont {J.}~\bibnamefont {Schwiening}}, \bibinfo {author} {\bibfnamefont {D.}~\bibnamefont {Scott}}, \bibinfo {author} {\bibfnamefont {F.}~\bibnamefont {Sefkow}}, \bibinfo {author} {\bibfnamefont {U.}~\bibnamefont {Seljak}}, \bibinfo {author} {\bibfnamefont {V.}~\bibnamefont {Sharma}}, \bibinfo {author} {\bibfnamefont {S.~R.}\ \bibnamefont {Sharpe}}, \bibinfo {author} {\bibfnamefont {V.}~\bibnamefont {Shiltsev}}, \bibinfo {author} {\bibfnamefont {G.}~\bibnamefont {Signorelli}}, \bibinfo {author} {\bibfnamefont {M.}~\bibnamefont {Silari}}, \bibinfo {author} {\bibfnamefont {F.}~\bibnamefont {Simon}}, \bibinfo {author} {\bibfnamefont {T.}~\bibnamefont {Sj\"ostrand}},
  \bibinfo {author} {\bibfnamefont {P.}~\bibnamefont {Skands}}, \bibinfo {author} {\bibfnamefont {T.}~\bibnamefont {Skwarnicki}}, \bibinfo {author} {\bibfnamefont {G.~F.}\ \bibnamefont {Smoot}}, \bibinfo {author} {\bibfnamefont {A.}~\bibnamefont {Soffer}}, \bibinfo {author} {\bibfnamefont {M.~S.}\ \bibnamefont {Sozzi}}, \bibinfo {author} {\bibfnamefont {C.}~\bibnamefont {Spiering}}, \bibinfo {author} {\bibfnamefont {A.}~\bibnamefont {Stahl}}, \bibinfo {author} {\bibfnamefont {Y.}~\bibnamefont {Sumino}}, \bibinfo {author} {\bibfnamefont {F.}~\bibnamefont {Takahashi}}, \bibinfo {author} {\bibfnamefont {M.}~\bibnamefont {Tanabashi}}, \bibinfo {author} {\bibfnamefont {J.}~\bibnamefont {Tanaka}}, \bibinfo {author} {\bibfnamefont {M.}~\bibnamefont {Ta\ifmmode~\check{s}\else \v{s}\fi{}evsk\'y}}, \bibinfo {author} {\bibfnamefont {K.}~\bibnamefont {Terao}}, \bibinfo {author} {\bibfnamefont {K.}~\bibnamefont {Terashi}}, \bibinfo {author} {\bibfnamefont {J.}~\bibnamefont {Terning}}, \bibinfo {author} {\bibfnamefont
  {U.}~\bibnamefont {Thoma}}, \bibinfo {author} {\bibfnamefont {R.~S.}\ \bibnamefont {Thorne}}, \bibinfo {author} {\bibfnamefont {L.}~\bibnamefont {Tiator}}, \bibinfo {author} {\bibfnamefont {M.}~\bibnamefont {Titov}}, \bibinfo {author} {\bibfnamefont {D.~R.}\ \bibnamefont {Tovey}}, \bibinfo {author} {\bibfnamefont {K.}~\bibnamefont {Trabelsi}}, \bibinfo {author} {\bibfnamefont {P.}~\bibnamefont {Urquijo}}, \bibinfo {author} {\bibfnamefont {G.}~\bibnamefont {Valencia}}, \bibinfo {author} {\bibfnamefont {R.}~\bibnamefont {Van~de Water}}, \bibinfo {author} {\bibfnamefont {N.}~\bibnamefont {Varelas}}, \bibinfo {author} {\bibfnamefont {L.}~\bibnamefont {Verde}}, \bibinfo {author} {\bibfnamefont {I.}~\bibnamefont {Vivarelli}}, \bibinfo {author} {\bibfnamefont {P.}~\bibnamefont {Vogel}}, \bibinfo {author} {\bibfnamefont {W.}~\bibnamefont {Vogelsang}}, \bibinfo {author} {\bibfnamefont {V.}~\bibnamefont {Vorobyev}}, \bibinfo {author} {\bibfnamefont {S.~P.}\ \bibnamefont {Wakely}}, \bibinfo {author} {\bibfnamefont
  {W.}~\bibnamefont {Walkowiak}}, \bibinfo {author} {\bibfnamefont {C.~W.}\ \bibnamefont {Walter}}, \bibinfo {author} {\bibfnamefont {D.}~\bibnamefont {Wands}}, \bibinfo {author} {\bibfnamefont {D.~H.}\ \bibnamefont {Weinberg}}, \bibinfo {author} {\bibfnamefont {E.~J.}\ \bibnamefont {Weinberg}}, \bibinfo {author} {\bibfnamefont {N.}~\bibnamefont {Wermes}}, \bibinfo {author} {\bibfnamefont {M.}~\bibnamefont {White}}, \bibinfo {author} {\bibfnamefont {L.~R.}\ \bibnamefont {Wiencke}}, \bibinfo {author} {\bibfnamefont {S.}~\bibnamefont {Willocq}}, \bibinfo {author} {\bibfnamefont {C.~L.}\ \bibnamefont {Woody}}, \bibinfo {author} {\bibfnamefont {R.~L.}\ \bibnamefont {Workman}}, \bibinfo {author} {\bibfnamefont {W.-M.}\ \bibnamefont {Yao}}, \bibinfo {author} {\bibfnamefont {M.}~\bibnamefont {Yokoyama}}, \bibinfo {author} {\bibfnamefont {R.}~\bibnamefont {Yoshida}}, \bibinfo {author} {\bibfnamefont {G.}~\bibnamefont {Zanderighi}}, \bibinfo {author} {\bibfnamefont {G.~P.}\ \bibnamefont {Zeller}}, \bibinfo {author}
  {\bibfnamefont {R.-Y.}\ \bibnamefont {Zhu}}, \bibinfo {author} {\bibfnamefont {S.-L.}\ \bibnamefont {Zhu}}, \bibinfo {author} {\bibfnamefont {F.}~\bibnamefont {Zimmermann}}, \bibinfo {author} {\bibfnamefont {P.~A.}\ \bibnamefont {Zyla}}, \bibinfo {author} {\bibfnamefont {J.}~\bibnamefont {Anderson}}, \bibinfo {author} {\bibfnamefont {M.}~\bibnamefont {Kramer}}, \bibinfo {author} {\bibfnamefont {P.}~\bibnamefont {Schaffner}},\ and\ \bibinfo {author} {\bibfnamefont {W.}~\bibnamefont {Zheng}} (\bibinfo {collaboration} {Particle Data Group Collaboration}),\ }\href {https://doi.org/10.1103/PhysRevD.110.030001} {\bibfield  {journal} {\bibinfo  {journal} {Phys. Rev. D}\ }\textbf {\bibinfo {volume} {110}},\ \bibinfo {pages} {030001} (\bibinfo {year} {2024}{\natexlab{b}})}\BibitemShut {NoStop}%
\bibitem [{\citenamefont {Fixsen}(2009)}]{Fixsen_2009}%
  \BibitemOpen
  \bibfield  {author} {\bibinfo {author} {\bibfnamefont {D.~J.}\ \bibnamefont {Fixsen}},\ }\href {https://doi.org/10.1088/0004-637X/707/2/916} {\bibfield  {journal} {\bibinfo  {journal} {The Astrophysical Journal}\ }\textbf {\bibinfo {volume} {707}},\ \bibinfo {pages} {916} (\bibinfo {year} {2009})}\BibitemShut {NoStop}%
\bibitem [{\citenamefont {Lewis}\ and\ \citenamefont {Bridle}(2002)}]{Lewis_2002}%
  \BibitemOpen
  \bibfield  {author} {\bibinfo {author} {\bibfnamefont {A.}~\bibnamefont {Lewis}}\ and\ \bibinfo {author} {\bibfnamefont {S.}~\bibnamefont {Bridle}},\ }\href {https://doi.org/10.1103/PhysRevD.66.103511} {\bibfield  {journal} {\bibinfo  {journal} {Phys. Rev. D}\ }\textbf {\bibinfo {volume} {66}},\ \bibinfo {pages} {103511} (\bibinfo {year} {2002})}\BibitemShut {NoStop}%
\bibitem [{\citenamefont {Lewis}\ \emph {et~al.}(2000)\citenamefont {Lewis}, \citenamefont {Challinor},\ and\ \citenamefont {Lasenby}}]{Lewis_2000}%
  \BibitemOpen
  \bibfield  {author} {\bibinfo {author} {\bibfnamefont {A.}~\bibnamefont {Lewis}}, \bibinfo {author} {\bibfnamefont {A.}~\bibnamefont {Challinor}},\ and\ \bibinfo {author} {\bibfnamefont {A.}~\bibnamefont {Lasenby}},\ }\href {https://doi.org/10.1086/309179} {\bibfield  {journal} {\bibinfo  {journal} {The Astrophysical Journal}\ }\textbf {\bibinfo {volume} {538}},\ \bibinfo {pages} {473} (\bibinfo {year} {2000})}\BibitemShut {NoStop}%
\bibitem [{\citenamefont {Lewis}(2019)}]{2019arXiv191013970L}%
  \BibitemOpen
  \bibfield  {author} {\bibinfo {author} {\bibfnamefont {A.}~\bibnamefont {Lewis}},\ }\href {https://arxiv.org/abs/1910.13970} {\bibinfo {title} {Getdist: a python package for analysing monte carlo samples}} (\bibinfo {year} {2019}),\ \Eprint {https://arxiv.org/abs/1910.13970} {arXiv:1910.13970 [astro-ph.IM]} \BibitemShut {NoStop}%
\bibitem [{\citenamefont {Sahni}\ \emph {et~al.}(2008)\citenamefont {Sahni}, \citenamefont {Shafieloo},\ and\ \citenamefont {Starobinsky}}]{PhysRevD.78.103502}%
  \BibitemOpen
  \bibfield  {author} {\bibinfo {author} {\bibfnamefont {V.}~\bibnamefont {Sahni}}, \bibinfo {author} {\bibfnamefont {A.}~\bibnamefont {Shafieloo}},\ and\ \bibinfo {author} {\bibfnamefont {A.~A.}\ \bibnamefont {Starobinsky}},\ }\href {https://doi.org/10.1103/PhysRevD.78.103502} {\bibfield  {journal} {\bibinfo  {journal} {Phys. Rev. D}\ }\textbf {\bibinfo {volume} {78}},\ \bibinfo {pages} {103502} (\bibinfo {year} {2008})}\BibitemShut {NoStop}%
\bibitem [{\citenamefont {Zunckel}\ and\ \citenamefont {Clarkson}(2008)}]{PhysRevLett.101.181301}%
  \BibitemOpen
  \bibfield  {author} {\bibinfo {author} {\bibfnamefont {C.}~\bibnamefont {Zunckel}}\ and\ \bibinfo {author} {\bibfnamefont {C.}~\bibnamefont {Clarkson}},\ }\href {https://doi.org/10.1103/PhysRevLett.101.181301} {\bibfield  {journal} {\bibinfo  {journal} {Phys. Rev. Lett.}\ }\textbf {\bibinfo {volume} {101}},\ \bibinfo {pages} {181301} (\bibinfo {year} {2008})}\BibitemShut {NoStop}%
\bibitem [{\citenamefont {Lodha}\ \emph {et~al.}(2025)\citenamefont {Lodha}, \citenamefont {Calderon}, \citenamefont {Matthewson}, \citenamefont {Shafieloo}, \citenamefont {Ishak}, \citenamefont {Pan}, \citenamefont {Garcia-Quintero}, \citenamefont {Huterer}, \citenamefont {Valogiannis}, \citenamefont {Ureña-López}, \citenamefont {Kamble}, \citenamefont {Parkinson}, \citenamefont {Kim}, \citenamefont {Zhao}, \citenamefont {Cervantes-Cota}, \citenamefont {Rohlf}, \citenamefont {Lozano-Rodríguez}, \citenamefont {Román-Herrera}, \citenamefont {Abdul-Karim}, \citenamefont {Aguilar}, \citenamefont {Ahlen}, \citenamefont {Alves}, \citenamefont {Andrade}, \citenamefont {Armengaud}, \citenamefont {Aviles}, \citenamefont {BenZvi}, \citenamefont {Bianchi}, \citenamefont {Brodzeller}, \citenamefont {Brooks}, \citenamefont {Burtin}, \citenamefont {Canning}, \citenamefont {Rosell}, \citenamefont {Casas}, \citenamefont {Castander}, \citenamefont {Charles}, \citenamefont {Chaussidon}, \citenamefont {Chaves-Montero},
  \citenamefont {Chebat}, \citenamefont {Claybaugh}, \citenamefont {Cole}, \citenamefont {Cuceu}, \citenamefont {Dawson}, \citenamefont {de~la Macorra}, \citenamefont {de~Mattia}, \citenamefont {Deiosso}, \citenamefont {Demina}, \citenamefont {Dey}, \citenamefont {Dey}, \citenamefont {Ding}, \citenamefont {Doel}, \citenamefont {Eisenstein}, \citenamefont {Elbers}, \citenamefont {Ferraro}, \citenamefont {Font-Ribera}, \citenamefont {Forero-Romero}, \citenamefont {Garrison}, \citenamefont {Gaztañaga}, \citenamefont {Gil-Marín}, \citenamefont {Gontcho}, \citenamefont {Gonzalez-Morales}, \citenamefont {Gutierrez}, \citenamefont {Guy}, \citenamefont {Hahn}, \citenamefont {Herbold}, \citenamefont {Herrera-Alcantar}, \citenamefont {Honscheid}, \citenamefont {Howlett}, \citenamefont {Juneau}, \citenamefont {Kehoe}, \citenamefont {Kirkby}, \citenamefont {Kisner}, \citenamefont {Kremin}, \citenamefont {Lahav}, \citenamefont {Lamman}, \citenamefont {Landriau}, \citenamefont {Guillou}, \citenamefont {Leauthaud},
  \citenamefont {Levi}, \citenamefont {Li}, \citenamefont {Magneville}, \citenamefont {Manera}, \citenamefont {Martini}, \citenamefont {Meisner}, \citenamefont {Mena-Fernández}, \citenamefont {Miquel}, \citenamefont {Moustakas}, \citenamefont {Santos}, \citenamefont {Muñoz-Gutiérrez}, \citenamefont {Myers}, \citenamefont {Nadathur}, \citenamefont {Niz}, \citenamefont {Noriega}, \citenamefont {Paillas}, \citenamefont {Palanque-Delabrouille}, \citenamefont {Percival}, \citenamefont {Pieri}, \citenamefont {Poppett}, \citenamefont {Prada}, \citenamefont {Pérez-Fernández}, \citenamefont {Pérez-Ràfols}, \citenamefont {Ramírez-Pérez}, \citenamefont {Rashkovetskyi}, \citenamefont {Ravoux}, \citenamefont {Ross}, \citenamefont {Rossi}, \citenamefont {Ruhlmann-Kleider}, \citenamefont {Samushia}, \citenamefont {Sanchez}, \citenamefont {Schlegel}, \citenamefont {Schubnell}, \citenamefont {Seo}, \citenamefont {Sinigaglia}, \citenamefont {Sprayberry}, \citenamefont {Tan}, \citenamefont {Tarlé}, \citenamefont
  {Taylor}, \citenamefont {Turner}, \citenamefont {Vargas-Magaña}, \citenamefont {Walther}, \citenamefont {Weaver}, \citenamefont {Wolfson}, \citenamefont {Yèche}, \citenamefont {Zarrouk}, \citenamefont {Zhou},\ and\ \citenamefont {Zou}}]{lodha2025extendeddarkenergyanalysis}%
  \BibitemOpen
  \bibfield  {author} {\bibinfo {author} {\bibfnamefont {K.}~\bibnamefont {Lodha}}, \bibinfo {author} {\bibfnamefont {R.}~\bibnamefont {Calderon}}, \bibinfo {author} {\bibfnamefont {W.~L.}\ \bibnamefont {Matthewson}}, \bibinfo {author} {\bibfnamefont {A.}~\bibnamefont {Shafieloo}}, \bibinfo {author} {\bibfnamefont {M.}~\bibnamefont {Ishak}}, \bibinfo {author} {\bibfnamefont {J.}~\bibnamefont {Pan}}, \bibinfo {author} {\bibfnamefont {C.}~\bibnamefont {Garcia-Quintero}}, \bibinfo {author} {\bibfnamefont {D.}~\bibnamefont {Huterer}}, \bibinfo {author} {\bibfnamefont {G.}~\bibnamefont {Valogiannis}}, \bibinfo {author} {\bibfnamefont {L.~A.}\ \bibnamefont {Ureña-López}}, \bibinfo {author} {\bibfnamefont {N.~V.}\ \bibnamefont {Kamble}}, \bibinfo {author} {\bibfnamefont {D.}~\bibnamefont {Parkinson}}, \bibinfo {author} {\bibfnamefont {A.~G.}\ \bibnamefont {Kim}}, \bibinfo {author} {\bibfnamefont {G.~B.}\ \bibnamefont {Zhao}}, \bibinfo {author} {\bibfnamefont {J.~L.}\ \bibnamefont {Cervantes-Cota}}, \bibinfo
  {author} {\bibfnamefont {J.}~\bibnamefont {Rohlf}}, \bibinfo {author} {\bibfnamefont {F.}~\bibnamefont {Lozano-Rodríguez}}, \bibinfo {author} {\bibfnamefont {J.~O.}\ \bibnamefont {Román-Herrera}}, \bibinfo {author} {\bibfnamefont {M.}~\bibnamefont {Abdul-Karim}}, \bibinfo {author} {\bibfnamefont {J.}~\bibnamefont {Aguilar}}, \bibinfo {author} {\bibfnamefont {S.}~\bibnamefont {Ahlen}}, \bibinfo {author} {\bibfnamefont {O.}~\bibnamefont {Alves}}, \bibinfo {author} {\bibfnamefont {U.}~\bibnamefont {Andrade}}, \bibinfo {author} {\bibfnamefont {E.}~\bibnamefont {Armengaud}}, \bibinfo {author} {\bibfnamefont {A.}~\bibnamefont {Aviles}}, \bibinfo {author} {\bibfnamefont {S.}~\bibnamefont {BenZvi}}, \bibinfo {author} {\bibfnamefont {D.}~\bibnamefont {Bianchi}}, \bibinfo {author} {\bibfnamefont {A.}~\bibnamefont {Brodzeller}}, \bibinfo {author} {\bibfnamefont {D.}~\bibnamefont {Brooks}}, \bibinfo {author} {\bibfnamefont {E.}~\bibnamefont {Burtin}}, \bibinfo {author} {\bibfnamefont {R.}~\bibnamefont {Canning}},
  \bibinfo {author} {\bibfnamefont {A.~C.}\ \bibnamefont {Rosell}}, \bibinfo {author} {\bibfnamefont {L.}~\bibnamefont {Casas}}, \bibinfo {author} {\bibfnamefont {F.~J.}\ \bibnamefont {Castander}}, \bibinfo {author} {\bibfnamefont {M.}~\bibnamefont {Charles}}, \bibinfo {author} {\bibfnamefont {E.}~\bibnamefont {Chaussidon}}, \bibinfo {author} {\bibfnamefont {J.}~\bibnamefont {Chaves-Montero}}, \bibinfo {author} {\bibfnamefont {D.}~\bibnamefont {Chebat}}, \bibinfo {author} {\bibfnamefont {T.}~\bibnamefont {Claybaugh}}, \bibinfo {author} {\bibfnamefont {S.}~\bibnamefont {Cole}}, \bibinfo {author} {\bibfnamefont {A.}~\bibnamefont {Cuceu}}, \bibinfo {author} {\bibfnamefont {K.~S.}\ \bibnamefont {Dawson}}, \bibinfo {author} {\bibfnamefont {A.}~\bibnamefont {de~la Macorra}}, \bibinfo {author} {\bibfnamefont {A.}~\bibnamefont {de~Mattia}}, \bibinfo {author} {\bibfnamefont {N.}~\bibnamefont {Deiosso}}, \bibinfo {author} {\bibfnamefont {R.}~\bibnamefont {Demina}}, \bibinfo {author} {\bibfnamefont {A.}~\bibnamefont
  {Dey}}, \bibinfo {author} {\bibfnamefont {B.}~\bibnamefont {Dey}}, \bibinfo {author} {\bibfnamefont {Z.}~\bibnamefont {Ding}}, \bibinfo {author} {\bibfnamefont {P.}~\bibnamefont {Doel}}, \bibinfo {author} {\bibfnamefont {D.~J.}\ \bibnamefont {Eisenstein}}, \bibinfo {author} {\bibfnamefont {W.}~\bibnamefont {Elbers}}, \bibinfo {author} {\bibfnamefont {S.}~\bibnamefont {Ferraro}}, \bibinfo {author} {\bibfnamefont {A.}~\bibnamefont {Font-Ribera}}, \bibinfo {author} {\bibfnamefont {J.~E.}\ \bibnamefont {Forero-Romero}}, \bibinfo {author} {\bibfnamefont {L.~H.}\ \bibnamefont {Garrison}}, \bibinfo {author} {\bibfnamefont {E.}~\bibnamefont {Gaztañaga}}, \bibinfo {author} {\bibfnamefont {H.}~\bibnamefont {Gil-Marín}}, \bibinfo {author} {\bibfnamefont {S.~G.~A.}\ \bibnamefont {Gontcho}}, \bibinfo {author} {\bibfnamefont {A.~X.}\ \bibnamefont {Gonzalez-Morales}}, \bibinfo {author} {\bibfnamefont {G.}~\bibnamefont {Gutierrez}}, \bibinfo {author} {\bibfnamefont {J.}~\bibnamefont {Guy}}, \bibinfo {author}
  {\bibfnamefont {C.}~\bibnamefont {Hahn}}, \bibinfo {author} {\bibfnamefont {M.}~\bibnamefont {Herbold}}, \bibinfo {author} {\bibfnamefont {H.~K.}\ \bibnamefont {Herrera-Alcantar}}, \bibinfo {author} {\bibfnamefont {K.}~\bibnamefont {Honscheid}}, \bibinfo {author} {\bibfnamefont {C.}~\bibnamefont {Howlett}}, \bibinfo {author} {\bibfnamefont {S.}~\bibnamefont {Juneau}}, \bibinfo {author} {\bibfnamefont {R.}~\bibnamefont {Kehoe}}, \bibinfo {author} {\bibfnamefont {D.}~\bibnamefont {Kirkby}}, \bibinfo {author} {\bibfnamefont {T.}~\bibnamefont {Kisner}}, \bibinfo {author} {\bibfnamefont {A.}~\bibnamefont {Kremin}}, \bibinfo {author} {\bibfnamefont {O.}~\bibnamefont {Lahav}}, \bibinfo {author} {\bibfnamefont {C.}~\bibnamefont {Lamman}}, \bibinfo {author} {\bibfnamefont {M.}~\bibnamefont {Landriau}}, \bibinfo {author} {\bibfnamefont {L.~L.}\ \bibnamefont {Guillou}}, \bibinfo {author} {\bibfnamefont {A.}~\bibnamefont {Leauthaud}}, \bibinfo {author} {\bibfnamefont {M.~E.}\ \bibnamefont {Levi}}, \bibinfo {author}
  {\bibfnamefont {Q.}~\bibnamefont {Li}}, \bibinfo {author} {\bibfnamefont {C.}~\bibnamefont {Magneville}}, \bibinfo {author} {\bibfnamefont {M.}~\bibnamefont {Manera}}, \bibinfo {author} {\bibfnamefont {P.}~\bibnamefont {Martini}}, \bibinfo {author} {\bibfnamefont {A.}~\bibnamefont {Meisner}}, \bibinfo {author} {\bibfnamefont {J.}~\bibnamefont {Mena-Fernández}}, \bibinfo {author} {\bibfnamefont {R.}~\bibnamefont {Miquel}}, \bibinfo {author} {\bibfnamefont {J.}~\bibnamefont {Moustakas}}, \bibinfo {author} {\bibfnamefont {D.~M.}\ \bibnamefont {Santos}}, \bibinfo {author} {\bibfnamefont {A.}~\bibnamefont {Muñoz-Gutiérrez}}, \bibinfo {author} {\bibfnamefont {A.~D.}\ \bibnamefont {Myers}}, \bibinfo {author} {\bibfnamefont {S.}~\bibnamefont {Nadathur}}, \bibinfo {author} {\bibfnamefont {G.}~\bibnamefont {Niz}}, \bibinfo {author} {\bibfnamefont {H.~E.}\ \bibnamefont {Noriega}}, \bibinfo {author} {\bibfnamefont {E.}~\bibnamefont {Paillas}}, \bibinfo {author} {\bibfnamefont {N.}~\bibnamefont
  {Palanque-Delabrouille}}, \bibinfo {author} {\bibfnamefont {W.~J.}\ \bibnamefont {Percival}}, \bibinfo {author} {\bibfnamefont {M.~M.}\ \bibnamefont {Pieri}}, \bibinfo {author} {\bibfnamefont {C.}~\bibnamefont {Poppett}}, \bibinfo {author} {\bibfnamefont {F.}~\bibnamefont {Prada}}, \bibinfo {author} {\bibfnamefont {A.}~\bibnamefont {Pérez-Fernández}}, \bibinfo {author} {\bibfnamefont {I.}~\bibnamefont {Pérez-Ràfols}}, \bibinfo {author} {\bibfnamefont {C.}~\bibnamefont {Ramírez-Pérez}}, \bibinfo {author} {\bibfnamefont {M.}~\bibnamefont {Rashkovetskyi}}, \bibinfo {author} {\bibfnamefont {C.}~\bibnamefont {Ravoux}}, \bibinfo {author} {\bibfnamefont {A.~J.}\ \bibnamefont {Ross}}, \bibinfo {author} {\bibfnamefont {G.}~\bibnamefont {Rossi}}, \bibinfo {author} {\bibfnamefont {V.}~\bibnamefont {Ruhlmann-Kleider}}, \bibinfo {author} {\bibfnamefont {L.}~\bibnamefont {Samushia}}, \bibinfo {author} {\bibfnamefont {E.}~\bibnamefont {Sanchez}}, \bibinfo {author} {\bibfnamefont {D.}~\bibnamefont {Schlegel}},
  \bibinfo {author} {\bibfnamefont {M.}~\bibnamefont {Schubnell}}, \bibinfo {author} {\bibfnamefont {H.}~\bibnamefont {Seo}}, \bibinfo {author} {\bibfnamefont {F.}~\bibnamefont {Sinigaglia}}, \bibinfo {author} {\bibfnamefont {D.}~\bibnamefont {Sprayberry}}, \bibinfo {author} {\bibfnamefont {T.}~\bibnamefont {Tan}}, \bibinfo {author} {\bibfnamefont {G.}~\bibnamefont {Tarlé}}, \bibinfo {author} {\bibfnamefont {P.}~\bibnamefont {Taylor}}, \bibinfo {author} {\bibfnamefont {W.}~\bibnamefont {Turner}}, \bibinfo {author} {\bibfnamefont {M.}~\bibnamefont {Vargas-Magaña}}, \bibinfo {author} {\bibfnamefont {M.}~\bibnamefont {Walther}}, \bibinfo {author} {\bibfnamefont {B.~A.}\ \bibnamefont {Weaver}}, \bibinfo {author} {\bibfnamefont {M.}~\bibnamefont {Wolfson}}, \bibinfo {author} {\bibfnamefont {C.}~\bibnamefont {Yèche}}, \bibinfo {author} {\bibfnamefont {P.}~\bibnamefont {Zarrouk}}, \bibinfo {author} {\bibfnamefont {R.}~\bibnamefont {Zhou}},\ and\ \bibinfo {author} {\bibfnamefont {H.}~\bibnamefont {Zou}},\ }\href
  {https://arxiv.org/abs/2503.14743} {\bibinfo {title} {Extended dark energy analysis using desi dr2 bao measurements}} (\bibinfo {year} {2025}),\ \Eprint {https://arxiv.org/abs/2503.14743} {arXiv:2503.14743 [astro-ph.CO]} \BibitemShut {NoStop}%
\bibitem [{\citenamefont {Cereskaite}\ \emph {et~al.}(2025)\citenamefont {Cereskaite}, \citenamefont {Mueller}, \citenamefont {Howlett}, \citenamefont {Davis}, \citenamefont {Aguilar}, \citenamefont {Ahlen}, \citenamefont {Bianchi}, \citenamefont {Brooks}, \citenamefont {Castander}, \citenamefont {Claybaugh}, \citenamefont {Cuceu}, \citenamefont {de~la Macorra}, \citenamefont {Ferraro}, \citenamefont {Font-Ribera}, \citenamefont {Forero-Romero}, \citenamefont {Gaztanaga}, \citenamefont {Gutierrez}, \citenamefont {Hahn}, \citenamefont {Honscheid}, \citenamefont {Huterer}, \citenamefont {Ishak}, \citenamefont {Joyce}, \citenamefont {Juneau}, \citenamefont {Kirkby}, \citenamefont {Kremin}, \citenamefont {Lahav}, \citenamefont {Lambert}, \citenamefont {Landriau}, \citenamefont {Guillou}, \citenamefont {Manera}, \citenamefont {Meisner}, \citenamefont {Miquel}, \citenamefont {Moustakas}, \citenamefont {Nadathur}, \citenamefont {Newman}, \citenamefont {Palanque-Delabrouille}, \citenamefont {Percival}, \citenamefont
  {Prada}, \citenamefont {Perez-Rafols}, \citenamefont {Rossi}, \citenamefont {Sanchez}, \citenamefont {Schlegel}, \citenamefont {Schubnell}, \citenamefont {Seo}, \citenamefont {Silber}, \citenamefont {Sprayberry}, \citenamefont {Tarle}, \citenamefont {Weaver}, \citenamefont {Zarrouk}, \citenamefont {Zhou},\ and\ \citenamefont {Zou}}]{cereskaite2025inferencematterpowerspectrum}%
  \BibitemOpen
  \bibfield  {author} {\bibinfo {author} {\bibfnamefont {R.}~\bibnamefont {Cereskaite}}, \bibinfo {author} {\bibfnamefont {E.}~\bibnamefont {Mueller}}, \bibinfo {author} {\bibfnamefont {C.}~\bibnamefont {Howlett}}, \bibinfo {author} {\bibfnamefont {T.~M.}\ \bibnamefont {Davis}}, \bibinfo {author} {\bibfnamefont {J.}~\bibnamefont {Aguilar}}, \bibinfo {author} {\bibfnamefont {S.}~\bibnamefont {Ahlen}}, \bibinfo {author} {\bibfnamefont {D.}~\bibnamefont {Bianchi}}, \bibinfo {author} {\bibfnamefont {D.}~\bibnamefont {Brooks}}, \bibinfo {author} {\bibfnamefont {F.~J.}\ \bibnamefont {Castander}}, \bibinfo {author} {\bibfnamefont {T.}~\bibnamefont {Claybaugh}}, \bibinfo {author} {\bibfnamefont {A.}~\bibnamefont {Cuceu}}, \bibinfo {author} {\bibfnamefont {A.}~\bibnamefont {de~la Macorra}}, \bibinfo {author} {\bibfnamefont {S.}~\bibnamefont {Ferraro}}, \bibinfo {author} {\bibfnamefont {A.}~\bibnamefont {Font-Ribera}}, \bibinfo {author} {\bibfnamefont {J.~E.}\ \bibnamefont {Forero-Romero}}, \bibinfo {author}
  {\bibfnamefont {E.}~\bibnamefont {Gaztanaga}}, \bibinfo {author} {\bibfnamefont {G.}~\bibnamefont {Gutierrez}}, \bibinfo {author} {\bibfnamefont {C.}~\bibnamefont {Hahn}}, \bibinfo {author} {\bibfnamefont {K.}~\bibnamefont {Honscheid}}, \bibinfo {author} {\bibfnamefont {D.}~\bibnamefont {Huterer}}, \bibinfo {author} {\bibfnamefont {M.}~\bibnamefont {Ishak}}, \bibinfo {author} {\bibfnamefont {R.}~\bibnamefont {Joyce}}, \bibinfo {author} {\bibfnamefont {S.}~\bibnamefont {Juneau}}, \bibinfo {author} {\bibfnamefont {D.}~\bibnamefont {Kirkby}}, \bibinfo {author} {\bibfnamefont {A.}~\bibnamefont {Kremin}}, \bibinfo {author} {\bibfnamefont {O.}~\bibnamefont {Lahav}}, \bibinfo {author} {\bibfnamefont {A.}~\bibnamefont {Lambert}}, \bibinfo {author} {\bibfnamefont {M.}~\bibnamefont {Landriau}}, \bibinfo {author} {\bibfnamefont {L.~L.}\ \bibnamefont {Guillou}}, \bibinfo {author} {\bibfnamefont {M.}~\bibnamefont {Manera}}, \bibinfo {author} {\bibfnamefont {A.}~\bibnamefont {Meisner}}, \bibinfo {author} {\bibfnamefont
  {R.}~\bibnamefont {Miquel}}, \bibinfo {author} {\bibfnamefont {J.}~\bibnamefont {Moustakas}}, \bibinfo {author} {\bibfnamefont {S.}~\bibnamefont {Nadathur}}, \bibinfo {author} {\bibfnamefont {J.~A.}\ \bibnamefont {Newman}}, \bibinfo {author} {\bibfnamefont {N.}~\bibnamefont {Palanque-Delabrouille}}, \bibinfo {author} {\bibfnamefont {W.~J.}\ \bibnamefont {Percival}}, \bibinfo {author} {\bibfnamefont {F.}~\bibnamefont {Prada}}, \bibinfo {author} {\bibfnamefont {I.}~\bibnamefont {Perez-Rafols}}, \bibinfo {author} {\bibfnamefont {G.}~\bibnamefont {Rossi}}, \bibinfo {author} {\bibfnamefont {E.}~\bibnamefont {Sanchez}}, \bibinfo {author} {\bibfnamefont {D.}~\bibnamefont {Schlegel}}, \bibinfo {author} {\bibfnamefont {M.}~\bibnamefont {Schubnell}}, \bibinfo {author} {\bibfnamefont {H.}~\bibnamefont {Seo}}, \bibinfo {author} {\bibfnamefont {J.}~\bibnamefont {Silber}}, \bibinfo {author} {\bibfnamefont {D.}~\bibnamefont {Sprayberry}}, \bibinfo {author} {\bibfnamefont {G.}~\bibnamefont {Tarle}}, \bibinfo {author}
  {\bibfnamefont {B.~A.}\ \bibnamefont {Weaver}}, \bibinfo {author} {\bibfnamefont {P.}~\bibnamefont {Zarrouk}}, \bibinfo {author} {\bibfnamefont {R.}~\bibnamefont {Zhou}},\ and\ \bibinfo {author} {\bibfnamefont {H.}~\bibnamefont {Zou}},\ }\href {https://arxiv.org/abs/2507.16590} {\bibinfo {title} {Inference of matter power spectrum at z=0 using desi dr1 full-shape data}} (\bibinfo {year} {2025}),\ \Eprint {https://arxiv.org/abs/2507.16590} {arXiv:2507.16590 [astro-ph.CO]} \BibitemShut {NoStop}%
\bibitem [{\citenamefont {Akaike}(1974)}]{1100705}%
  \BibitemOpen
  \bibfield  {author} {\bibinfo {author} {\bibfnamefont {H.}~\bibnamefont {Akaike}},\ }\href {https://doi.org/10.1109/TAC.1974.1100705} {\bibfield  {journal} {\bibinfo  {journal} {IEEE Transactions on Automatic Control}\ }\textbf {\bibinfo {volume} {19}},\ \bibinfo {pages} {716} (\bibinfo {year} {1974})}\BibitemShut {NoStop}%
\bibitem [{\citenamefont {Schwarz}(1978)}]{10.1214/aos/1176344136}%
  \BibitemOpen
  \bibfield  {author} {\bibinfo {author} {\bibfnamefont {G.}~\bibnamefont {Schwarz}},\ }\href {https://doi.org/10.1214/aos/1176344136} {\bibfield  {journal} {\bibinfo  {journal} {The Annals of Statistics}\ }\textbf {\bibinfo {volume} {6}},\ \bibinfo {pages} {461 } (\bibinfo {year} {1978})}\BibitemShut {NoStop}%
\bibitem [{\citenamefont {Li}\ \emph {et~al.}(2023)\citenamefont {Li}, \citenamefont {Zhang},\ and\ \citenamefont {Liang}}]{10.1093/mnras/stad838}%
  \BibitemOpen
  \bibfield  {author} {\bibinfo {author} {\bibfnamefont {Z.}~\bibnamefont {Li}}, \bibinfo {author} {\bibfnamefont {B.}~\bibnamefont {Zhang}},\ and\ \bibinfo {author} {\bibfnamefont {N.}~\bibnamefont {Liang}},\ }\href {https://doi.org/10.1093/mnras/stad838} {\bibfield  {journal} {\bibinfo  {journal} {Monthly Notices of the Royal Astronomical Society}\ }\textbf {\bibinfo {volume} {521}},\ \bibinfo {pages} {4406} (\bibinfo {year} {2023})},\ \Eprint {https://arxiv.org/abs/https://academic.oup.com/mnras/article-pdf/521/3/4406/49688935/stad838.pdf} {https://academic.oup.com/mnras/article-pdf/521/3/4406/49688935/stad838.pdf} \BibitemShut {NoStop}%
\end{thebibliography}
%
\end{document}